%% file: steering-review.tex
\documentclass[a4paper,aps,twocolumn,nobibnotes,rmp,superscriptaddress,longbibliography]{revtex4-1}
\usepackage{color}
\usepackage{graphicx}
\usepackage{amssymb}
\usepackage{amsmath}
\usepackage{bm}
\usepackage{verbatim}
\usepackage[utf8]{inputenc}
\usepackage{bbm}

\usepackage[mathscr,mathcal]{euscript}
\newcommand{\mean}[1]{\ensuremath{\langle#1\rangle}}
\newcommand{\ket}[1]{\ensuremath{|#1\rangle}}
\newcommand{\bra}[1]{\ensuremath{\langle#1|}}
\newcommand{\braket}[2]{\ensuremath{\langle #1|#2\rangle}}
\newcommand{\ketbra}[2]{\ensuremath{| #1 \rangle \langle #2 |}}
\newcommand{\tr}{{\rm Tr}}

\newcommand{\dprod}[2]{\left\langle #1, #2\right\rangle}

\newcommand{\Tr}{\operatorname{Tr}}

\newcommand{\be}{\begin{eqnarray}}
\newcommand{\ee}{\end{eqnarray}}

\newcommand{\PP}{\ensuremath{\mathcal{P}}}

\usepackage{tikz}

\newcommand{\abs}[1]{{| #1 |}}
\newcommand{\norm}[1]{{\Vert #1 \Vert}}

\newcommand{\new}[1]{{#1}}

\begin{document}

\title{Quantum Steering
%: Correlations, Applications, and the Connection to Measurement Theory
}
%Quantum Steering: EPR correlations, their applications and the connection to measurement theory
%
\author{Roope Uola}%\email{roope.uola@gmail.com}
\affiliation{Naturwissenschaftlich-Technische Fakult\"at, Universit\"at Siegen, Walter-Flex-Stra{\ss}e 3, 57068 Siegen, Germany}
\affiliation{D\'{e}partement de Physique Appliqu\'{e}e, Universit\'{e}  de Gen\`{e}ve, CH-1211 Gen\`{e}ve, Switzerland}
\author{Ana C. S. Costa}%\email{ana.costa@physik.uni-siegen.de}
\affiliation{Naturwissenschaftlich-Technische Fakult\"at, Universit\"at Siegen, Walter-Flex-Stra{\ss}e 3, 57068 Siegen, Germany}
\affiliation{Department of Physics, Federal University of Paran\'a, 81531-980 Curitiba, PR, Brazil}
\author{H. Chau Nguyen}
\affiliation{Naturwissenschaftlich-Technische Fakult\"at, Universit\"at Siegen, Walter-Flex-Stra{\ss}e 3, 57068 Siegen, Germany}
\author{Otfried G\"uhne}%\email{otfried.guehne@uni-siegen.de}
\affiliation{Naturwissenschaftlich-Technische Fakult\"at, Universit\"at Siegen, Walter-Flex-Stra{\ss}e 3, 57068 Siegen, Germany}
\date{\today}

\begin{abstract}
Quantum correlations between two parties are essential for the argument 
of Einstein, Podolsky, and Rosen in favour of the incompleteness of quantum 
mechanics. Schrödinger noted that an essential point is the fact that one 
party can influence the wave function of the other party by performing 
suitable measurements. He called this phenomenon quantum steering and studied
its properties, but only in the last years this kind of quantum correlation
attracted significant interest in quantum information theory.
In this paper the theory of quantum steering is reviewed. 
First, the basic concepts of steering and local hidden state 
models are presented and their relation to entanglement and 
Bell nonlocality is explained. Then various criteria for 
characterizing steerability and structural results on the 
phenomenon are described. A detailed discussion is given 
on the connections between steering and incompatibility 
of quantum measurements. Finally, applications of steering 
in quantum information processing and further related 
topics are reviewed.
\end{abstract}

\maketitle

\tableofcontents

\section{Introduction}
\label{sec-intro}
\input{Intro/intro}

%------------------------------------------------------------------------------------
\section{Detection of steering}
\label{sec-detection}

In this part of the review, we discuss how steering can be verified 
in different scenarios. Mainly three cases can be distinguished. 
First, given some expectation values of the form $\mean{A_i \otimes B_j},$
one can ask whether these correlations can prove steerability. Second, 
one can consider the case where Bob's assemblage $\{\varrho_{a|x}\}$ is
given, and ask whether or not it can be explained by an LHS model. Finally, 
one can take a complete state $\varrho_{AB}$ and ask whether this state 
allows \new{seeing} the phenomenon of steering if Alice makes appropriate 
measurements. 

\subsection{Steering detection from correlations}
\label{sec-correlations}
\input{Detection/Correlations}

\subsection{Steering detection from state assemblages}
\label{sec-assemblages}
\input{Detection/State-assemblages}

\subsection{Steering detection from full information}
\label{sec-full-information}
\input{Detection/Full-information}

%------------------------------------------------------------------------------------
\section{Conceptual aspects of steering}
\label{sec-properties}
In this section we review results on the general properties and structures
of quantum steering. We start with a detailed discussion on the connection
between steering, entanglement, and Bell nonlocality. We also present in some
detail LHS models for different families of states. Then, we explain properties
like one-way steering, steering of bound entangled states, steering maps and 
the superactivation of steering.

	\subsection{Hierarchy of correlations}
	\label{sec-hierarchy}
	\input{Properties/Hierarchy}

	\subsection{Special states and their local hidden state models}
	\label{sec-special-states}
	\input{Properties/Special-states}

	\subsection{Steering and local filtering}
	\label{sec-filtering}
	\input{Properties/Filtering}
	
	\subsection{One-way steerable states}
	\label{sec-one-way}
	\input{Properties/One-way-steerable-states}

	\subsection{Steering with bound entangled states}
	\label{sec-bound-entangled}
	\input{Properties/Bound-entangled-states}

	\subsection{Steering maps and dimension-bounded steering}
	\label{sec-steering-maps}
	\input{Properties/Dimension-bounded-steering}

	\subsection{Superactivation of steering}
	\label{sec-superactivation}
	\input{Properties/Superactivation}

%------------------------------------------------------------------------------------
\section{Joint measurability and steering}
\label{sec-joint-measurability}
In this section we discuss the problem of joint measurability to which steering 
is related in a many-to-one manner~\cite{quintino14,uola14,uola15,kiukas17}. 
Joint measurability is a natural extension of commutativity for general measurements. 
Operationally it corresponds to the possibility of deducing the statistics of several 
measurements from the statistics of a single one. {The connection between the concepts of joint measurability and steering unlocks the technical machinery developed within the framework of quantum measurement theory to be used in the context of quantum correlations. More precisely, joint measurability has been studied extensively already a few decades before steering was formulated in its modern form.} We review the connection on three levels: joint measurability on Alice's side (pure states), on Bob's side (mixed states), and on the level of the incompatibility breaking quantum channels (Choi isomorphism). Moreover, we discuss in detail how known results on one field can be mapped to new ones on the other.

	\subsection{Measurement incompatibility}
	\input{Joint-measurability/Measurement-incompatibility}

	\subsection{Joint measurability on Alice's side}
	\label{sec-JMAlice}
	\input{Joint-measurability/JM-Alice}

	\subsection{Joint measurability on Bob's side}
	\label{sec-JMBob}
	\input{Joint-measurability/JM-Bob}
	
	\subsection{Incompatibility breaking quantum channels}
	\input{Joint-measurability/breaking-quantum-channels}
	
	\subsection{Further topics on incompatibility}
	\label{sec-Further}
	\input{Joint-measurability/Further-JM}

%------------------------------------------------------------------------------------
\section{Further topics and applications of steering}
\label{sec-applications}
In this section we discuss further aspects and applications of steering. 
We start with multipartite steering, steering of Gaussian states and 
temporal steering. Then we discuss applications of steering such as 
quantum key distribution, randomness certification or channel discrimination.
Finally, we review the resource theory of steering and the phenomenon of
post-quantum steering. 
	
	\subsection{Multipartite steering}
	\label{sec-multipartite-steering}
	\input{Applications/Multipartite-steering}

	\subsection{The steering ellipsoid}
	\label{sec-steering-ellipsoids}
	\input{Applications/Steering-ellipsoid}

	\subsection{Gaussian steering}
	\label{sec-gaussian}
	\subsubsection{A criterion for steering of Gaussian states}
	\input{Applications/Gaussian-steering}
	
	\subsubsection{Refining Gaussian steering with EPR-type observables}
	\label{sec-epr-revisited}
	\input{Applications/EPR-revisited}

	\subsection{Temporal and channel steering}
	\label{sec-temporal-steering}
	\input{Applications/Temporal-channel-steering}

	\subsection{Quantum key distribution}
	\label{sec-qkd}
	\input{Applications/Quantum-key-distribution}

	\subsection{Randomness certification}
	\label{sec-randomness}
	\input{Applications/Randomness-verification}

	\subsection{Subchannel discrimination}
	\label{sec-subchannel}
	\input{Applications/Subchannel-discrimination}

	\subsection{One-sided device-independent and device-independent quantification of measurement incompatibility}
	\label{sec-selftesting}
	\input{Applications/Self-testing}

	\subsection{Secret sharing}
	\label{sec-secretsharing}
	\input{Applications/Secret-sharing}

	\subsection{Quantum teleportation}
	\label{sec-teleportation}
	\input{Applications/Quantum-teleportation}

	\subsection{Resource theory of steering}
	\label{sec-resource}
	\input{Applications/Resource-theory}

	\subsection{Post-quantum steering}
	\label{sec-postquantum}
	\input{Applications/Post-quantum-steering}

	\new{
	\subsection{Historical aspects of steering}
	\label{sec-history}
	\input{Applications/History-steering}

	}
%------------------------------------------------------------------------------------
\section{Conclusion}
\input{Conclusion/Conclusion}

%\nocite{*} 

\bibliographystyle{apsrmp4-1}
\bibliography{references}

\end{document}

%% file: Intro/intro.tex
\subsection{Overview}

In 1935, Einstein, Podolsky and Rosen (EPR) presented their 
famous argument against the completeness of quantum 
mechanics \cite{epr}. In this argument, a two-particle state 
is considered, where one party can measure the position or 
momentum and the correlations of the state allow one to predict
the results of these measurements on the other party if the
same measurement is performed there. The EPR argument led to
long-lasting discussions, but already directly after its 
publication, Schrödinger noted an interesting phenomenon in 
the argument: The first party can, by choosing the measurement, 
steer the state on the other side into an eigenstate of position 
or momentum. This can not be used
to transmit information, but still Schrödinger considered
it to be {\it magic}. 

\new{The early works of Schrödinger \cite{schroedinger35discussion, 
schrodinger36} did not receive much attention (see also Section 
\ref{sec-history}). This changed in 2007, 
when a formulation in the language of quantum information processing 
was given and systematic criteria were developed \cite{wiseman07}.} In the modern 
view, steering denotes the impossibility to describe the conditional 
states at one party by a local hidden state model. As such, steering 
denotes a quantum correlation situated between entanglement and Bell 
nonlocality. In the following years, the theory of steering developed 
rapidly. It was noted that steering provides the natural formulation 
for discussing quantum information processing, if for some of the 
parties the measurement devices are not well characterized. Also, 
the concept of steering helped to understand and answer open questions 
in quantum information theory. An important example is here the 
construction of counterexamples to the Peres conjecture, which states
that certain weakly entangled states do not violate any Bell 
inequality. Finally, steering turned out to be closely related 
to the concept of joint measurability of generalized measurements 
in quantum mechanics. \new{More precisely, measurements that are not jointly
measurable are exactly the measurements that are useful to reveal the 
steering phenomenon.} This has sparked interest in the question 
in which sense measurements in quantum mechanics can be considered as
a resource. 

This review article aims to give an introduction to the concept and 
applications of quantum steering. Starting from the basic definitions,
we explain steering criteria and structural results on quantum steering.
We also discuss in some detail related concepts, such as quantum 
entanglement or the joint measurability of observables. \new{We focus 
on the conceptual and theoretical issues and on the finite-dimensional 
case} and mention experiments only very shortly. For discussing quantum 
steering, the tool of semidefinite programming has 
turned out to be useful. Concerning this, we only discuss the main formulations, 
concrete examples and algorithms can be found in a different review article
\cite{cavalcanti17}.

As mentioned, quantum steering is related to several other central concepts
in quantum theory, so it may be useful to the reader to mention related 
relevant literature here. First, a review on the quantitative aspects of 
the EPR argument can be found in \cite{reid09}. The phenomenon of entanglement 
is extensively discussed in \cite{horodecki09} and methods to characterize it 
in \cite{guehne09}. A detailed overview on Bell inequalities and their 
applications is given in \cite{brunner14}. Finally, the theory of quantum 
measurements is in depth developed in \cite{busch16}.

\new{
The structure of the current article is the following: In the remainder 
of this introduction we explain the idea of quantum steering and the 
main definitions. We also provide a short comparison with quantum 
entanglement and Bell nonlocality, as this is central for the further 
discussion. 

Section II presents different methods for the detection of quantum 
steering. We discuss in detail how steerability can be inferred, if 
some correlations or the complete quantum state is known. These methods 
are then used in Section III, where we describe key conceptual aspects of 
steering. This includes the discussion of one-way steering, the 
superactivation of steering, the steerability of bound entangled 
states and the construction of steering maps. In addition, we can then
present the relation to other types of quantum correlations in detail. 

Section IV deals with the connections between steering and the joint 
measurability of observables. We explain the concept of joint measurability 
and its various connections to steering. These connections allow one to 
transfer results from one topic to the other. Section V describes 
different applications of steering as well as further topics. This includes 
applications in quantum key distribution and randomness certification, but 
also topics like multiparticle steering, steering of Gaussian states, the steering
ellipsoid and historical aspects of steering. Finally, Section VI presents 
the conclusion and some open questions. 
}

%%%%%%%%%%%%%%%%%%%%%%%%%%%%%%%%%%%%%%%%%%%%%%%%%%%%%%%%%%%%%
\subsection{Steering as a formalization of the EPR argument}
%%%%%%%%%%%%%%%%%%%%%%%%%%%%%%%%%%%%%%%%%%%%%%%%%%%%%%%%%%%%%
\label{sec:formepr}

\begin{figure}[t!]
\includegraphics[width=0.99\columnwidth]{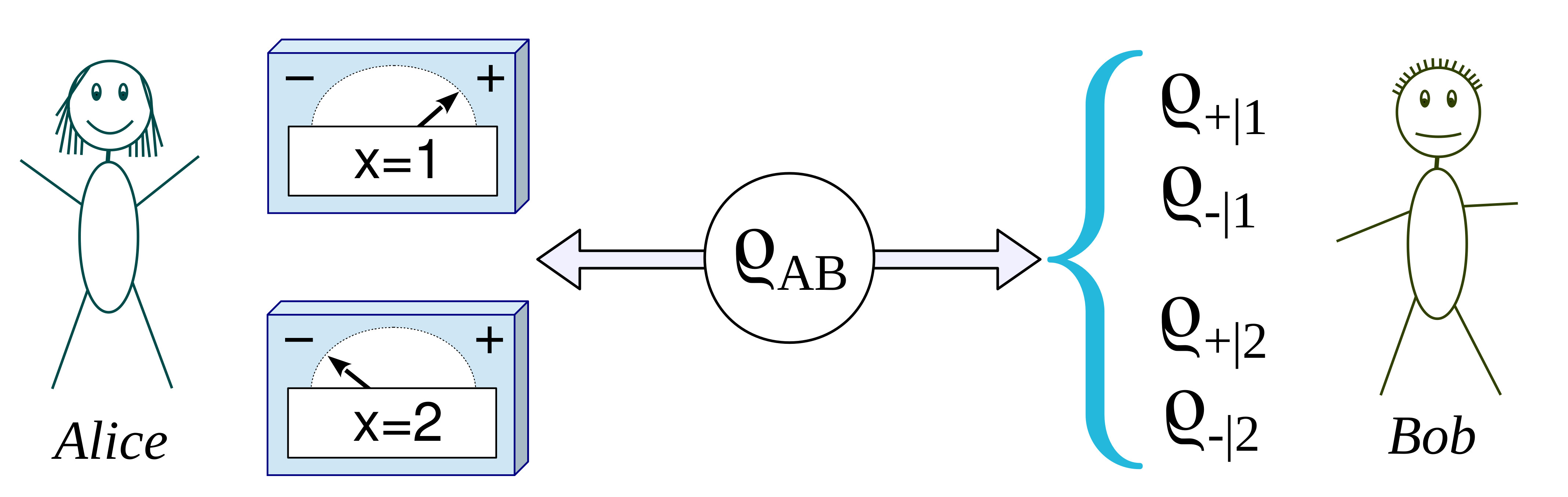}
\caption{Schematic description of the steering phenomenon: A state
$\varrho_{AB}$ is distributed between two parties. Alice 
performs a measurement (labeled by $x \in \{1,2 \}$) on her particle and obtains 
the result $a=\pm$. Bob receives the corresponding unnormalized conditional states
$\varrho_{a|x}.$ If Bob cannot explain this assemblage of states by 
assuming pre-existing states at his location, he has to believe that 
Alice can influence his state from a distance.}
\label{fig:steeringscheme}
\end{figure}

Let us start by recalling the EPR argument. Originally, EPR used
the position and momentum of two particles to explain their 
line of reasoning \cite{epr}, but in the simplest setting, the argument 
can be explained with two spin-$\frac{1}{2}$ particles or qubits 
\cite{bohm51}. Consider two particles that are in different locations 
and are controlled by 
Alice and Bob, see Fig.~\ref{fig:steeringscheme}. They are in the so-called 
singlet state, 
\begin{equation}
\ket{\psi}_{AB} = 
\frac{1}{\sqrt{2}}(\ket{01}-\ket{10}),
\label{eq-singlet}
\end{equation}
where $\ket{0}= \ket{z^+}$ and $\ket{1}= \ket{z^-}$ denote the two possible 
spin orientations in the $z$-direction. If Alice measures the spin of her 
particle in the $z$-direction and obtains the result $+1$ (or $-1$) then, 
due to the perfect anti-correlations of the singlet state, Bob's state will 
be either in the state $\ket{1}$ (or $\ket{0}$). Similarly, if Alice measures 
the spin in the $x$-direction, Bob's conditional states are given by 
$\ket{x^+} = (\ket{0}+\ket{1})/{\sqrt{2}}$, if Alice's result is $-1$ 
and $\ket{x^-} = (\ket{0}-\ket{1})/{\sqrt{2}}$ for the result $+1$.

So, by choosing her measurement setting, Alice can predict with certainty 
the values of a $z$- or $x$-measurement on Bob's side. 
According to EPR, this means that both observables must correspond to 
``elements of reality'', as each of them can be predicted in principle 
with certainty and without disturbing the system. This raises
problems if one assumes that the wave function is a complete description
of the physical situation, since the corresponding observables do not commute  
and the quantum mechanical formalism does not allow one to assign simultaneously 
definite values to both of them. Consequently, EPR concluded that quantum mechanics 
is incomplete.

Alice cannot transfer any information to Bob by choosing her measurement 
directions since Bob's reduced state is independent of this choice. But, 
as Schr\"odinger noted, she can determine whether the wave function on his 
side is in an eigenstate of the Pauli matrix $\sigma_x$ or $\sigma_z$. This 
{\it steering} of the wave function is, in Schr\"odinger's own words, 
``magic'', as it forces Bob to believe that Alice can influence his 
particle \new{from a distance, see also Section \ref{sec-history} for details.}

The situation for general quantum states other than the singlet state
can be formalized as follows \cite{wiseman07}: Alice and Bob share a 
bipartite quantum state $\varrho_{AB}$ and Alice performs different 
measurements, which do not need to be projective. For 
each of Alice's measurement setting $x$ and result $a$, Bob remains 
with an unnormalized conditional state $\varrho_{a|x}$. The set of these states is
called the steering assemblage, and the conditional states obey the 
condition $\sum_a \varrho_{a|x} = \varrho_B$, meaning that the reduced 
state $\varrho_B=\tr_A(\varrho_{AB})$ on Bob's side is independent of 
Alice's choice of measurements. 

After characterizing the states $\varrho_{a|x}$, Bob may try to explain 
their appearance as follows: He assumes that initially his particle was 
in some \new{hidden} state $\sigma_\lambda$ with probability $p(\lambda)$, 
parametrized by some parameter \new{(or hidden variable)} $\lambda$. Then, 
Alice's measurement and result just gave him additional information on the 
probability of the states. This leads to states of the form \cite{wiseman07}
\begin{align}
\varrho_{a|x} & 
= p(a|x) \int d\lambda p(\lambda|a,x) \sigma_\lambda 
\nonumber
\\
&= \int d\lambda p(\lambda) p(a|x, \lambda) \sigma_\lambda.
\label{eq-lhsmodel}
\end{align}
\new{The equivalence between these two expressions is easy 
to verify if the setting $x$ can be chosen freely and does 
not depend on the parameter $\lambda$, i.e., $p(x,\lambda) = p(x)p(\lambda).$
The two representations, however, point at different 
interpretations.

The first representation can be interpreted as if the 
probability distribution $p(\lambda)$ is just updated 
to $p(\lambda|a,x)$, depending on the classical 
information about the result $a$ and setting $x$. Here, 
Bob does not need to believe that Alice has control over
his state, her measurements and results just gave him 
additional information about the distribution of the 
states $\sigma_\lambda.$ 

The second representation can be interpreted as a simulation
task. Here, Alice can simulate the state $\varrho_{a|x}$ by 
drawing the states $\sigma_\lambda$ according to the distribution 
$p(\lambda)$ and, at the same time, announcing the result $a$ 
depending on the known setting $x$ and the parameter $\lambda$. 
Consequently, Bob does not need to believe that the initial state
shared by him with Alice was entangled.}

Generally, if a representation as in Eq.~(\ref{eq-lhsmodel}) exists, 
Bob does not need to assume any kind of action at a distance to explain 
the post-measurement states $\varrho_{a|x}$. Consequently, he does not 
need to believe that Alice can steer his state by her measurements and 
one also says that the state $\varrho_{AB}$ is  \emph{unsteerable} or 
has a local hidden state (LHS) model. If such a model does not exist, 
Bob is required to believe that Alice can steer the state in his 
laboratory by some ``action at a distance''. In this case, the 
state is said to be \emph{steerable}. \new{Note that steerability is an 
inherently asymmetric correlation, there are states where Alice can
steer Bob but not the other way round, see also Section \ref{sec-one-way}.}

For the wave function in Eq.~(\ref{eq-singlet}) the corresponding assemblage
is formed by the states $\ketbra{0}{0}/2$, $\ketbra{1}{1}/2$, $\ketbra{x^+}{x^+}/2$, 
and $\ketbra{x^-}{x^-}/2$ and one can directly see that no LHS model exists: 
The four conditional states are, up to normalization, pure and thus cannot be 
mixtures of other states. Thus, the occurring normalized $\sigma_\lambda$ have 
to be proportional to the four conditional states. So
Eq.~(\ref{eq-lhsmodel}) implies that $\ketbra{\eta}{\eta}/2 = 
\int d\lambda p(\lambda) p(a|x,\lambda) \sigma_\lambda$ for all four 
$\ketbra{\eta}{\eta}$ and $\sigma_\lambda$ coming from the set 
$\{ \ketbra{0}{0}, \ketbra{1}{1}, \ketbra{x^+}{x^+},\ketbra{x^-}{x^-}\}$. 
As mixtures are excluded, one must have $p(a|x,\lambda)=1$ if the 
$\sigma_\lambda$ corresponds to $\varrho_{a|x}$ and therefore 
$p(\lambda)=1/2$ for all $\lambda$. But then the probability distribution $p(\lambda)$ 
cannot be normalized.

For general states and measurements, however, the existence of an LHS model is not 
straightforward to decide. This leads to the question of how one can decide for a
given state $\varrho_{AB}$ or a given assemblage $\{\varrho_{a|x}\}$ whether it
is steerable or not, and this is one of the central questions of the present review
article.

%%%%%%%%%%%%%%%%%%%%%%%%%%%%%%%%%%%%%%%%%%%%%%%%%%%%%%%%%%
\subsection{Steering, Bell \new{nonlocality} and entanglement}
%%%%%%%%%%%%%%%%%%%%%%%%%%%%%%%%%%%%%%%%%%%%%%%%%%%%%%%%%%
\label{sec:steerbell}

There is another way to motivate the definition of steering and steerable 
correlations as in Eq.~(\ref{eq-lhsmodel}). For that, we shortly have to
explain the notions of local hidden variable (LHV) models and entanglement.

In a general Bell experiment, Alice and Bob perform measurements on their
particles, denoted by $A_x$ and $B_y$ and labeled by $x$ and $y$.  For the 
obtained results $a,b$ one asks whether their probabilities can be written as
\begin{equation}
p(a,b|x,y) =\int d\lambda p(\lambda) p(a|x,\lambda) p(b|y,\lambda).
\label{eq-lhvmodel}
\end{equation}
Such a description is known as an LHV model: The hidden variable $\lambda$
occurs with probability $p(\lambda)$ and Alice and Bob can compute the 
occurring joint probabilities with local response functions $p(a|x,\lambda)$ 
and $p(b|y,\lambda).$ For a given finite number of settings $x,y$ and outcomes $a,b$
the probabilities that can be written as in Eq.~(\ref{eq-lhvmodel}) form a 
high-dimensional polytope. The facets of the polytope are described by linear
inequalities, the so-called Bell inequalities. Quantum states can result in
probabilities that violate the Bell inequalities, but deciding, whether a given
state violates a Bell inequality or not is not straightforward and subject
of an entire field of research \cite{brunner14}. 

Let us now describe the notion of entanglement. In general, a state 
on a two-particle system is called separable, if it can be written 
as a convex combination of product states,
\begin{equation}
\varrho_{AB} = \sum_k p_k \varrho_k^A \otimes \varrho_k^B,
\label{eq-separabilitydefinition}
\end{equation}
otherwise it is called entangled. The separability of a quantum state is not
easy to decide, except for systems consisting of two qubits or one qubit and a 
qutrit, where the method of the partial transposition gives a necessary and sufficient
criterion, see also Section \ref{sec-bound-entangled}. 

For our discussion, it is important that the measurements on separable states 
clearly can be explained by an LHV model. A general measurement $M_x$ is given 
by a positive operator-valued measure (POVM). This means that one considers a
set of effects $E_{a|x}$ that are positive operators, $E_{a|x} \geq 0$, summing 
up to the identity $\sum_a E_{a|x} = \openone.$ The probability of the result
$a$ in a state $\varrho$  is computed according to $p(a) = \tr(\varrho E_{a|x}).$
Applying this to a separable state, one directly sees that the probabilities
of distributed measurements can be written as
\begin{equation}
p(a,b|x,y) = \sum_k  p_k \tr(E_{a|x} \varrho_k^A) \tr(E_{b|y} \varrho_k^B).
\label{eq-lhvmodel-separable}
\end{equation}
This is clearly an LHV model as in Eq.~(\ref{eq-lhvmodel}), with the extra condition
that the response functions $p(a|x,\lambda)$ and $p(b|y,\lambda)$ are coming from the
quantum mechanical description of measurements. 

Having Eqs.~(\ref{eq-lhvmodel}) and (\ref{eq-lhvmodel-separable}) in mind, one may 
ask whether the probabilities can also be described by a hybrid model, where Alice
has a general response function, while Bob's function is derived from the quantum
mechanical measurement rule. That is, one considers probabilities of the form
\begin{align}
 p(a,b|x,y) &= \int d\lambda
 p(\lambda) 
 p(a|x,\lambda) \tr(E_{b|y} \sigma_\lambda^B).
 \label{eq-lhvmodel-hybrid}
\end{align}
The point is that such probabilities are exactly the ones that occur in the
steering scenario. By linearity we can rewrite Eq.~(\ref{eq-lhvmodel-hybrid})
as
\begin{align}
 p(a,b|x,y) &= \tr(E_{b|y} \varrho_{a|x}),
\end{align}
where $\varrho_{a|x}=\int d\lambda p(\lambda) p(a|x,\lambda) \sigma_\lambda^B$
are the conditional states, allowing for an LHS model as in Eq.~(\ref{eq-lhsmodel}).

\begin{figure}[t!]
\includegraphics[width=0.9\columnwidth]{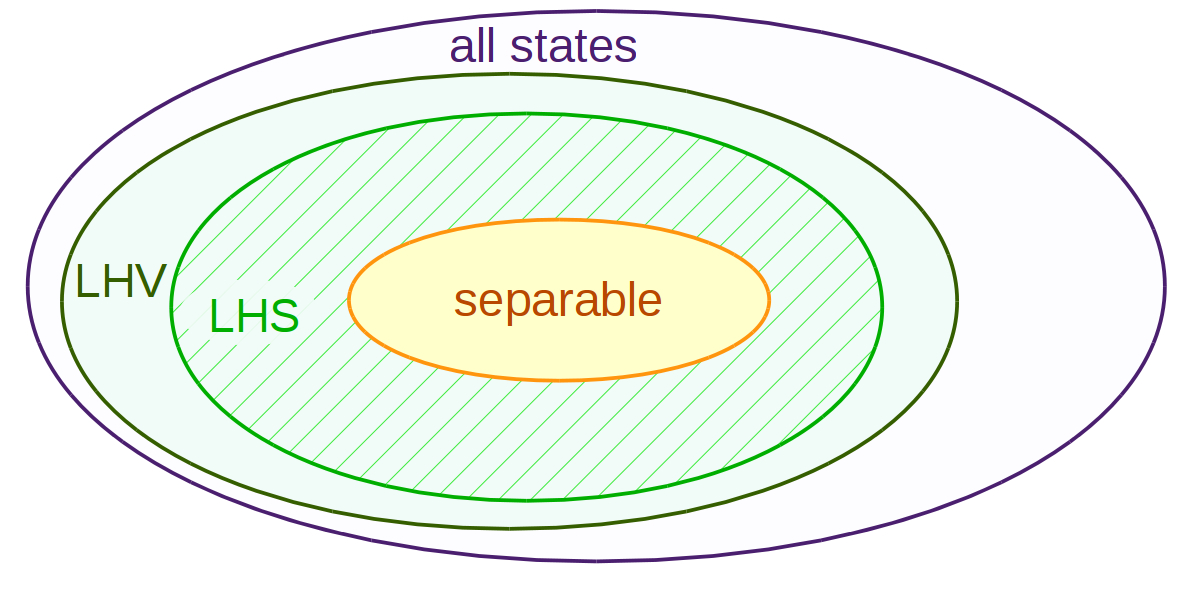}
\caption{Inclusion relation between entanglement, steering, and Bell inequality
violations. The set of all states is convex. The states which have an LHV model and
therefore do not violate any Bell inequality form a convex subset. The states with
an LHS model are unsteerable and form a convex subset of the LHV states. Finally, 
the separable states are a convex subset of the LHS states.}
\label{fig-entanglement-steering-bell-intro}
\end{figure}

We can conclude that the steering phenomenon relies on quantum correlations which
are between entanglement and violation of a Bell inequality. In fact, any state that
violates a Bell inequality can be used for steering, and any steerable state is 
entangled (see also Fig.~\ref{fig-entanglement-steering-bell-intro}). These 
inclusions are strict in the sense that there are entangled states that cannot 
be used for steering, and there are steerable states that do not violate any Bell 
inequality. In Section \ref{sec-hierarchy} we will discuss in detail the relation 
and the known examples of states in the various subsets.

\new{
It is important to note that the indicated hierarchy represents also different 
levels of trust in the measurement devices in entanglement verification. In 
general, in quantum information processing tasks such as cryptography, it makes
a difference whether or not one assumes that the measurement devices are well 
characterized. Completely uncharacterized devices can be seen as a black box, 
giving just some  measurement results without any knowledge about the quantum
description. There can also be situations where the devices are partly 
characterized, e.g., if the dimension of the quantum system is known, but not
the precise form of the measurement operators. 

Entanglement, steering, and Bell nonlocality correspond to different
levels of trust in the following sense. The usual schemes of entanglement 
verification, such as quantum state tomography or entanglement witnesses, 
require well-characterized measurement devices. The violation of
a Bell inequality, however, certifies the presence of entanglement 
without any assumption on the measurements or dimension of the system. 
Steering is between the two scenarios: If a state is steerable, its 
entanglement can be verified in a one-sided device-independent scenario, 
where Bob's measurements are characterized, but Alice's not. In some 
cases, the assumptions on Bob's system can even be relaxed, see 
Section \ref{sec-steering-maps} for an example.}

%% file: Detection/Correlations.tex
The simplest way to detect steering is to formulate criteria 
for the correlations between Alice's and Bob's measurement 
statistics. These can then be directly evaluated in experiments, 
without the need of reconstructing the whole assemblage. 

This approach of detecting steerability has a natural connection 
to the task of entanglement verification \cite{guehne09} and many 
concepts are similar to the case of entanglement detection. This 
includes linear criteria that are similar to entanglement witnesses, 
criteria based on variances or entropic uncertainty relations, and 
criteria similar to Bell inequalities. 

%%%%%%%%%%%%%%%%%%%%%%%%%%%%%%%%%%%%%%%%%%%%%%%%%
\subsubsection{Linear \new{and nonlinear} steering criteria}
%%%%%%%%%%%%%%%%%%%%%%%%%%%%%%%%%%%%%%%%%%%%%%%%%
Some of the typical ideas for deriving steering criteria are best 
explained with an example. Consider two qubits and the operator
\begin{equation}
Q = \sigma_x \otimes \sigma_x + \sigma_y \otimes \sigma_y +
\sigma_z \otimes \sigma_z.
\end{equation}
The question is which values $\mean{Q}$ can have for separable states. 
If one tries to maximize or minimize $\mean{Q}$ over separable states,
it suffices to consider product states of the form $\varrho_A \otimes 
\varrho_B$, as these are the extreme points of the separable states.
But for product states the single expectation values factorize and 
one has \cite{toth05a}
\begin{align}
|\mean{Q}|
& = 
|
\mean{\sigma_x}_A \mean{\sigma_x}_B +
\mean{\sigma_y}_A \mean{\sigma_y}_B +
\mean{\sigma_z}_A \mean{\sigma_z}_B 
|
\nonumber
\\
&
\leq
\Vert \vec{a} \Vert
\Vert \vec{b} \Vert
\leq 1,
\end{align}
with $ \vec{a} = (\mean{\sigma_x}_A, \mean{\sigma_y}_A, \mean{\sigma_z}_A)$
and $\vec{b}$ defined \new{analogously}. Here, first the Cauchy-Schwarz 
inequality was used, and then the fact that for single qubit states 
$\mean{\sigma_x}^2 + \mean{\sigma_y}^2+\mean{\sigma_z}^2 \leq 1$ holds. 
For the singlet state, however, $\mean{Q}=-3$. So the operator
$\mathcal{W}= \openone + \new{Q}$ is an entanglement witness, as it 
has a positive mean value on all separable states, but is negative on 
some entangled states.

If one wishes to estimate $\mean{Q}$ for unsteerable states, then, in view
of Eq.~(\ref{eq-lhvmodel-hybrid}), it suffices to consider product 
distributions again. This time, however, only Bob's results are 
described by quantum mechanics, so only the norm $\Vert \vec{b} \Vert \leq 1$
is bounded, while $\Vert \vec{a} \Vert = \sqrt{3}$ is possible. So,
$|\mean{Q}| \leq \sqrt{3}$ is a valid steering inequality that allows \new{detecting} the steerability of the singlet state \cite{cavalcanti09}.

A possible modification and generalization is the following: Consider
$N$ measurements $A_k$ on Alice's side which can take the two values 
$\pm 1$ and arbitrary observables $B_k$ on Bob's side. Then, for 
unsteerable states \cite{saunders10}
\begin{equation}
\sum_{k=1}^N 
\left|\langle A_k \otimes B_k \rangle\right| 
\leq 
\max_{\{a_k\}}\Big[\lambda_{\max}\big(\sum_{k=1}^N a_k B_k\big)\Big],
\label{eq-corr-linear}
\end{equation}
where $\lambda_{\max}(X)$ denotes the largest eigenvalue of $X$ and $a_k = \pm 1$. 
To prove this bound, it suffices to consider a product distribution as above, 
then each $A_k$ can just change the signs of the $B_k$, and the mean value 
of the resulting sum is bounded by the maximal eigenvalue.

A different kind of generalization uses expectation values on Bob's side that 
are conditioned on Alice's outcome. If Alice makes a measurement $A_k$ with 
possible results labeled by $a$, one can denote with $\mean{B}|_a$ the mean 
value of a measurement on Bob's side, conditioned on the outcome $a$. Then one 
can consider the nonlinear expression
\begin{equation}
T_x^{(k)} = \sum_a p(a|k) (\mean{\sigma_x}|_a)^2
\label{eq-steer-c1}
\end{equation}
and summing this up for three Pauli measurements gives the bound
for unsteerable states \cite{wittmann12}
\begin{equation}
T_x^{(1)} + T_y^{(2)} + T_z^{(3)} \leq 1.
\label{eq-steer-c2}
\end{equation}
This follows by considering product distributions and 
$\mean{\sigma_x}^2 + \mean{\sigma_y}^2+\mean{\sigma_z}^2 \leq 1.$ 
Note that similar bounds on sums of squared mean values are known 
for many cases of anticommuting observables or mutually unbiased 
bases \cite{toth05b, wu09, wehner10}, so one can directly generalize 
the criteria from above to broader classes of observables on Bob's 
side \cite{evans13}. \new{With mutually unbiased bases as a generalization 
of the Pauli matrices one can even find steering inequalities with an 
unbounded violation \cite{marciniak15, rutkowski17}.}

For all the criteria from above the question arises, which are the best 
measurement directions for a given state in order to detect steering.
For the criterion in Eq.~(\ref{eq-corr-linear}) this has been studied
in \cite{evans14}. For criteria using Pauli matrices, one can still ask for
the best orientation of the coordinate system. In \cite{paternostro17} it
has been observed that often, but not always, the measurements that 
correspond to the semiaxes of the so-called steering ellipsoid (i.e., 
the ellipsoid of the potential conditional states in Bob's Bloch sphere 
considering all possible measurements for Alice, see Section 
\ref{sec-steering-ellipsoids}) give 
strong criteria. For higher-dimensional systems, a systematic study of 
optimal measurements in restrictive scenarios, i.e., in the context of 
$N$ measurements of $k$ outcomes, has been performed in~\cite{bavaresco17}.

So far, we have considered criteria that were motivated by concepts in
entanglement theory. A different method to design steering inequalities
for a given special scenario comes from the theory of semidefinite 
programs (SDPs). As we will see in Section \ref{sec-assemblages}, the question of whether
a given assemblage $\{\varrho_{a|x}\}$ is steerable can be decided via 
an SDP. The corresponding dual problem can then be considered as a linear
steering inequality. Further details are given in Section \ref{sec:assemblage_steering_inequalities}.

\new{
The discussed criteria or small variations thereof have been 
used in several experiments \cite{saunders10, wittmann12, 
smith12, bennet12, weston18}. In the experimental works, it 
is also important to close loopholes, such as the one arising 
from inefficient detectors. Theoretical aspects of this issue 
are in detail discussed in \cite{evans13, vallone13,jeon19} 
and experimentally studied in \cite{wittmann12, 
smith12, bennet12, weston18}. 
}

%%%%%%%%%%%%%%%%%%%%%%%%%%%%%%%%%%%%%%%%%%%%%%%%%%%%%%%%%%%%%%%%%%%%%%%
\subsubsection{Steering criteria from uncertainty relations}
%%%%%%%%%%%%%%%%%%%%%%%%%%%%%%%%%%%%%%%%%%%%%%%%%%%%%%%%%%%%%%%%%%%%%%%%

Steering inequalities based on uncertainty relations have been proposed
already long before the formal definition of steerability in the context
of the EPR argument \cite{reid89, reid09}. Also the criterion in 
Eqs.~(\ref{eq-steer-c1}, \ref{eq-steer-c2}) can be seen as a criterion
in terms of conditional variances. 

A systematic approach using entropic uncertainty relations (EURs) has been 
proposed in \cite{walborn11} for continuous variable systems and tailored
to discrete systems  by \cite{schneeloch13}. Here, we focus on the discrete
version. In general, a measurement $M$ results in a probability distribution 
$\PP = (p_1,\cdots,p_n)$ of the outcomes, for which one can consider the 
Shannon entropy $S(\PP) = - \sum_i p_i\log (p_i)$ as the entropy of the 
measurement $S(M).$ For two projective measurements given by the corresponding
hermitian operators $B_1=\sum_i \lambda_i
\ketbra{v_i}{v_i}$ and 
$B_2=\sum_i \mu_i\ketbra{w_i}{w_i}$  on Bob's side, one has
the general EUR \cite{maassen88}
\begin{equation}
\label{sec-corr-esc2}
S(B_1) + S(B_2) \geq - \ln (\Omega_B),
\end{equation}
where $\Omega_B = \max_{i,j} (|\langle v_i|w_j\rangle|^2)$ is the maximal
overlap between the eigenstates. This and similar EURs are central to 
quantum information theory and quantum cryptography \cite{coles17}.

For product measurements $A \otimes B$ on two particles, one can consider the 
joint distribution and the conditional Shannon entropy $S(B|A) = S(A,B) - S(A).$
Then, for unsteerable states the relation
\begin{equation}
S(B_1|A_1)+ S(B_2|A_2) \geq -\ln (\Omega_B)
\end{equation}
holds.  The intuition behind this criterion is that if Alice can predict from her 
measurement data Bob's measurement results better than the EUR allows, then there 
cannot be local quantum states for Bob that reproduce such measurement results.

A generalization of this criterion to other entropies has been 
developed~\cite{costa17, costa18}. The general approach works for 
any entropy with the following properties: (i) the entropy is (pseudo-)additive 
for independent distributions; (ii) one has a state independent EUR; and (iii) 
the corresponding relative entropy is jointly convex. The resulting criteria 
based on Tsallis entropy are typically stronger than the ones from Shannon 
entropy. In addition, \citet{krivachy18} obtained tight steering inequalities 
in terms of the R\'enyi entropy \new{for scenarios with two measurements per party}, and \citet{jia17} developed methods of detecting 
entanglement and steering based on universal uncertainty relations and fine-grained 
uncertainty relations using majorization. 

In the case of continuous variable systems, the entropic criteria proposed 
in \cite{walborn11} are connected to one of the first criteria 
mentioned above. In \cite{reid89} the question is addressed to which 
extent Alice can infer the value of Bob's position $X_B$ or momentum 
$P_B$ by measuring her own canonical variables. The best estimator 
of $X_B$ has as uncertainty the minimal variance 
$\delta^2_{\rm min}(X_B) = \int dx_A p(x_A) \delta^2(x_B|x_A)$, 
where $\delta^2(x_B|x_A)$ is the variance of the conditional probability 
distribution. Then, for quantum states that do not give rise to an EPR 
argument, the condition
\begin{equation}
\delta^2_{\rm min}(X_B)\delta^2_{\rm min}(P_B)\geq \frac{1}{4}
\label{eq-reid89}
\end{equation}
holds. Later, \citet{walborn11} showed the criterion 
$
S(X_B|X_A) + S(P_B|P_A) \geq \ln (\pi e)
$
and demonstrated that this implies Eq.~(\ref{eq-reid89}). In addition, \citet{walborn11}
reports the experimental observation of states, which can be detected by the entropic 
criterion, but not by Eq.~(\ref{eq-reid89}).

\new{Other experiments involving steering criteria from uncertainty 
relations have been reported in the case of continuous variable 
systems in \cite{bowen03}, and recently in the case of discrete 
systems \cite{wollmann19,yang19}.}

%%%%%%%%%%%%%%%%%%%%%%%%%%%%%%%%%%%%%%%55
\subsubsection{Steering and the CHSH inequality}
%%%%%%%%%%%%%%%%%%%%%%%%%%%%%%%%%%%%%%%%%
Given a certain set of measurements, one can ask for the optimal 
inequalities for detecting quantum correlations. For Bell nonlocality 
and the case of two measurements with two outcomes each, it is known 
that the \new{probabilities} allow an LHV description, if and only if the 
Clauser-Horne-Shimony-Holt (CHSH) inequality
\begin{equation}
\mean{A_1 B_1} +  \mean{A_1 B_2} +  \mean{A_2 B_1} -  \mean{A_2 B_2}  \leq 2
\end{equation}
holds \cite{fine82}, where also permutations of the measurements and 
outcomes have to be taken into account. \new{More precisely, the 
inequality implies that the probabilities of all outcomes for the 
measurements $A_i B_j$ can be explained by a LHV model. Note that 
these probabilities include more information than the full correlations 
$\mean{A_i B_j}$ only, as the marginals $\mean{A_i}$ and $\mean{B_j}$ 
are independent of the full correlations. In other words, if the CHSH 
inequality is fulfilled, there is also no two-setting Bell inequality with 
marginal terms that is violated.}

\new{Similar statements are known from entanglement theory. For 
instance, one can consider the situation of two qubits, where Alice 
and Bob perform each the two measurements $\sigma_x$ and $\sigma_z$
only, and not full tomography. For this scenario, all relevant 
entanglement witnesses have been characterized \cite{curty04}.}

For quantum steering one can consider also two measurements with two 
outcomes per party, where only the measurements of Bob may be characterized. 
First, one can consider the case that Bob has a qubit 
and performs two mutually unbiased measurements (e.g., two Pauli 
measurements). For this scenario, it was shown by \cite{cavalcanti15b} 
that the full correlations $\mean{A_i B_j}$ admit an LHS model, iff the 
inequality
\begin{align}
&\sqrt{\langle (A_1+A_2)B_1\rangle^2 +  \langle (A_1+A_2)B_2\rangle^2} 
\nonumber \\
&+ \sqrt{\langle (A_1-A_2)B_1\rangle^2 + \langle (A_1-A_2)B_2\rangle^2} \leq 2
\label{eq-steer-chsh}
\end{align}
holds. Note that the resulting inequality and the underlying problem has some 
similarity to Bell inequalities for orthogonal measurement directions for one 
or both parties \cite{uffink08}.

For the more general scenario, one has to distinguish carefully whether the 
LHS model should explain the full correlations $\mean{A_i B_j}$ only, or in 
addition the marginal distributions $\mean{A_i}$ and $\mean{B_j}$.

Concerning full correlations, in \cite{girdhar16} the case of uncharacterized 
projective measurements on Bob's qubit has been considered. First, two projective 
measurements $B_1$ and $B_2$ on a qubit define a plane on the Bloch sphere, and in 
this plane one can always find a third measurement $B_3$ such that $B_1$ and $B_3$ 
are mutually unbiased; moreover, the mean values of $B_1$ and $B_2$ can be obtained 
from the mean values of $B_1$ and $B_3$ and vice versa. Then, it was shown that 
$B_1$ and $B_2$ allow an LHS model iff Eq.~(\ref{eq-steer-chsh}) holds for $B_1$ 
and $B_3$. In addition, it was shown that if Eq.~(\ref{eq-steer-chsh}) is violated, 
then the state violates also the original CHSH inequality and is thus nonlocal, but
possibly for a different set of measurements (see also \cite{quan17} for an 
independent proof). Finally, also a characterization of POVMs with two outcomes 
has been given in \cite{girdhar16}.

At the same time, \citet{quan17} considered the question of whether full correlations
and marginals of two dichotomic measurements can be explained via an LHS model. 
For this case, the equivalence is not true anymore: There are two-qubit states, 
which do not violate any CHSH inequality, nevertheless no LHS model can explain
the full correlations and marginals of certain $A_1, A_2, B_1, B_2.$ One can also
find two-qubit states, for which steerability from Alice to Bob can be proved by
two measurements on each side, but steering from Bob to Alice is not possible
(see also Section \ref{sec-one-way}).
Finally, the interplay between steering and Bell inequality violation for
specific families of states has been discussed in \cite{costa16,quan16}.

%%%%%%%%%%%%%%%%%%%%%%%%%%%%%%%%%%%%%%%%%%%%
\subsubsection{Moment matrix approach}
\label{sec-moment-matrix}
%%%%%%%%%%%%%%%%%%%%%%%%%%%%%%%%%%%%%%%%%%%%
Another method that can be used for the characterization of quantum 
correlations consists of moment matrices or expectation value matrices. In 
general, one considers a set of operators of the form 
$M_k= \{A_{i_k} \otimes B_{j_k}\}$ 
and builds the matrix of expectation values 
\begin{equation}
\Gamma_{kl}=\langle M_k^\dagger M_l\rangle.
\end{equation}
The remaining task is to characterize the possible matrices $\Gamma$ that 
origin from unsteerable or separable states. Clearly, $\Gamma \geq 0,$ i.e., 
it has no negative eigenvalues.

This approach of moment matrices is a well-known tool in entanglement 
theory \cite{shchukin05,miranowicz09,haeseler08,moroder08}. There one can argue that the matrix $\Gamma$ inherits 
a separable structure so that approaches using the partial transposition 
can be applied. This also allows {characterization} of entanglement if some of the
entries $\Gamma_{kl}$ are not known or if the measurement devices are 
not trusted \cite{moroder13}.

Concerning steering, it follows from Eqs.~(\ref{eq-lhvmodel-separable}, 
\ref{eq-lhvmodel-hybrid}) that the correlations of unsteerable states
can be explained by an underlying separable state, \new{where} the measurements
of Alice are commuting \cite{kogias15b, moroder16}. The commutativity of Alice's 
measurements together with possible exploitation of the structure of 
Bob's characterised measurements (e.g., an algebraic structure such as 
the one of the Pauli spin operators) results in constraints on the moment 
matrix. In the end, for a given set of product operators $\{M_k\}$, one 
needs to check whether there exist (complex) parameters for the unknown 
entries of the moment matrix (such as squares of Alice's measurements) that 
make the matrix positive. As any moment matrix is positive, proving that 
such an assignment of parameters is not possible implies that the underlying 
state is steerable. The main result of \cite{kogias15b} can be then 
formulated as follows. For any unsteerable correlation experiment
\begin{equation}
\Gamma_R \geq 0\ \text{for some } R,
\end{equation}
where $\Gamma_R$ is the moment matrix $\Gamma$ for a set of parameters 
$R$ (fulfilling the requirements inherited from commutativity on Alice's 
side and possible further structure on Bob's side) as the unknown entries. 
In \cite{kogias15b} the authors further pointed out that checking 
the existence of such parameters forms a semidefinite program and provided 
various examples. Note that this approach can still be augmented by using
the separable structure of $\Gamma$.

In \cite{chen16b} the concept of a moment matrix has been used to characterize
steerability in a more refined way. Namely, one can also consider the moment
matrices $\Gamma_{a|x}$ for each state in the assemblage $\{\varrho_{a|x}\}.$
Using the methods from \cite{moroder13} this allows then to characterize and 
quantify steerability in a device independent way. 

%%%%%%%%%%%%%%%%%%%%%%%%%%%%%%%%%%%%%%%%%%%%%%%%%%%
\subsubsection{Steering criteria based on local uncertainty relations}
%%%%%%%%%%%%%%%%%%%%%%%%%%%%%%%%%%%%%%%%%%%%%%%%%%%
Local uncertainty relations (LURs) are a common tool for entanglement 
detection and the underlying idea can directly be generalized to steering 
detection. For the case of entanglement, the idea is the following: Consider 
observables $A_k$ on Alice's side, obeying an uncertainty relation 
$
\sum_k \delta^2 (A_k) \geq C_A,
$
where $\delta^2(X) = \langle X^2\rangle + \langle X\rangle^2$ denotes
the variance. An example of such a relation is 
$\sum_{i=x,y,z} \delta^2(\sigma_i)\geq 2$, for general observables 
such bounds can be computed systematically \cite{schwonnek17, huang12,maccone14}. 
Similarly, one can consider observables $B_k$ for Bob, fulfilling 
$
\sum_k \delta^2 (B_k) \geq C_B,
$
and the global observables $M_k = A_k\otimes \openone + \openone \otimes B_k.$
Then, for separable states, the bound
$
\sum_k \delta^2 (M_k) \geq C_A + C_B
$
holds \cite{hofmann03}. This is a very strong entanglement criterion and its 
properties have been studied in detail \cite{guehne06, zhang07, gittsovich08}.

For steering detection, we can use the same construction, the only difference is 
that Alice's measurements are not characterized, so no uncertainty relation for
them is available \cite{ji15, zhen16}. Consequently, unsteerable states obey
\begin{equation}
\sum_k \delta^2 (M_k) \geq  C_B.
\end{equation}

The criterion of the LURs can be formulated in terms of covariance matrices \cite{guehne07}, 
and this also works for steering \cite{ji15}. For a given quantum state $\varrho$ and 
observables $\{X_k\}$ the symmetric covariance matrix $\gamma$ is defined by their 
elements
$
\gamma_{ij} = (\langle X_i X_j\rangle + \langle X_j X_i\rangle)/2 
- \langle X_i\rangle \langle X_j\rangle.
$
If one considers in a composite system the set of observables 
$\{X_k\} = \{A_{i_k} \otimes \openone, \openone \otimes B_{j_k}\}$
then the covariance matrix has a block structure
\be
\gamma_{AB} = 
\left[
\begin{array}{cc}
A & C \\
C^T & B
\end{array}
\right],
\ee
where $A = \gamma (\varrho_A,\{A_i\})$ and $B = \gamma (\varrho_B,\{B_j\})$ 
are covariance matrices for the reduced states and $C$ is the correlation 
matrix with elements 
$C_{ij} = \langle A_i\otimes B_j\rangle - \langle A_i \rangle \langle B_j 
\rangle$. 

Given this type of covariance matrices for unsteerable states it holds that
\be
\gamma_{AB} \geq \mathbf{0}_A\oplus \kappa_B,
\label{eq-cm-steering}
\ee 
with $\kappa_B = \sum_k p_k \gamma (\ket{b_k}\bra{b_k})$ being a convex 
combination of covariance matrices of pure states on Bob's system. Here, 
$\mathbf{0}_A$ is a $m \times m$ null matrix, where $m$ is the number 
of observables on Alice's side. The characterization of the possible
$\kappa_B$ has been discussed for typical cases, such as qubit states
or local orthogonal observables \cite{gittsovich08}.

Finally, it should be noted that the criterion in Eq.~(\ref{eq-cm-steering}) 
is the discrete analog to a criterion for the continuous variable case, 
see also Section \ref{sec-gaussian}.

% [OK, INCorporated ana & chau]

%% file: Detection/State-assemblages.tex
\label{sec:assemblages}

When full knowledge of the unnormalized conditional ensembles $\varrho_{a|x}$ 
on Bob's side is available, steerability can be detected 
more efficiently. As already mentioned, a set of conditional 
ensembles on Bob's side corresponding to certain measurement 
settings from Alice is called a steering assemblage. As Alice's 
choice of measurement cannot be detected on Bob's side, such 
assemblages are non-signalling in that 
$\sum_a\varrho_{a|x}=\sum_a\varrho_{a|x'}$ for different settings 
$x,x'$. \new{It is interesting to note that,} for the bipartite case, 
any non-signalling assemblage can be prepared with some shared state 
and some measurements on Alice's side \cite{schrodinger36}, see also 
Sections \ref{sec-postquantum} and \ref{sec-history}.
\new{Also}, any unsteerable assemblage can be prepared using a separable 
state and commuting measurements on Alice's side \cite{kogias15b,moroder16}. 

The main point for steering detection is that for \new{\textit{finite}} steering assemblages, the question whether there exists an LHS model described by Eq.~\eqref{eq-lhsmodel} 
can be decided via the so-called semidefinite programming (SDP) technique~\cite{pusey13}. 
The SDP approach also allows one to derive steering inequalities and to quantify 
the steerability of finite assemblages.

%%%%%%%%%%%%%%%%%%%%%%%%%%%%%%%%%%%%%%%%%%%%%%%%%%%%%%%%
\subsubsection{Formulation of the semidefinite program}
\label{sec:assemblage_steering_sdp}
%%%%%%%%%%%%%%%%%%%%%%%%%%%%%%%%%%%%%%%%%%%%%%%%%%%%%%%%

The crucial idea is that for a finite steering assemblage 
$\{\varrho_{a|x}\}_{a,x}$, it is sufficient to consider a 
finite LHS ensemble $\sigma_\lambda$~\cite{ali09, pusey13}. 
Moreover, the response functions in Eq.~\eqref{eq-lhsmodel} 
can also be chosen to be fixed. So the remaining problem is to 
construct a finite number of positive operators $\sigma_\lambda$. 
We focus on the conceptual aspects of the SDP formulation, a detailed 
review on computational aspects can be found in \cite{cavalcanti17}.

Consider a set of $m$ measurement settings on Alice's side, $x \in \{1,2,\ldots,m\}$, 
each has $q$ outcomes $a \in \{1,2,\ldots,q\}$. \new{Given the shared state $\varrho$, this gives rise to an assemblage $\{\varrho_{a|x}\}_{a,x}$ of $m$ ensembles, each consisting of $q$ conditional states.} The space of the 
hidden variables $\lambda$ can be constructed as follows. The variable
$\lambda$ can take $q^m$ values, each can be thought of as a string 
of outcomes ordered according to the measurements, 
$(a_{x=1},a_{x=2},\cdots,a_{x=m})$. For such a string $\lambda$, we 
denote by $\lambda(x)$ the value of the outcome at position $x$. 
Then $D(a|x,\lambda)$ denotes the deterministic response function 
defined by $D(a|x,\lambda)=\delta_{a,\lambda(x)}$.
This means that $D(a|x,\lambda)$ equals one for strings 
$\lambda$ which predict the outcome $a$ for the measurement 
$x$ and zero otherwise. 

The crucial statement is the following: a finite steering assemblage 
admits an LHS model described by Eq.~\eqref{eq-lhsmodel} if and only 
if it also admits an LHS model with the constructed set of strings 
$\lambda$  as the LHV and the {fixed} deterministic functions 
$D(a|x,\lambda)$ as the response functions. The latter means that there 
exists a set of (unnormalized) operators $\sigma_{\lambda}$ satisfying 
\begin{align}
\varrho_{a|x} = & \sum_\lambda D(a|x,\lambda)\sigma_\lambda \qquad 
\mbox{ for all } a,x,
\nonumber\\
\textrm{s.t.:} \quad 
& \sigma_\lambda \geq 0 \quad 
\mbox{ for all }\lambda.
\label{eq:sdp_lhs_model}
\end{align}
\new{Writing with the explicit definitions of the hidden variable $\lambda$ and the deterministic response function $D(a|x,\lambda)$, the equality in the above equation is simply
\begin{equation}
\varrho_{a|x}=\sum_{\{a_i\}} \delta_{a,a_x} \sigma_{a_1,a_2,\ldots,a_m}.
\label{eq:sdp_lhs_model_explicit}
\end{equation}
Intuitively, one can think of the hidden states $\sigma_{a_1,a_2,\ldots,a_m}$ as being indexed by $m$ variables. The conditional state $\varrho_{a|x}$ is obtained by summing the function over the values of all variables except for the $x$-th one, which is fixed $a_x=a$.
}

To give an example, if one considers the case where Alice performs 
two measurements $x \in \{1,2\}$ with two possible outcomes 
$a \in \{\pm\}$, the steering assemblage $\{\varrho_{a|x}\}$ is 
unsteerable if and only if it is possible to find four positive 
semidefinite operators $\omega_{ij}$, with $i,j=\pm$ such that
\begin{align}
\varrho_{+|1} = \omega_{++} + \omega_{+-}, \qquad \varrho_{+|2} = \omega_{++} + \omega_{-+}, \nonumber \\
\varrho_{-|1} = \omega_{-+} + \omega_{--}, \qquad \varrho_{-|2} = \omega_{+-} + \omega_{--}
\label{eq-steeringmaps-cond1}.
\end{align}

\new{It is remarkable that} in passing from Eq.~\eqref{eq-lhsmodel} to 
Eq.~\eqref{eq:sdp_lhs_model}, we have passed from an arbitrary 
hidden variable to a finite discrete hidden variable and, at 
the same time, fixed the response functions. One notices that 
the finiteness of the set of measurements plays a crucial role 
in this approach.

Given an assemblage $\{\varrho_{a|x}\}_{a,x}$, determining the existence 
of $\sigma_\lambda$ satisfying Eq.~\eqref{eq:sdp_lhs_model} is in fact a 
well-known problem in convex optimization. More precisely, it is known 
as a feasibility problem in semidefinite programming (SDP)~\cite{boyd04}, \new{which can be solved straightforwardly by an appropriate ready-to-use software.} \new{Furthermore, it has been shown that using the so-called 
order-monotonic functions, the SDPs can be approximated by simpler ones~\cite{zhu16}.}

%%%%%%%%%%%%%%%%%%%%%%%%%%%%%%%%%%%%%%%%%%%%%%%%%%%%%%%
\subsubsection{Steering inequalities from the SDP}
\label{sec:assemblage_steering_inequalities}
%%%%%%%%%%%%%%%%%%%%%%%%%%%%%%%%%%%%%%%%%%%%%%%%%%%%%%%

The feasibility SDP~\eqref{eq:sdp_lhs_model} can be used to construct 
steering inequalities. First, one can convert such a feasibility problem 
to an explicit convex maximization,
\begin{align}
\max \quad & \mu 
\nonumber \\
\textrm{w.r.t} \quad & \mu, \{\sigma_\lambda\} 
\nonumber \\
\textrm{s.t.} \quad & \varrho_{a|x} = \sum_\lambda D(a|x,\lambda)\sigma_\lambda \quad 
\mbox{ for all } a,x 
\nonumber\\
& \sigma_\lambda \geq \mu \mathbbm{1} \quad \mbox{ for all } \lambda.
\label{eq:feasible_sdp_lhs_model}
\end{align}
If the optimal value of $\mu$ turns to be negative, then the problem in Eq.~\eqref{eq:sdp_lhs_model} is infeasible, indicating that the assemblage 
is steerable. 

To analyze this maximization, there is a powerful tool in convex 
optimization known as duality theory. In a nutshell, the maximization problem in Eq.~\eqref{eq:feasible_sdp_lhs_model} is coupled to a so-called dual 
minimization  problem, 
\begin{align}
\min \quad & \tr \sum_{a,x} F_{a|x}\varrho_{a|x} 
\nonumber \\
\textrm{w.r.t} \quad & \{F_{a|x}\} 
\nonumber \\
\textrm{s.t.} \quad & \sum_{a,x} F_{a|x} D(a|x,\lambda) \geq 0 \quad \mbox{ for all } \lambda \nonumber\\
& \tr \sum_{a,x,\lambda} F_{a|x}D(a|x,\lambda) = 1.
\label{eq:dual_sdp_lhs_model}
\end{align}
The two problems are dual in the sense that the optimal value of the 
minimization in Eq.~\eqref{eq:dual_sdp_lhs_model} is an upper bound 
for the maximization in Eq.~\eqref{eq:feasible_sdp_lhs_model}. This is 
known as weak duality. Under weak additional conditions, strong 
duality also holds: the two optimal values are equal~\cite{boyd04}.   

The duality theory implies that if there exists a collection of 
observables $\{F_{a|x}\}$ satisfying the constraints in 
problem~\eqref{eq:dual_sdp_lhs_model} and if  
\new{$\tr\sum_{a,x} F_{a|x} \varrho_{a|x} \le 0$} then the assemblage 
is steerable. So, the dual problem naturally defines a steering 
inequality. The minimizer of the problem~\eqref{eq:dual_sdp_lhs_model} 
thus yields optimal steering inequalities for \new{a} steerable 
assemblage.  Such steering inequalities found applications in several
scenarios, see also Sections \ref{sec-bound-entangled} and
\ref{sec-randomness}.

%%%%%%%%%%%%%%%%%%%%%%%%%%%%%%%%%%%%%%%%%%%%%%%%%%%%%%%%%%%
\subsubsection{Quantification of steerability with SDPs}
\label{sec:assemblage_quantification}
%%%%%%%%%%%%%%%%%%%%%%%%%%%%%%%%%%%%%%%%%%%%%%%%%%%%%%%%%%%

The SDP approach also allows one to quantify the steerability of 
an assemblage $\{\varrho_{a|x}\}$. There are several different quantification schemes. We selectively discuss some of those; 
for an extensive discussion we refer to \cite{cavalcanti17}. 

The idea of quantifying the steerability of an assemblage is as 
follows. Let us fix the number of measurements $m$ and the number of outcomes $q$ per measurement at Alice's side. Then the space 
of all  assemblages, i.e., all different $m$ decompositions of 
Bob's reduced states to $q$-component ensembles, admits a 
natural convex structure. To be more precise, let 
$\{\varrho_{a|x}\}$ and $\{\tilde\varrho_{a|x}\}$ be 
two assemblages, then for $0 \le p \le 1$, the set
$\{p \varrho_{a|x} + (1-p) \tilde \varrho_{a|x}\}$ is also an 
assemblage. Within the set of all assemblages, the unsteerable 
assemblages form a subset, which is clearly convex. Now, how 
much steerable an assemblage is can be measured by \new{some kind of relative distance} to the set of unsteerable assemblages. \new{More precisely, as long as only the linear structure of the state assemblages is concerned, the absolute distance is not meaningful, and one can only consider relative ratios of distances on a line. In practice, one therefore compares the distances between the considered assemblage, the boundary of unsteerable assemblages and the boundary of all assemblages through a certain line.} Different \new{ratios constructed from these distances} give rise to different steerability quantifiers.

\begin{figure}[t]
\includegraphics[width=0.23\textwidth]{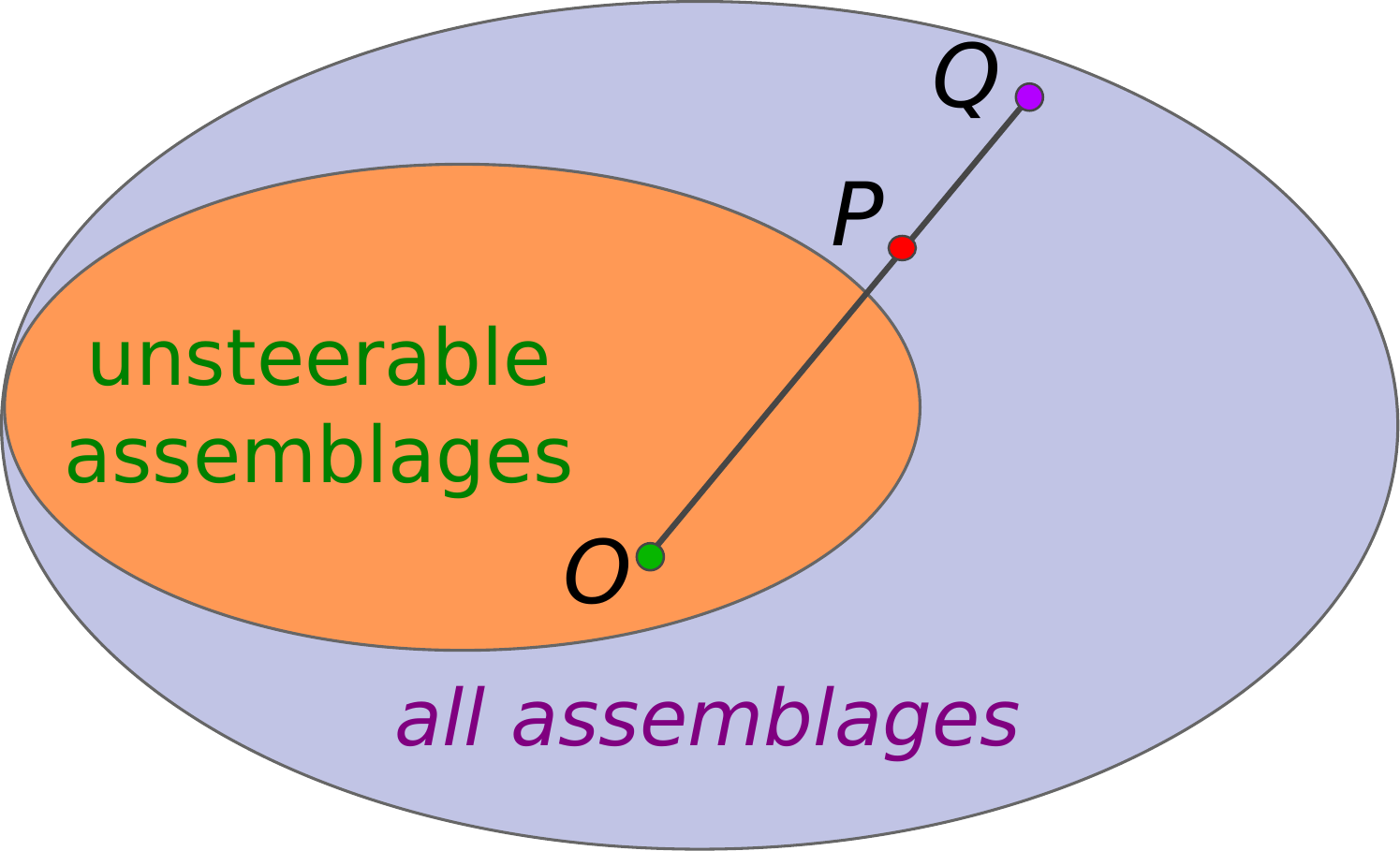}
\hspace{5pt}
\includegraphics[width=0.23\textwidth]{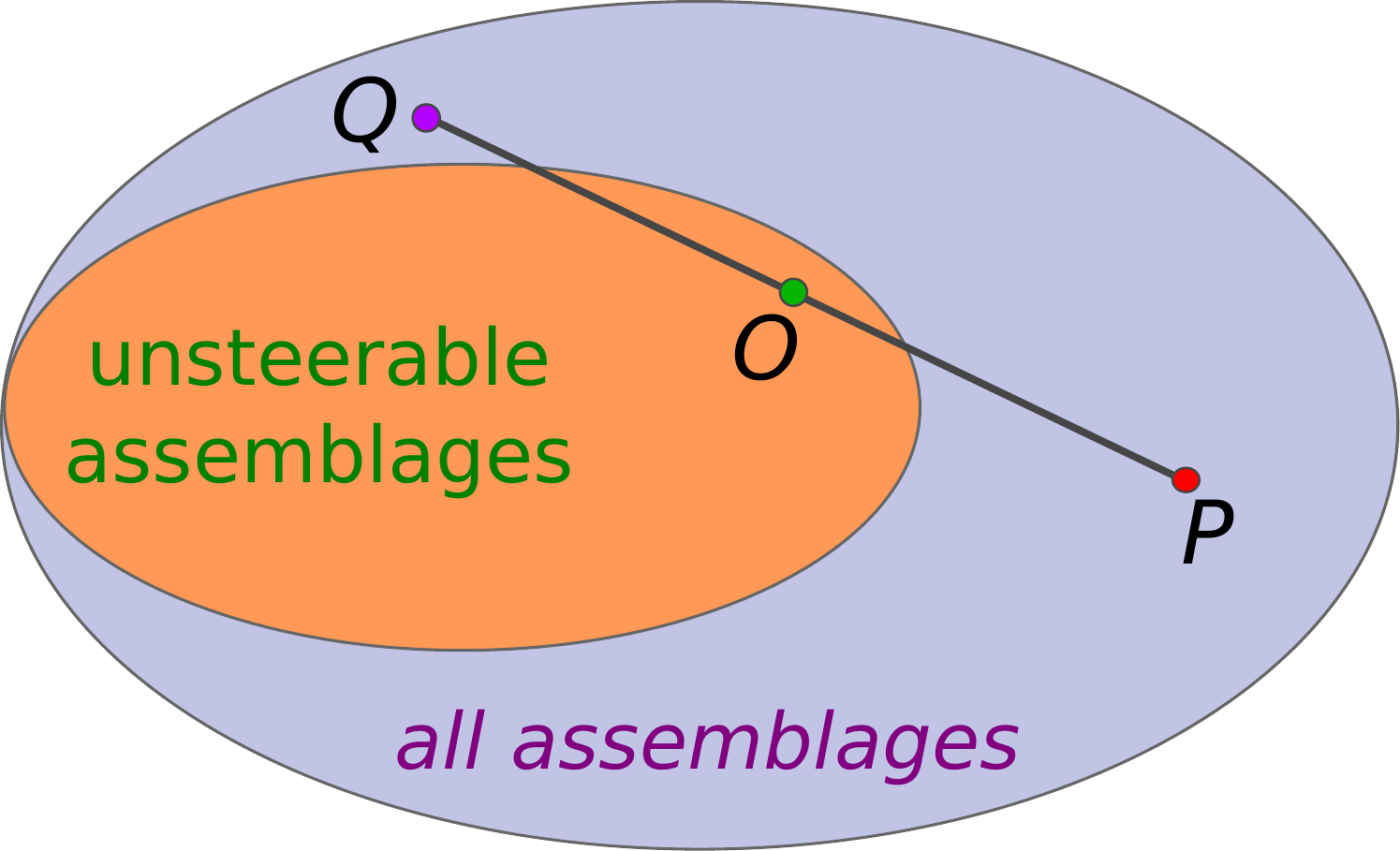}
\caption{Geometrical illustrations of the steering weight (left) and the steering 
robustness (right). Here $P$ denotes the assemblage $\{\varrho_{a|x}\}$ under 
consideration, $Q$ denotes a general realizable assemblage $\{\gamma_{a|x}\}$ 
and $O$ denotes an unsteerable assemblage $\{\varrho^{\textrm{LHS}}_{a|x}\}$. 
The steering weight (left) seeks to minimize $p=PO/OQ$ for varying $Q$ and $O$. 
The steering robustness (right) seeks to minimize $t/(1+t)=PO/PQ$ for varying $Q$ and $O$.}
\label{fig:sdp_quantification}
\end{figure}

The first quantification of steerability of an assemblage was proposed 
by~\cite{skrzypczyk14}, known as {\sl steering weight}. An assemblage 
$\{\varrho_{a|x}\}$ is first written as a convex combination of an 
unsteerable assemblage $\{\varrho_{a|x}^{\textrm{LHS}}\}$ and a general 
assemblage $\{\gamma_{a|x}\}$,
\begin{equation}
\varrho_{a|x} = p \gamma_{a|x} + (1-p)\varrho_{a|x}^{\textrm{LHS}} \quad \mbox{ for all } a,x.
\end{equation}
The steering weight of $\{\varrho_{a|x}\}$, denoted by $\operatorname{SW}(\{\varrho_{a|x}\})$, 
is the minimal weight $p$ in such decomposition with respect to all possible choices of 
the general assemblage $\{\gamma_{a|x}\}$ and the unsteerable assemblage $\{\varrho_{a|x}^{\textrm{LHS}}\}$. A geometrical illustration of the steering 
weight is given in Fig.~\ref{fig:sdp_quantification} (left).  As the set of all assemblages 
and unsteerable assemblages can be characterized via SDPs, the steering weight can 
also be determined by an SDP. More precisely, $\operatorname{SW}(\{\varrho_{a|x}\})$ 
is given by
\begin{align}
\min \quad & 1- \tr \sum_{\lambda} \sigma_{\lambda} 
\nonumber \\
\textrm{w.r.t.} \quad &  \{ \sigma_{\lambda} \} 
\nonumber \\
\textrm{s.t.}\quad &  \varrho_{a|x} - \sum_{\lambda} D (a|x,\lambda)\sigma_\lambda \ge 0 \quad \mbox{ for all } a,x 
\nonumber \\
\quad& \sigma_{\lambda} \ge 0 \quad \mbox{ for all } \lambda .
\label{eq:steering_weight}
\end{align}
A similar quantification of steerability is {\sl steering robustness}, 
\new{first defined} in the context of subchannel discrimination~\cite{piani15b}; see also Section \ref{sec-subchannel}. Here, the steering robustness 
$\operatorname{SR}(\{\varrho_{a|x}\})$ is given by the minimal weight on 
a general assemblage $\{\gamma_{a|x}\}$ considered as noise one needs 
to mix to the assemblage $\{\varrho_{a|x}\}$ so that it becomes 
unsteerable. The geometrical illustration is given in 
Fig.~\ref{fig:sdp_quantification} (right). Like the steering
weight, the steering robustness $\operatorname{SR}(\{\varrho_{a|x}\})$ 
can be computed via a simple SDP~\cite{piani15b,cavalcanti16},
\begin{align}
\min \quad & \tr \sum_{\lambda} \sigma_{\lambda} -1 
\nonumber \\
\textrm{w.r.t.}  \quad & \{ \sigma_{\lambda} \} \nonumber \\
\textrm{s.t.} \quad & \sum_{\lambda} D (a|x,\lambda)\sigma_\lambda -\varrho_{a|x} \ge 0 \quad \mbox{ for all } a,x \nonumber \\
\quad &\sigma_{\lambda} \ge 0 \quad \mbox{ for all } \lambda .
\label{eq:steering_robustness}
\end{align}
\new{To compare the two quantifiers, we note that whereas the steering weight measures the unsteerable fraction in a given assemblage, the generalized robustness measures the noise tolerance of an assemblage in terms of mixing. Also, it can be shown that whereas robustness relates to the task of subchannel discrimination (Section~\ref{sec-subchannel}), steering weight relates to the task of subchannel exclusion~\cite{uola19c}. It also turns out that the steering weight can also be understood as the maximal violation of all possible steering inequalities with an appropriate normalization~\cite{hsieh16a}. Moreover, one can make a general comment on the relation between the quantifiers that applies to general resource theories: any extremal point has maximal weight, but the robustnesses can vary among different extremal points.}

Beyond the steering weight and steering robustness,~\citet{ku18} defined a geometric quantifier based on the trace distance between a given assemblage and its corresponding closest assemblage admitting an LHS model. Also, a device-independent quantification of steerability 
was proposed by~\citet{chen16b}. This method is based on assemblage moment 
matrices, a collection of matrices of expectation values, each associated 
with a conditional quantum state; see also 
Section~\ref{sec-moment-matrix}. Finally, a different quantifier is given
by the critical radius, as explained in Section~\ref{sec:full_information_two_qubit}.

%%%%%%%%%%%%%%%%%%%%%%%%%%%%%%%%%%%%%%%%%%%%%%%%%%%%%%%%%%%%%%%%
\subsubsection{From finite to infinite number of measurements}
\label{sec:assemblage_infinite}
%%%%%%%%%%%%%%%%%%%%%%%%%%%%%%%%%%%%%%%%%%%%%%%%%%%%%%%%%%%%%%%%
While the SDP approach was originally designed to construct LHS 
models when Alice is limited to a finite set of measurements, 
one can also draw certain conclusions for the case where Alice 
has an infinite number of measurements~\cite{hirsch16,cavalcanti16a}. 
The idea is as follows. One starts with a finite set of measurements 
on Alice's side and constructs an LHS model as described above. In 
fact, one obtains a bit more. The outcome is an LHS model not 
only for the original finite set of measurements but for all 
measurements in its convex hull. 

Then, one considers the set of all measurements, typically limited 
to projective ones. One can add certain noise to the set of 
measurements, e.g., by sending them through a depolarizing 
channel and obtains a new set of noisy measurements. For 
a certain level of noise added, the set of noisy measurements 
will shrink to fit inside the convex hull of the original 
finite set of measurements, for which we have an LHS model. 
One thus also has an LHS model for  the set of noisy 
measurements. Alternatively, the noise in the measurements 
can be also put onto the state instead of the measurement 
set~\cite{hirsch16,cavalcanti16a}. Thus one can conclude 
an LHS model for the set of all measurements, but for a 
noisier version of the considered state. This construction 
works similarly for Bell nonlocality~\cite{hirsch16,cavalcanti16a}.     

While the SDP approach has proven to be useful in algorithmically 
constructing certain LHS and LHV 
models~\cite{hirsch16,cavalcanti16a,fillettaz18,cavalcanti17}, 
it has a significant computational drawback. To reduce the 
noise needed to add to the state, the original finite set 
of measurements needs to be sufficiently large. However, the size 
of the SDP, as one observes, increases exponentially with 
respect to the number of measurement settings. 
\new{ As clearly illustrated in a systematic study~\cite{fillettaz18}, 
this often imposes a significant computational difficulty 
on the problem of deciding the steerability with high 
accuracy even for two-qubit states.}

%% file: Detection/Full-information.tex
\label{sec:full_information}

When the complete density matrix $\varrho_{AB}$ is exploited, one might expect 
to have a more complete characterization, i.e., a necessary and sufficient 
condition for steerability. Like entanglement detection or Bell nonlocality 
detection, this is a difficult question. There were exact results only for 
entanglement detection of low-dimensional or special states. It is thus very encouraging that some exact results can also be derived for quantum steering. It was recognized already by~\citet{wiseman07} that for certain highly symmetric states, the problem of determining steerability with projective measurements can be solved completely, see Section~\ref{sec-special-states}. Recently, a complete characterization of quantum steering has been also achieved for two-qubit states and projective measurements~\cite{chau18,jevtic15,chau16a}. 

%--------------------------------------------------------------------------------------
\subsubsection{Two-qubit states and projective measurements}
\label{sec:full_information_two_qubit}

\new{From Eq.~\eqref{eq-lhsmodel}, one sees that in order to determine the steerability of a given state one has to consider all possible LHS ensembles $\{p (\lambda),\sigma_\lambda\}$, and for each measurement, solve for the response functions $p(a|x,\lambda)$. The source of difficulty is that the possible choice of the indexing hidden variable $\lambda$ seems to be arbitrary: it can be a discrete variable, a real-valued variable or a  multidimensional variable, etc. It is now worth re-examining how the SDP approach discussed in Section~\ref{sec-assemblages} works: one assumes that Alice can only make a finite number of measurements, which implies the finiteness of a necessary LHS ensemble -- a unique choice of the hidden variable is thus singled out. When Alice's set of measurements is not finite, this approach breaks down. Fortunately, one can show that~\cite{chau18,chau18a} for quantum steering, there is a canonical choice of the indexing hidden variable, namely Bob's pure states. This is also true for higher-dimensional systems. In fact, an LHS ensemble can be identified with a probability distribution (or to be more precise, a probability measure) $\mu$ over Bob's pure states $\mathcal{S}_B$.} 

\new{Our above discussion implies that the LHS model Eq.~\eqref{eq-lhsmodel} can be written as
\begin{equation}
	\varrho_{a|x}= \int_{\mathcal{S}_B} d \mu (\sigma) \tilde{p}(a|x,\sigma) \sigma,
\label{eq:lhs-model-simplified}
\end{equation} 
for a certain choice of $\tilde{p}(a|x,\sigma)$~\cite{chau18a}. Note that the response function $\tilde{p}(a|x,\sigma)$ may have no simple relation to $p(a|x,\lambda)$ in Eq.~\eqref{eq-lhsmodel} if the association $\lambda$ to $\sigma_\lambda$ is not injective; see~\cite{chau18a} for the detailed discussions.}% To emphasize: as passing  from~Eq.~\eqref{eq-lhsmodel} to~Eq.~\eqref{eq:lhs-model-simplified} one has singled out a specific choice of the LHV (Bob's pure states). This seemingly mid-simplification in fact is the basis for all exact characterizations of quantum steering; see also Section~\ref{sec-special-states}. In constrast, for Bell nonlocality Eq.~\eqref{eq-lhvmodel}, no canonical choice of the LHV variable can be made, making the problem highly intractable.} 

\new{Consider now a system of two qubits. To proceed, let us for now fix an LHS ensemble $\mu$.} Given an LHS ensemble $\mu$, one still faces with the problem of solving Eq.~\eqref{eq:lhs-model-simplified} for $\tilde{p} (a|x,\sigma)$ for all possible measurements $x$. The next step is to abandon this constructive approach; instead, one only determines a condition for this equation to have a solution. 
\new{To this end, for a given LHS ensemble $\mu$, one defines~\cite{chau16a,chau16b,chau18}
\begin{equation}
r(\varrho_{AB},\mu)= \min_{C} \frac{\int_{\mathcal{S}_B} \!\! d \mu (\sigma)  |\Tr_B({C}{\sigma})|}{\sqrt{2}\Vert{\Tr_B[\bar{\varrho}_{AB} (\openone_A \otimes C)]}\Vert},
\label{eq:principal-radius}
\end{equation}
where $\bar{\varrho}_{AB}=\varrho_{AB}-(\openone_{A} \otimes \varrho_B)/2$,
the norm is given by $\norm{X}=\sqrt{\Tr(X^\dagger X)}$, and  the
minimization runs over all single-qubit observables 
$C$ on Bob's space. In fact, the quantity $r(\varrho_{AB},\mu)$  characterizes the geometry of the set of conditional states Alice can simulate from the LHS ensemble $\mu$, and is called the \emph{principal radius} of $\mu$~\cite{chau16b}. It can then be shown that~\cite{chau16b,chau18} Eq.~\eqref{eq:lhs-model-simplified} has a solution $\tilde{p} (a|x,\sigma)$ for $x$ running over all possible projective measurements if and only if $r(\varrho_{AB},\mu) \ge 1$.  }

\begin{figure}[t]
\begin{center}
\includegraphics[width=0.25\textwidth]{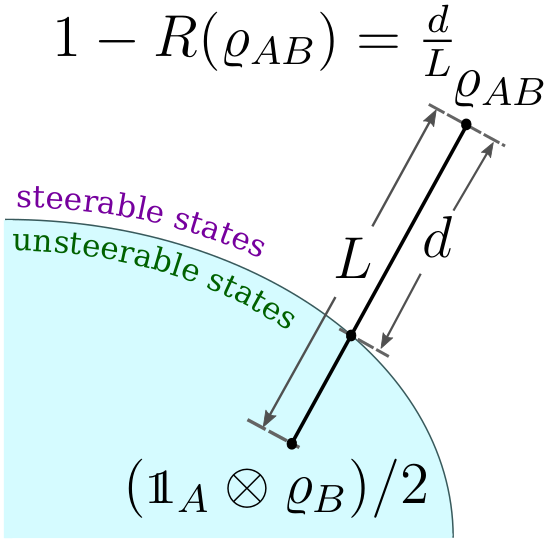}
\end{center}
\caption{The operational meaning of the critical radius: $1-R(\varrho_{AB})$ measures the distance from $\varrho$ to the surface of unsteerable/steerable states relatively to $({\openone_A} \otimes \varrho_B)/2$. Figure taken from~\cite{chau18}.}
\label{fig:geometry2}
\end{figure}

\new{So far, a fixed choice of LHS ensemble $\mu$ is made.} One now defines the \emph{critical radius} as the maximum of the principal radius~\eqref{eq:principal-radius} over all LHS ensembles, 
\begin{equation}
R(\varrho_{AB}) = \max_{\mu} r(\varrho_{AB},\mu).
\label{eq:critical-radius}
\end{equation}
Then a two-qubit state is unsteerable if and only if $R(\varrho_{AB}) \ge 1$. 

\new{Remarkably}, let $\varrho_{AB}^{(\alpha)}=\alpha \varrho_{AB} + (1-\alpha) (\openone_A \otimes \varrho_B)/2$, then $R(\varrho_{AB}^{(\alpha)})=\alpha^{-1} R(\varrho_{AB})$. 
This relation also gives an operational meaning to the critical radius, namely $1-R(\varrho_{AB})$ measures the distance from the given state to the surface that separates steerable states from unsteerable states; see Fig.~\ref{fig:geometry2}. \new{As a consequence, one can in fact equivalently define the critical radius as
\begin{equation}
R(\varrho_{AB}) = \max\{\alpha \ge 0: \mbox{$\varrho_{AB}^{(\alpha)}$ is unsteerable} \},
\label{eq:critical_radius_operational}
\end{equation}
where unsteerability is considered with respect to projective measurements. We will see that this definition can be naturally generalized to generalized  measurements and higher-dimensional systems.}  

\begin{figure}[!t]
\begin{center}
\includegraphics[width=0.21\textwidth]{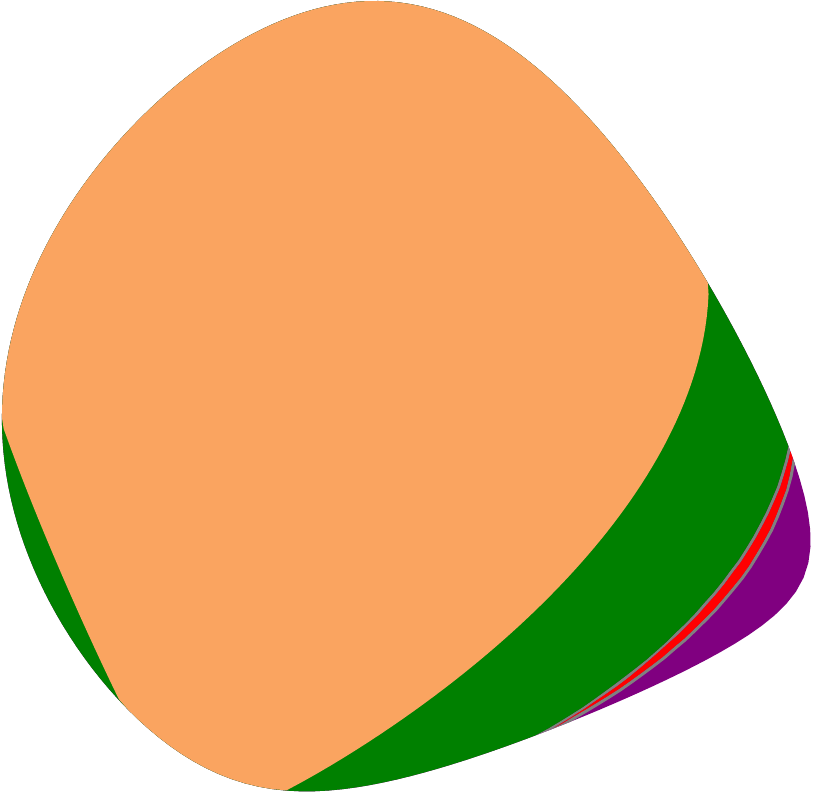}
\hspace{0.3cm}
\includegraphics[width=0.21\textwidth]{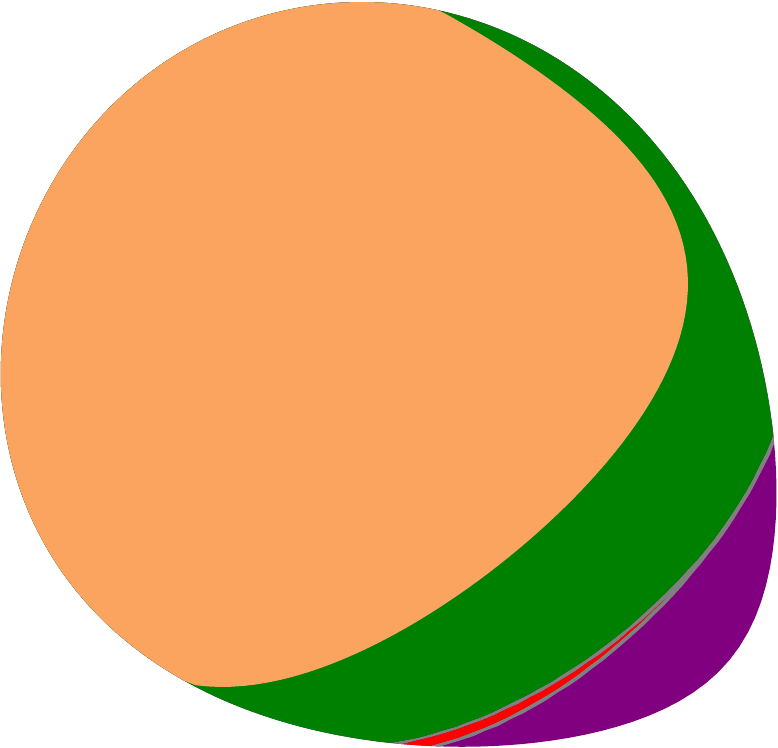}
\\ {\ } \\
\includegraphics[width=0.47\textwidth]{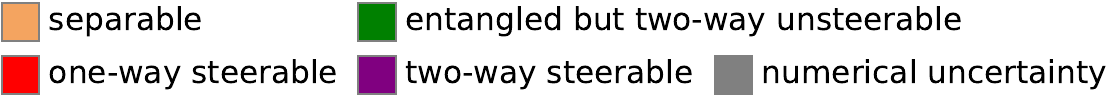}
\end{center}
\caption{Two two{-dimensional} random cross-sections of the set 
of all two-qubit states. 
\new{From the inner most to the outer most, different areas} with different colors denote the set of separable states
characterized by the partial transposition \cite{peres96,horodecki96}, 
entangled states that are unsteerable, one-way steerable states (Alice to 
Bob or vice versa), and two-way steerable states (Alice to Bob and vice 
versa). \new{The very thin grey lines at the two boundaries of the area corresponding to the one-way steerable states} denote those states where the used numerical precision
was not sufficient to make an unambiguous decision. Figure taken from~\cite{chau18}.}
\label{fig:cross_sections}
\end{figure}

\new{The definition of the critical radius by Eq.~\eqref{eq:critical-radius} also allows for its evaluation. It is shown that~\cite{chau18a}} by replacing Bob's Bloch sphere by polytopes from inside and from outside, one obtains rigorous upper and lower bounds for the critical radius. Both the computation of the upper and lower bounds are linear programs, of which the sizes scale cubically with respect to the numbers of vertices of the polytopes. Upon increasing the numbers of vertices, the two bounds quickly converge to the actual value of the critical radius. This approach has been used to access the geometry of the set of unsteerable states  via its two-dimensional random cross-sections~\cite{chau18}; see Fig.~\ref{fig:cross_sections}.

Notably, for the so-called Bell-diagonal states, or $T$-states, an explicit formula for the critical radius has been obtained,
\begin{equation}
R(\varrho_T)= 2 \pi N_T \abs{\det (T)}, 
\end{equation}
where $T$ is the correlation matrix of the $T$-state, $T_{ij}=\operatorname{Tr}(\varrho_T \sigma_i \otimes \sigma_j)$ for $i,j=1,2,3$, and the normalization factor $N_T$ is given by an integration over the Bloch sphere $N_T^{-1}=\int d S(\vec{n}) [\vec{n}^T T^{-2} \vec{n}]^{-2}$~\cite{jevtic15,chau16a}. Based on this solution for $T$-states, analytical bounds for the critical radius of a general state can also be derived~\cite{chau18}.

\new{For further discussions on LHS models for two-qubit states in special cases, see~\cite{Yu18a,Yu18b,Miller18a,Zhang19b}.}

\subsubsection{Steering of higher-dimensional systems and with generalized measurements}
\label{sec:full-information-POVM}

\new{Characterizing quantum steering of higher-dimensional systems and with generalized measurements (POVMs) is difficult.}  Most of the results on quantum steering, in this case, rely on the idea of adding sufficient noise to the state such that an LHS for simpler measurements (e.g., projective measurements) can be turned into an LHS model for POVMs~\cite{hirsch13,quintino15,tischler18}. More specifically, if a state $\varrho_{AB}$ of dimension $d_A \times d_B$ is unsteerable with respect to two-outcome POVMs, then the state
\begin{equation}
\tilde{\varrho}_{AB}=\frac{1}{d_A} \varrho_{AB} + \frac{d_A-1}{d_A} \sigma_A \otimes \varrho_B
\label{eq:povm_shrinking}
\end{equation}
with an arbitrary choice of state $\sigma_A$ and $\varrho_B= \tr_A(\varrho_{AB})$ is unsteerable for arbitrary POVMs. One observes that in Eq.~\eqref{eq:povm_shrinking}, the weight $(d_A-1)/d_A$ of the separable noise $\sigma_A \otimes \varrho_B$ is close to $1$ if the dimension is high. Yet, this technique has played an important role in demonstrating the hierarchy of nonlocality under generalized measurements (see Section~\ref{sec-hierarchy}), superactivation of nonlocality by local filtering (see Section~\ref{sec-superactivation}), and  one-way steering with POVMs (see Section \ref{sec-one-way}).

Remarkably, the critical radius approach explained \new{in the previous subsection} gives a promising framework to generally analyze quantum steering with POVMs and higher-dimensional systems. \new{In fact, in any dimension, one can define the critical radius with respect to a certain class of measurements in the same way as Eq.~\eqref{eq:critical_radius_operational}. In particular, considering the set of generalized measurements (POVMs) of $n$ outcomes, one can define the critical radius for a bipartite state $\rho_{AB}$ of dimension $d_A \times d_B$ by
\begin{equation}
R_n(\varrho_{AB})=\max \{ \alpha \ge 0: \mbox{$\varrho_{AB}^{(\alpha)}$ is unsteerable}\},
\label{eq:general_critical_radius_operational}
\end{equation}
with $\varrho_{AB}^{(\alpha)}= \alpha \varrho_{AB} + (1-\alpha) (\openone_A \otimes \varrho_B)/d_A$, $\varrho_B= \Tr_A(\varrho_{AB})$ and unsteerability being considered with respect to POVMs of $n$ outcomes on Alice's side.
Defined in this way, $1-R_n(\varrho_{AB})$ can still be interpreted as measuring the distance from $\varrho_{AB}$ to the surface separating unsteerable/steerable states, here defined with respect to POVMs of $n$ outcomes on Alice's side; see again Fig.~\ref{fig:geometry2}. However, direct evaluation of the critical radius from the definition~Eq.~\eqref{eq:general_critical_radius_operational} is clearly not possible. 

Interestingly, an alternative formula for the critical radius in similarity to Eq.~\eqref{eq:principal-radius} and Eq.~\eqref{eq:critical-radius}~\cite{chau18a,chau18} can also be found for high dimensional systems.} 
To this end, for a finite dimensional bipartite state $\varrho_{AB}$, one can define the principal radius for a given LHS ensemble $\mu$ by 
\begin{equation}
r^{-1}_n (\varrho_{AB},\mu)= \sup_{Z,E} F^{-1} (\varrho_{AB},\mu,Z,E),
\label{eq:general_principal_radius}
\end{equation}
with $F^{-1}(\varrho_{AB},\mu,Z,E)$ defined to be
\begin{equation}
\frac{\sum_{i=1}^{n} \tr [\varrho_{AB} (E_i \otimes Z_i) ] - \frac{1}{d_A}  \sum_{i=1}^{n} \tr(E_i) \tr (\varrho_B Z_i)}{\int d \mu (\sigma) \max_i \{\dprod{Z_i}{\sigma}\} - \frac{1}{d_A} \sum_{i=1}^{n}  \tr(E_i) \tr (\varrho_B Z_i)},
\label{eq:fraction_function_general}
\end{equation}
where the supremum is taken over all possible $n$-POVMs $E=(E_1,E_2,\ldots,E_n)$ on Alice's side and all possible $n$ observables $Z=(Z_1,Z_2,\ldots,Z_n)$ on Bob's side. \new{The critical radius as defined by~Eq.~\eqref{eq:general_critical_radius_operational} can be computed as~\cite{chau18,chau19b}}
\begin{equation}
R^{-1}_n(\varrho_{AB})= \min_{\mu} r^{-1}_n (\varrho_{AB},\mu).
\label{eq:general_crictical_radius}
\end{equation}   
%Defined in this way, one can show that the state $\varrho_{AB}$ is unsteerable with POVMs of $n$ outcomes if and only if $R_n (\varrho_{AB}) \ge 1$~\cite{chau18,chau18a}. Most of the important properties of the critical radius described in Section~\ref{sec:full_information_two_qubit} still hold~\cite{chau18}. 
\new{In this way, the problem of computing the critical radius and the princpal radius is in principle an optimization problem. Unfortunately,  even in this form, a deterministic algorithm to compute the principal radius and the critical radius with $n \ge 3$ is still unknown, and one has to invoke heuristic techniques in practice~\cite{chau18,chau18a}. }

\new{One observes that to study quantum steering, the set of generalized measurements was stratified  according to their number of outcomes. This calls for an investigation of the relation between them.} Since POVMs of $n$ outcomes form a natural subset of POVMs of $n+1$ outcomes, one has a decreasing chain $R_2 (\varrho_{AB}) \ge R_3 (\varrho_{AB}) \ge R_4 (\varrho_{AB}) \ge \cdots$. As extreme POVMs have at most $d_A^2$ outcomes with $d_A$ being Alice's dimension, this chain turns to equality at $n=d_A^2$. Using the evidence from a heuristic computation for the principal radius, it has been conjectured~\cite{chau18} that for two-qubit states ($d_A^2=4$), the chain in fact consists of a single number, namely $R_2 (\varrho_{AB}) = R_3 (\varrho_{AB}) = R_4 (\varrho_{AB})$. \new{In words, this conjecture implies that measurements of two outcomes (dichotomic measurements) are sufficient to fully demonstrate the quantum steerability of a two-qubit system; measurements of more outcomes are not necessary. Unfortunately, extrapolating this conjecture to higher-dimensional systems fails; it is later shown~\cite{chau19b} that this equality breaks down already for a system of two qutrits (see also Section~\ref{sec-special-states}).}

%------------------------------------------------------------------------------------------------------------
\subsubsection{Full information steering inequality}
\label{sec:full_information_inqualities}
As we discussed, for high-dimensional systems, even when full information about a state is available, a computable necessary and sufficient condition for quantum steerability is not available. In this case, the detection of steerability still relies on steering inequalities. An example of steering inequalities based on full information of the state is given by~\citet{zhen16} in terms of the so-called local orthogonal observables. Embedding in a higher-dimensional space if necessary, we can assume that Alice and Bob have the same local dimension $d$. One can choose a set of $d^2$ orthogonal operators $\{G_k\}$ which serves as a basis for the local observable space, i.e., $\Tr (G_iG_j)=\delta_{ij}$ and $\{G_k\}$ spans the space of operators~\cite{yu05}. The Pauli matrices are a familiar example of such orthogonal operators for a qubit system.  By means of the Schmidt decomposition in the operator space, one can choose the orthonormal observables for the local spaces at Alice and Bob, $\{G_k^A\}$ and $\{G_k^B\}$, such that the joint state $\varrho_{AB}$ can be written as
\begin{equation}
\varrho_{AB}= \sum_{k=1}^{d^2} \lambda_k G^A_k \otimes G^B_k,
\end{equation}
where $\lambda_k \ge 0$. Then using the local uncertainty relations (see Section~\ref{sec-correlations}),~\citet{zhen16} showed that the state $\varrho_{AB}$ is steerable from $A$ to $B$ if 
\begin{equation}
\sum_{k=1}^{d^2} \delta^2 (g_k G_k^A + G_k^B) < d-1
\label{eq:fullinfo_ineq1}
\end{equation}
for some choice of $g_k$, where $\delta^2(X)$ denotes the variance of operator $X$. By a particular choice of $g_k$, one can easily show that if
\begin{equation}
\sum_{k} \lambda_k > \sqrt{d},
\label{eq:fullinfo_ineq2}
\end{equation}
the state is steerable~\cite{zhen16}. This elegant inequality resembles the familiar computable cross norm or realignment (CCNR) entanglement criterion~\cite{chen03,rudolph05}, where 
$\sum_{k} \lambda_k > 1$
implies that the state is entangled.

Note that the steering inequalities~\eqref{eq:fullinfo_ineq1} and~\eqref{eq:fullinfo_ineq2} are different from various inequalities discussed in Section~\ref{sec-correlations} in the sense that they exploit the full information about the state.

%% file: Properties/Hierarchy.tex
We explained already in Section \ref{sec:steerbell} that 
there is a hierarchy between Bell nonlocality, steering, and 
entanglement in the sense that one implies the other, but not 
the other ways around. In this section, we first discuss in some
detail the known examples of states where the notions differ. 
Then we explain how the relations between the three concepts 
can be exploited to characterize one via another. Detailed
LHS models are discussed in Section \ref{sec-special-states}.

When discussing the existence of an LHV or LHS model for a given 
quantum state, one has to distinguish whether the model should 
explain the results for all projective measurements, or, more 
generally, for all POVMs. Let us start our discussion with 
projective measurements. The inequivalence between the notion of 
entanglement and Bell nonlocality was, in fact, one of the starting 
points of entanglement theory \cite{werner89}. For that, one may 
consider the so-called two-qubit Werner state
\begin{equation}
\varrho(p)= p \ketbra{\psi^-}{\psi^-} + (1-p)\frac{\openone}{4} 
\end{equation}
where $\ket{\psi^-}=(\ket{01}-\ket{10})/\sqrt{2}$ is the singlet state. Using
the PPT criterion (see Section \ref{sec:pptsteering}) one can directly verify
that this state is entangled iff $p>1/3.$ In \cite{werner89}, however, an LHS 
model for projective measurements was constructed for all values $p\leq 1/2.$ 
Moreover, in \cite{acin06, hirsch17} it was shown that an LHV model exists up 
to $p \leq 1/K_G(3)\approx 0.6829$, where $K_G(3)$ is the Grothendieck constant 
of order three, so up to this value no Bell inequality can be violated. These 
results demonstrate that there are entangled states which do not show Bell 
nonlocality. Using the fact that the Werner states is steerable for $p>1/2$
\cite{wiseman07}, this also proves that steering and Bell nonlocality are 
inequivalent for projective measurements. \new{States of this type have 
been prepared experimentally and their steerability has been 
demonstrated in \cite{saunders10}.}

It remains to discuss the more general case of POVMs. First, in \cite{barrett02} 
an LHV model for Werner states with $p \leq 5/12$  has been constructed which 
explains all the measurement probabilities for arbitrary POVMs. In fact, this 
model can directly be converted into an LHS model \cite{quintino15}. Consequently, 
there are entangled states for which all correlations for POVMs can be explained
by an LHV model. In addition, \cite{quintino15} presented examples of states 
in a $3\times3$ system which are steerable in both directions, but nevertheless
an LHV model for all POVMs can be found. This proves the inequivalence of Bell
nonlocality and steering for POVMs.

As mentioned, a local model that explains the results of all projective measurements 
does not necessarily explain all the correlations for POVMs. It is not clear, however, 
that POVMs give an advantage in the detection of steering or Bell nonlocality. As 
discussed in Section \ref{sec:full-information-POVM}, there is numerical evidence 
that two-qubit states which are unsteerable for projective measurements are also 
unsteerable for POVMs \cite{chau18}. Concerning Bell nonlocality, in 
\cite{vertesi10, gomez16} Bell inequalities have been presented, for which the 
maximal violation requires POVMs, but this does not imply that the states leading 
to this violation do not violate also some Bell inequality for projective measurements.

Given the similarity in the definitions of Bell nonlocality, quantum steerability 
and nonseparability, one may expect that some methods to characterize the different 
notions can be related to each other. Specifically, given a state that admits an LHV 
model as in Eq.~\eqref{eq-lhvmodel}, one may expect that by adding suitable separable 
noise to the state, one can obtain a state that admits an LHS model \eqref{eq-lhsmodel}. 
This has been shown to be the case~\cite{chen18}. The authors showed that if a bipartite 
qudit-qubit state $\varrho_{AB}$ admits an LHV model, then the state 
\begin{equation}
\tilde{\varrho}_{AB} = \mu \varrho_{AB} + (1-\mu) \varrho_{A} \otimes \frac{\openone_B}{2},
\end{equation} 
with $\mu=1/\sqrt{3}$ is unsteerable from Alice to Bob. Turning the logic around, 
if $\tilde{\varrho}_{AB}$ is steerable, then $\varrho_{AB}$ must be Bell nonlocal. 
A similar statement between steerability and nonseparability has also been obtained~\cite{chen18,das18}. Namely, if a bipartite qubit-qudit state 
$\varrho_{AB}$ is unsteerable from Alice to Bob, then the state 
\begin{equation}
\tilde{\varrho}_{AB} = \mu \varrho_{AB} + (1-\mu) \frac{\openone_A}{2} \otimes \varrho_B,
\end{equation} 
with $\mu=1/\sqrt{3}$ is separable. Or, if the latter is entangled, the former is 
steerable. Detailed applications of this approach to detecting different nonlocality 
notions from the others can be found in~\cite{chen16,chen18,das18}.

%% file: Properties/Special-states.tex
\new{As already highlighted in Section~\ref{sec-full-information}, the fact that for quantum steering, there is a canonical choice for the hidden variable, namely Bob's pure states, turns out to have far-reaching consequences. 
The point is that, having simplified the LHS model from Eq.~\eqref{eq-lhsmodel} to the form of Eq.~\eqref{eq:lhs-model-simplified}, the symmetry of the state has stronger implications on the choice of the LHS ensemble~\cite{chau18a}.}
For certain highly symmetric states such as the Werner states and the isotropic states, the symmetry is then enough to uniquely single out an optimal choice of the LHS ensemble, rendering their exact characterizations of quantum steering with projective measurements possible~\cite{wiseman07,jones07}. 
\new{This is in contrast to Bell nonlocality: in  Eq.~\eqref{eq-lhvmodel}, no canonical choice of the LHV is possible. Thus even for highly symmetric states such as the isotropic states and the Werner states, no exact characterization of Bell nonlocality is known.}

\subsubsection{Werner states}

Suppose Alice and Bob share the Werner state of dimension $d \times d$~\cite{werner89}, defined by
\begin{equation}
W^\eta_d= \frac{d-1+\eta}{d-1} \frac{\openone}{d^2} - \frac{\eta}{d-1} \frac{V}{d},
\label{eq:werner_state_general}
\end{equation}
where $\openone$ is the bipartite identity operator and $V$ is the flip operator given by $V \ket{\phi,\psi} = \ket{\psi,\phi}$. Here we follow the parameterization by~\citet{wiseman07} so that $W^\eta_d$ is a product state if the mixing parameter $\eta=0$ and is a state at all only if $\eta \le 1$. The Werner state is entangled if and only if $\eta > \frac{1}{d+1}$~\cite{werner89}.

In fact, the Werner state was constructed in a way such that it is invariant under the same local unitary transformation at Alice's and Bob's side~\cite{werner89}, that is, for any unitary operator $U$ acting in dimension $d$, $W^{\eta}_d =(U \otimes U) W^\eta_d (U^\dagger \otimes U^\dagger)$. This implies that the optimal LHS ensemble on Bob's Bloch sphere can be chosen to be symmetric under the unitary group $U(d)$, i.e., the Haar measure~\cite{wiseman07,chau18a}.  

\begin{figure}[!t]
\begin{minipage}{0.235\textwidth}
\includegraphics[width=\textwidth]{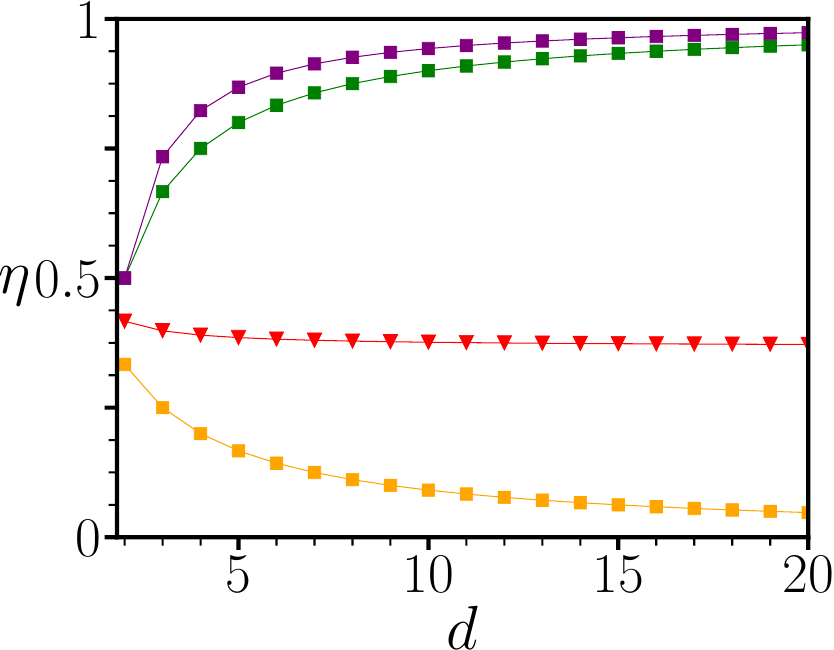}
\end{minipage}
\begin{minipage}{0.225\textwidth}
\includegraphics[width=\textwidth]{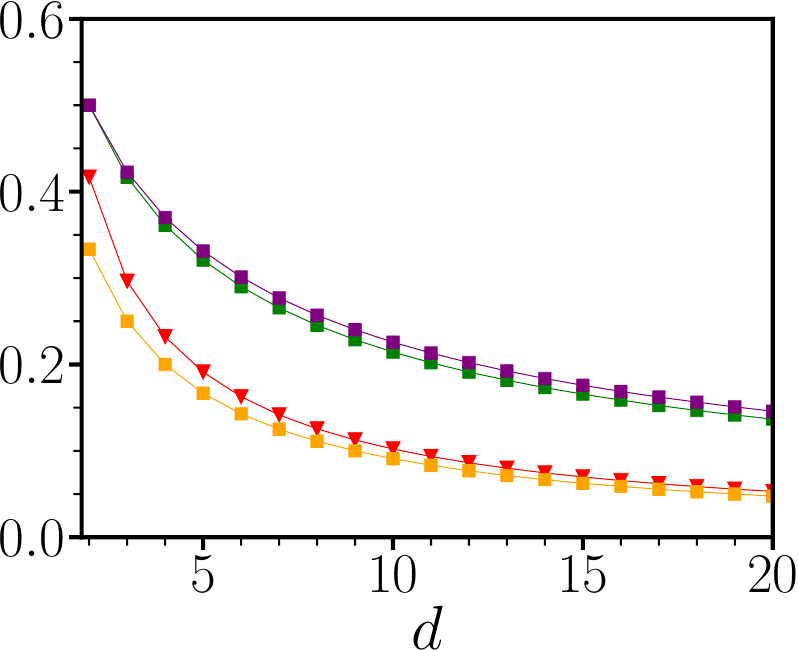}
\end{minipage}
\caption{\new{The exact noise thresholds for for the Werner states (left) and the isotropic states (right) to be steerable with dichotomic measurements (square, violet, upper), projective measurements (square, green, middle), and to be separable (square, orange, lower). The lower bounds for the noise thresholds for them to be  unsteerable with all generalized measurements obtained from Eqs.~\eqref{eq:barrett_isotropic} and~\eqref{eq:barrett_isotropic} are also presented (triangle, red). Figure adapted from~\cite{chau19b}.}}
	\label{fig:exact_thresholds}
\end{figure}

Identifying the hidden variable $\lambda$ indexing the LHS with Bob's pure states $\ket{\lambda}$, it remains to construct the response function $p(a \vert x, \lambda)$ for a projective measurement $\{E_{a|x}\}=\{P_{a \vert x}\}$ to complete an LHS  model. Note that for a projection outcome $P_{a \vert x}= \ketbra{a}{a}$ at Alice's side, Bob's conditional state is
\begin{equation}
\varrho_{a \vert x} = \frac{d-1+\eta}{d(d-1)} \frac{\openone}{d} - \frac{\eta}{d (d-1)} \ketbra{a}{a}.
\label{eq:werner_state_outcomes}
\end{equation}
The minus sign in front of the last term indicates that the two parties in the Werner state are anti-correlated.
To construct the response function, it is natural then to associate a pure state $\ket{\lambda}$ to the outcome that has the \emph{least} overlap with $P_{a \vert x}$. The resulting response function is
\begin{equation}
p(a \vert x, \lambda) = 
\left\{ 
\begin{array}{l}
1 \mbox{ if $\abs{\braket{\lambda}{a}}  < \abs{\braket{\lambda}{a'}} \quad \forall a' \ne a$} \\
0 \mbox{ otherwise.}  
\end{array}
\right.
\end{equation}
With the LHS ensemble and this choice of the response function, it is straightforward~\cite{wiseman07} to show that the Werner state is unsteerable for 
\begin{equation}
\eta \le 1-\frac{1}{d}.
\label{eq:werner_state_unsteerable}
\end{equation} 
It can also be shown that for the mixing parameter $\eta > 1-\frac{1}{d}$, no construction of response function is possible~\cite{wiseman07}. This threshold together with the threshold for the Werner state to be separable is presented in Fig.~\ref{fig:exact_thresholds} (left). Thus, whereas the exact threshold of $\eta$ for which the Werner state is Bell nonlocal is still unknown even in dimension $d=2$, the threshold for steerability has an analytical expression in all dimensions.~\citet{wiseman07} also noted that the construction given above was actually the original construction by~\citet{werner89} to show that Bell nonlocality and entanglement are distinct notions.

\new{Projective measurements are however not the only case where the quantum steerability of the Werner states can be exactly characterised. Specifically, it was shown that~\cite{chau19b}  when Alice is limited to making dichotomic measurements, the threshold upto which the Werner state is unsteerable can also be derived in practically closed form,
\begin{equation}
\eta \le (d-1)^2 [1-(1-1/d)^{1/(d-1)}],
\label{eq:dichotomic_werner_threshold}
\end{equation}  
for $d \le 10^5$; see Fig.~\ref{fig:exact_thresholds} (left). The threshold is also conjectured to hold for all dimensions~\cite{chau19b}. Interestingly, for dimension $d \ge 3$, the threshold Eq.~\eqref{eq:werner_state_unsteerable} is strictly stronger than that of Eq.~\eqref{eq:dichotomic_werner_threshold}. These are thus concrete examples illustrating that quantum steering with dichotomic measurements is strictly weaker than that of measurements with more outcomes for higher-dimensional systems, contrasting with the conjecture on their equivalence for two-qubit systems (see Section~\ref{sec-full-information}).
}

\subsubsection{Isotropic states}

Another important family of states that allows for exact characterization of quantum steering is that of isotropic states~\cite{wiseman07}. The isotropic state of dimension $d \times d$ at mixing parameter $\eta$, $0 \le \eta \le 1$, is defined by
\begin{equation}
S^\eta_d= (1-\eta) \frac{\openone}{d^2} + \eta \ketbra{\psi_+}{\psi_+},
\end{equation}
where $\ket{\psi_+}=1/\sqrt{d} \sum_{i=1}^{d} \ket{i,i}$. From the definition, one notes that the isotropic state is defined with respect to a particular choice of basis. The isotropic state is entangled if and only if $\eta > 1/(d+1)$~\cite{horodecki99}. In similarity to the Werner state, the isotropic state also has a unitary symmetry, namely, $S^\eta_d= (\bar{U} \otimes U) S^\eta_q (\bar{U}^{\dagger} \otimes U^\dagger)$ for any $d \times d$ unitary matrix $U$, with $\bar{U}$ being its complex conjugate. This again implies that the optimal choice of LHS ensemble is the uniform  Haar measure over Bob's Bloch sphere. 

For a projection outcome $P_{a \vert x}= \ketbra{a}{a}$ on her side, with an isotropic state, Alice steers Bob's system to the conditional state
\begin{equation}
	\varrho_{a \vert x} = \frac{1-\eta}{d} \frac{\openone}{d} + \frac{\eta}{d} \ketbra{\bar{a}}{\bar{a}},
\label{eq:isotropic_state_general}
\end{equation}
where $\ket{\bar{a}}$ is the complex conjugate of state $\ket{a}$. 
In contrast to Eq.~\eqref{eq:werner_state_outcomes}, the plus sign in front of the last term in the above equation indicates that parties sharing an isotropic state are correlated upto a complex conjugation.
This motivates the following choice of the response function
\begin{equation}
p(a \vert x, \lambda ) = 
\left\{ 
\begin{array}{l}
1 \mbox{ if $\abs{\braket{\lambda}{\bar{a}}}  > \abs{\braket{\lambda}{\bar{a}'}} \quad \forall a' \ne a$} \\
0 \mbox{ otherwise,}  
\end{array}
\right.
\end{equation}
where we have also again identified the hidden variable $\lambda$ indexing the LHS ensemble with Bob's pure states $\ket{\lambda}$.
This construction leads to an LHS model for the isotropic state with 
\begin{equation}
\eta \le \frac{H_d-1}{d-1},
\label{eq:isotropic_state_unsteerable}
\end{equation} 
where $H_d=1+1/2+1/3+\cdots+1/d$. It can again be shown that this threshold is optimal; for $\eta > {(H_d-1)}/{(d-1)}$ no construction for the response function is possible~\cite{wiseman07}. Remarkably,~\citet{almeida07} also obtained this threshold in an attempt to construct an LHV model for the isotropic states before learning of the definition of quantum steering. This threshold is presented in Fig.~\ref{fig:exact_thresholds} (right) together with that for separability.

\new{
Like the Werner state, the quantum steerability of the isotropic states with dichotomic measurements can also be exactly characterized. It was shown in~\cite{chau19b} that for $d \le 10^5$, if 
\begin{equation}
\eta \le 1 - d^{-1/(d-1)},
\end{equation}
the isotropic state is unsteerable when Alice's measurements are limited to dichotomic ones; otherwise, it is steerable (see Fig.~\ref{fig:exact_thresholds} (right)). The threshold is also conjectured to hold for all dimensions~\cite{chau19b}.  
}

\subsubsection{LHS model for POVMs}

As quantum steering with projective and \new{dichotomic} measurements is well-understood for the Werner states and the isotropic states, one may hope that certain LHS models with \new{general} POVMs for them can also be constructed. This is indeed the case. By an explicit construction,~\citet{barrett02} demonstrated that sufficiently weakly entangled Werner states do admit an LHV model for all POVMs. \new{Under the light of the formal definition of quantum steering~\cite{wiseman07}, the LHV model turns out to be an LHS model~\cite{quintino15}. The model was revised recently~\cite{chau19b}, and it can be shown that Barrett's original construction works best for the isotropic states; for the Werner states, a better model can be constructed. Further, for two-qubit systems, the construction can also be extended to Bell-diagonal states~\cite{chau19a}.} 
%However, the model works only for $\eta$ strictly smaller than that required by the model for projective measurements in Eq.~\eqref{eq:werner_state_unsteerable}. It thus does not demonstrate the equivalence between POVMs and projective measurements for these highly symmetric states.    

To construct the model, it is sufficient to consider only POVMs with rank-$1$ effects, $\{E_{a|x}\} =\{\alpha_{a|x} \ketbra{a}{a}\}$, where $\ketbra{a}{a}$ are rank-$1$ projections and $0 \le \alpha_{a|x} \le 1$~\cite{barrett02}. This is because other POVMs can be post-processed from these (see also Section~\ref{sec-joint-measurability}). The optimal choice for the LHS ensemble is again the uniform distribution over Bob's Bloch sphere~\cite{chau18a,wiseman07}. It is then left to construct the response functions $p(a|x,\lambda)$ for the mentioned measurements. 

\new{For the isotropic state, the response function can be given as~\cite{barrett02,almeida07}
\begin{align}
& p(a \vert x, \lambda ) = \alpha_{a|x} \abs{\braket{\lambda}{\bar{a}}}^2 \Theta (\abs{\braket{\lambda}{\bar{a}}}^2 -1/d) \nonumber \\
&\qquad + \frac{\alpha_{a|x}}{d} \left(1 - \sum_{a}  \alpha_{a|x} \abs{\braket{\lambda}{\bar{a}}}^2 \Theta (\abs{\braket{\lambda}{\bar{a}}}^2 -1/d) \right). 
\label{eq:barrett_isotropic_response}
\end{align}
With} this choice of the response function, direct computation shows that the isotropic state is unsteerable for arbitrary POVMs on Alice's side if~\cite{barrett02,almeida07} 
\begin{equation}
\eta \le \frac{3d-1}{d+1} (d-1)^{d-1}d^{-d}.
\label{eq:barrett_isotropic}
\end{equation}
\new{As we mentioned, this construction was originally suggested as an LHS model for Werner states and the same threshold Eq.~\eqref{eq:barrett_isotropic} was found~\cite{barrett02,quintino15}. However, it was shown~\cite{chau19b} that for the Werner state, a better choice of the response functions is possible, namely
\begin{align}
p(a|x,\lambda) &=\frac{\alpha_{a|x}}{d-1}(1-\abs{\braket{\lambda}{a}}^2) \Theta (1/d - \abs{\braket{\lambda}{a}}^2) \nonumber \\
&\!\!\!\!\!\!\!\!\!\!\!\!\!\!\! +\frac{\alpha_{a|x}}{d} \left(1- \sum_{a} \frac{\alpha_{a|x}}{d-1}(1-\abs{\braket{\lambda}{a}}^2) \Theta (1/d - \abs{\braket{\lambda}{a}}^2) \right).
\label{eq:barrett_werner_response}
\end{align}
The Werner state was then shown to be unsteerable for arbitrary POVMs on Alice's side if~\cite{chau19b} \begin{equation}
\eta \le \frac{1+(d-1)^{d+1}d^{-d}}{d+1}.
\label{eq:barrett_werner}
\end{equation}

The two bounds Eq.~\eqref{eq:barrett_isotropic} and Eq.~\eqref{eq:barrett_werner} are also presented in Fig.~\ref{fig:exact_thresholds}. For the Werner states, the bound Eq.~\eqref{eq:barrett_werner} is strictly better than the bound given by Eq.~\eqref{eq:barrett_isotropic} for $d \ge 3$. However both bounds Eq.~\eqref{eq:barrett_isotropic} and Eq.~\eqref{eq:barrett_werner} are strictly within the respective thresholds for the isotropic states and the Werner states to be unsteerable with projective measurements, Eq.~\eqref{eq:isotropic_state_unsteerable} and Eq.~\eqref{eq:werner_state_unsteerable}. On the other hand, the constructions~Eq.~\eqref{eq:barrett_isotropic_response} and Eq.~\eqref{eq:barrett_werner_response} are by no means optimal; in fact, it is expected to be not optimal~\cite{chau19a}. Thus it is still unclear at the moment if steering with projective measurements is equivalent to steering with generalized measurements even for these highly symmetric states.
}

%This condition is strictly stronger than the condition for the Werner state to admit an LHS model with projective measurements, $\eta \le 1-1/d$. 

{The case of two-qubit Werner states ($d=2$) is slightly better understood. In this case, both the bounds Eq.~\eqref{eq:barrett_isotropic} and Eq.~\eqref{eq:barrett_werner} show that for $\eta \le \frac{5}{12}$, the Werner state is unsteerable for arbitrary POVMs on Alice's side~\cite{barrett02,quintino15}.} In the range $\frac{5}{12} \le p \le \frac{1}{2}$, the state is also known to be unsteerable if the POVMs are limited to those with $3$ outcomes~\cite{werner14}. For most general POVMs, numerical evidences based on the critical radius approach are available, which indicate that the state is also unsteerable in this range~\cite{chau18a,chau18}.  

To conclude this section, we refer the interested readers to~\cite{augusiak14} for further constructions of LHV and LHS models.

%% file: Properties/Filtering.tex
For characterizing steerability and other correlations in quantum 
states, it is relevant to study their behaviour under local operations. 
Given a general quantum state $\varrho_{AB}$ one can consider states 
of the type
\begin{equation}
\tilde{\varrho}_{AB} = 
\frac{1}{\mathcal{N}} 
(T_A \otimes T_B)  \varrho_{AB} (T_A^\dagger \otimes T_B^\dagger), 
\end{equation}
where $T_{A/B}$ are some transformation matrices and $\mathcal{N}$ denotes 
a potential renormalization.

Then, one can ask whether the correlations in the state $\varrho_{AB}$
are related to those of the state $\tilde \varrho_{AB}.$ Clearly, this 
depends on the properties of the matrices $T_{A/B}$. For them there
are mainly two possible choices: Either one restricts them to be unitary 
$T_{A/B}= U_{A/B}$ and therefore one considers local unitary transformations.
Or one considers general invertible matrices $T_{A/B}
= F_{A/B}$, which are the so-called local filtering operations and are 
more general than local unitaries. 

For the case of entanglement one can directly see from the definition
in Eq.~(\ref{eq-separabilitydefinition}) that local filtering 
operations keep the property of a state being separable or entangled.
Filtering operations can, nevertheless, change the amount of entanglement. 
Any full rank state can be brought into a normal form under filtering
operations, where the reduced states $\tilde{\varrho}_{A}$ and 
$\tilde{\varrho}_{B}$ are maximally mixed \cite{verstraete03, leinaas06}. 
In this form, certain entanglement measures are maximized \cite{verstraete03}
and bringing a state in this normal form can improve many entanglement
criteria \cite{gittsovich08}. For Bell nonlocality, it can be seen that
local unitary transformations keep the property of a state having an LHV model.
But local filtering operations are already too general, there are two-qubit
states that do not violate the CHSH inequality, but after local filtering,
they do \cite{gisin96, popescu95}.

Steering is a notion between entanglement and nonlocality, so a mixed behaviour
under local transformations can be expected. In fact, it was noted in 
\cite{gallego15, uola14, quintino15} that local unitaries on Alice's side and local 
filtering on Bob's side, 
\begin{equation}
\tilde{\varrho}_{AB} = 
\frac{1}{\mathcal{N}} 
(U_A \otimes F_B)  \varrho_{AB} (U_A^\dagger \otimes F_B^\dagger), 
\end{equation}
do not change the steerability of a state. Also the critical radius
as a steering parameter (see Section \ref{sec:full_information_two_qubit}) 
is not affected. With these transformations one can achieve that 
$\tilde{\varrho}_{B}$ is maximally mixed on its support. \new{If 
a state is in this form, this can simplify calculations and therefore 
it is good starting point to study the steerability of a state \cite{chau18}.}

%% file: Properties/One-way-steerable-states.tex
\begin{figure}[!t]
\includegraphics[width=0.40\textwidth]{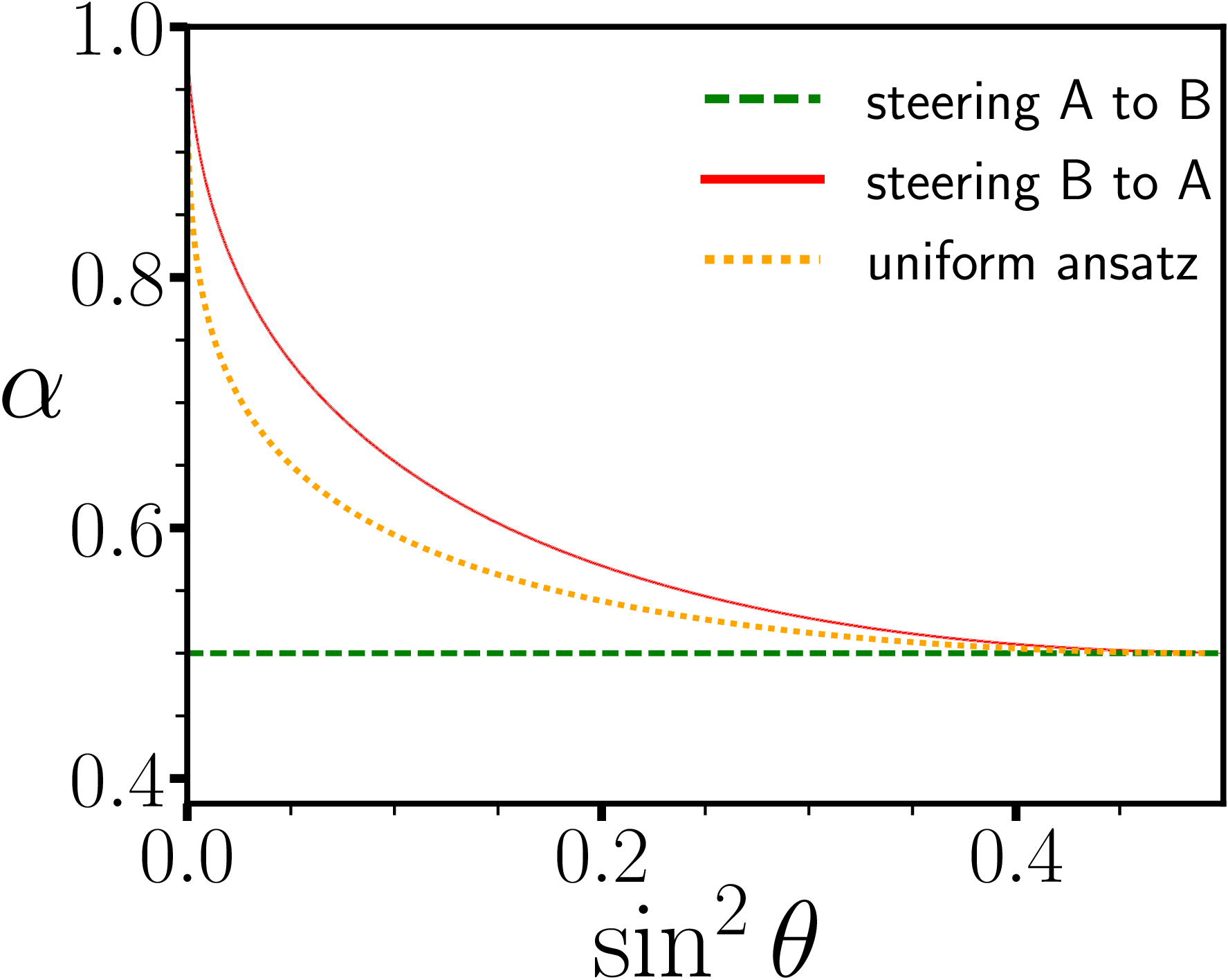}
\caption{The border of the one-way steerable area  for the family of states given by Eq.~\eqref{eq:oneway}. \new{The thickness of the border for steering from $B$ to $A$ indicates the uncertain area}. The inner bound for the border of steering from \new{$B$ to $A$} with the uniform LHS ensemble as an ansatz is also included (dotted line).}
\label{fig:oneway}
\end{figure}

The asymmetry between the two parties in the definition of quantum steering immediately strikes one with the question whether there is a state where Alice can steer Bob, but not the other way around~\cite{wiseman07}.  Such one-way steerable states were first constructed for continuous variable systems~\cite{midgley10,olsen13}. One-way steerable states for discrete systems were studied later by~\citet{evans14,bowles14,skrzypczyk14}. More recently a simpler family of one-way steerable two-qubit states was identified in~\cite{bowles16}. This family of states is given by
\begin{equation}\label{eq:oneway}
\varrho_{AB}(\alpha,\theta) = \alpha \ket{\psi_\theta}\bra{\psi_\theta} + (1-\alpha)\frac{\openone}{2} \otimes \varrho_B,
\end{equation}
where $\ket{\psi_\theta} = \cos \theta\ket{00}+\sin \theta \ket{11}$ and $\varrho_B = \textrm{Tr}_A\ket{\psi_\theta}\bra{\psi_\theta}$, with $0\le \alpha \le 1$ and $0< \theta \le \pi/4$. The states can be brought into the two-qubit Werner states with the same mixing probability $\alpha$ by a local filtering on Bob's side and a local unitary on Alice's side. Therefore it is steerable from Alice to Bob if and only if $\alpha > 1/2$ (see Section~\ref{sec-filtering}). Using the uniform distribution as an ansatz for the LHS ensemble,~\cite{bowles16} showed that the state is unsteerable from Bob to Alice for $\cos^2(2\theta)\geq \frac{2\alpha-1}{(2-\alpha)\alpha^3}$. With the complete characterization of steerability for two-qubit states described in Section~\ref{sec:full_information_two_qubit}, the boundary of the set of unsteerable states from Bob to Alice has been obtained with high accuracy~\cite{chau18}; see Fig.~\ref{fig:oneway}. It is clearly visible from the figure that $\varrho_{AB}(\alpha,\theta)$ is one-
way steerable for a large range of parameters. 

The one-way steering phenomenon can also be shown to persist when the measurements are extended to POVMs~\cite{quintino15}. The idea to construct an example is as follows. One first embeds a state which is unsteerable from Alice to Bob with respect to projective measurements, but steerable for the other direction, into a higher dimension on Alice's side. One then constructs a state that admits an LHS model for all POVMs performed on Alice's side using Eq.~\eqref{eq:povm_shrinking} with the state $\sigma_A$ chosen to be supported only in the extended dimension on Alice's side. With this choice of $\sigma_A$, it is easy to show that the state is still steerable from Bob to Alice~\cite{quintino15}. The constructed state is thus one-way steerable also when one considers all POVMs.

The one-way steering phenomenon also attracts attention from the experimental side. Early experiments demonstrating one-way steering were carried out for continuous variable systems and Gaussian measurements~\cite{handchen12}. {The effects of various types of noise on the direction of steering were later analyzed and probed experimentally by~\citet{qin17}}. Experiments demonstrating one-way steering for discrete systems were performed by~\citet{sun16,xiao17,wollmann16}. \citet{sun16} and~\citet{xiao17} concentrated on demonstrating one-way steering when measurements are limited to two and three settings.~\citet{wollmann16} also demonstrated the persistence of the phenomena for POVMs. Most recently it was realized~\cite{baker18} that existing experiments demonstrating one-way steering committed certain assumptions on the states or the measurements and were therefore inconclusive. A conclusive experiment~\cite{tischler18} was then performed shortly after.

%% file: Properties/Bound-entangled-states.tex
\label{sec:pptsteering}

Now we  discuss the steerability of so-called bound 
entangled states. This provides a relevant example for the fact that 
the characterization of steering gives new insights into old problems 
in entanglement theory. 

Before presenting the result, we have to recall some facts about the entanglement
criterion of the positivity of the partial transpose (PPT criterion) and entanglement 
distillation. Let us start with the PPT criterion \cite{peres96, horodecki96}. Generally, 
for a two-particle state $\varrho = \sum_{ij,kl}\varrho_{ij,kl} \ket{i}\bra{j}\otimes\ket{k}\bra{l}$ 
the partial transposition with respect to Bob is defined as
\begin{equation}
\varrho^{T_B} = \sum_{ij,kl}\varrho_{ij,lk} \ket{i}\bra{j}\otimes\ket{k}\bra{l}.
\end{equation}
Similarly, one can define a partial transposition with respect to Alice which obeys
$\varrho^{T_A}= (\varrho^{T_B})^T.$ Note that the partial transposition may change
the eigenvalues of a matrix, contrary to the full transposition. 

The PPT criterion states that for separable states the partial transposition has no 
negative eigenvalues, $\varrho^{T_B} \geq 0$, such states are also called PPT states. 
It was further proven that for systems consisting of two qubits ($2 \times 2$-systems) 
or one qubit and one qutrit ($2\times3$-system) this criterion is sufficient for
separability and all PPT states are separable. For all other dimensions, PPT 
entangled states exist, these states are, in some sense, weakly entangled, as they 
cannot be used for certain quantum information tasks. 

The main quantum information task where PPT entangled states are useless is the task 
of entanglement distillation. Entanglement distillation is the process where many 
copies of some noisy entangled state are distilled to few highly entangled pure 
states via local operations and classical communication \cite{horodecki98}.
Surprisingly, not all entangled states can be used for distillation, and these
undistillable states are called bound entangled. It was shown that PPT entangled 
states are bound entangled, but there are some NPT states, for which it has been
conjectured that they are also bound entangled \cite{pankowski10}.

In 1999, Peres  formulated the conjecture that bound entangled states do not 
violate any Bell inequality \cite{peres99}. This conjecture was based on an analogy
between a general distillation protocol and Bell inequalities for many observers,
but for a long time no proof could be found. In 2013 the conjecture was made that
bound entangled states are also useless for steering \cite{pusey13,skrzypczyk14}. 
This so-called stronger Peres conjecture could potentially open a way to prove the 
original Peres conjecture, especially as the PPT criterion and the question of 
steerability are closely related to SDPs.

It was shown by \cite{moroder14}, however, that some bound entangled states can be used 
for steering, and an explicit example for two qutrits was given. The idea to find 
the counterexample is the following. For a given state of two qutrits and two 
measurements with three outcomes each one can decide the steerability of the 
assemblage $\{\varrho_{a|x}\}$ with an SDP (see also Section 
\ref{sec:assemblage_steering_sdp}). Considering the dual formulation of the SDP, 
one finds that the operator
\begin{align}
W= & 
A_{1|1}\otimes Z_{13} + A_{2|1}\otimes Z_{23} + A_{1|2}\otimes Z_{31} + 
A_{2|2}\otimes Z_{32} 
\nonumber
\\
&
+(A_{3|1}+A_{3|1}-\openone)\otimes Z_{33}  
\end{align}
\new{defines} a steering inequality, that is, $\tr(\varrho W) \geq 0$ for unsteerable 
states. Here, the $A_{a|x}$ are arbitrary measurement 
operators for Alice and the set $\{Z_{13}, Z_{23},Z_{31},Z_{32},Z_{33}\}$ 
consists of five positive operators, obeying the four semidefinite constraints 
$Z_{i3}+Z_{3j}-Z_{33}\geq 0$ for $i,j \in \{1,2\}.$

Given this steering inequality, one can look for steerable PPT states by an 
iteration of SDPs: One starts with a random initial steerable state 
$\varrho$ and fixes Alice's measurements $A_{a|x}$ to be measurements in 
two mutually unbiased bases. Then, by optimizing the $Z_{ij}$ via an SDP 
one can minimize the mean value $\tr(\varrho W)$ and find the optimal steering 
inequality $W$. Given this $W$, one can ask for the minimal expectation value 
of it with respect to PPT states, this is again an SDP. Having found the 
PPT state with the smallest $\tr(\varrho W)$ one can optimize over the $Z_{ij}$
again and  then iterate. In practice this procedure converges quickly towards 
PPT states which are steerable, delivering the desired counterexamples
to the stronger Peres conjecture.

\begin{figure}[t!]
\includegraphics[width=0.9\columnwidth]{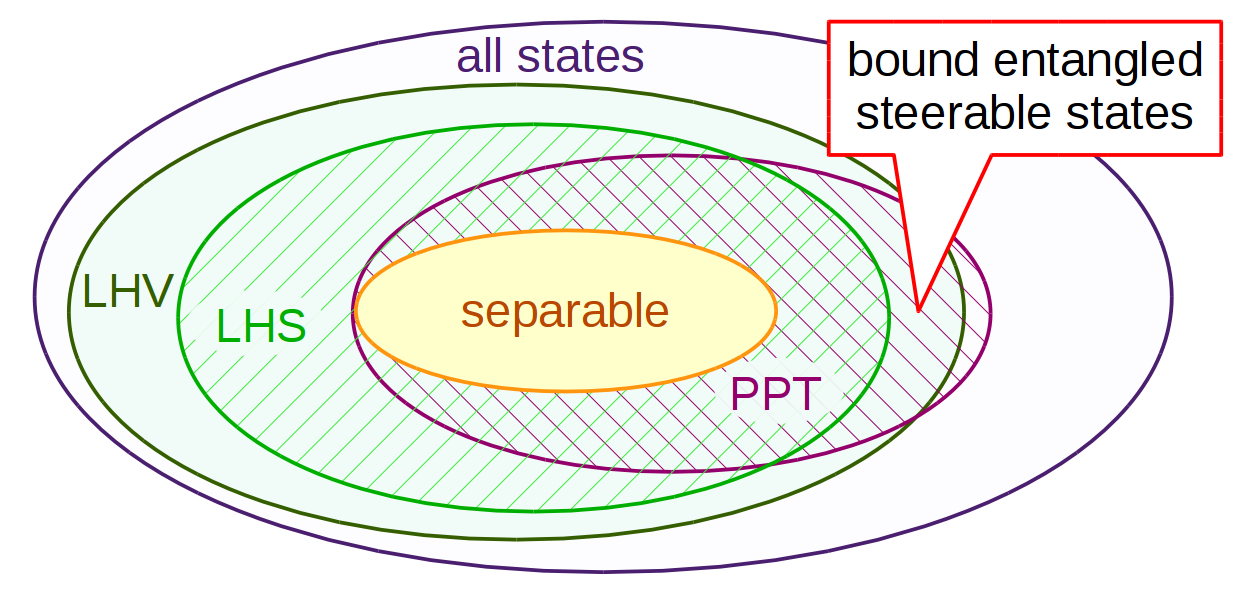}
\caption{Inclusion relation between the PPT states and entanglement, steering, 
and Bell inequality violations. Separable states are PPT, but some entangled 
states are PPT as well. PPT entangled states are bound entangled, as no pure 
state entanglement can be distilled from them. There exist, however, PPT states 
that can be used for steering and also PPT states that violate Bell inequalities.
These states are counterexamples to the Peres conjecture.
}
\label{fig-entanglement-steering-bell}
\end{figure}

Having found the counterexamples, it is a natural question whether these states
also violate a Bell inequality. Indeed, as has been shown by \cite{vertesi14},
these states are also counterexamples to the original Peres conjecture. Finally, 
\cite{yu17} presented an analytical approach, giving explicit families of PPT
entangled states in any dimension $d\geq3$ which, for appropriate parameters, 
violate Bell inequalities or can be used for steering, see Fig.~\ref{fig-entanglement-steering-bell}.

%To demonstrate the power of the hierarchy in the above section, we consider briefly 
%a special class of steerable states. Namely, we review the results of 
%Refs.~\cite{moroder14,yu17} showing that there exists steerable and, 
%by extension, nonlocal quantum states which are bound entangled~\cite{vertesi14}. These results constitute a resolution to the Peres conjecture \cite{peres99,pusey13}.

%% file: Properties/Dimension-bounded-steering.tex
In this section, we describe how the steering problem can be viewed as
a certain kind of separability problem \cite{moroder16}. This allows to 
apply the powerful techniques of entanglement theory \cite{guehne09, horodecki09}, and  
study problems such as the detection of steering, if Bob's system is not 
well characterized and only its dimension is known. 

To formulate the main idea it suffices to consider the case of two measurements 
($x\in\{1,2\}$) with two outcomes ($\pm$) on Alice's side. As discussed in 
Section \ref{sec:assemblage_steering_sdp}, steerability of the assemblage 
$\{\varrho_{a|x}\}$ can be decided by an SDP. More precisely,  
Eqs.~(\ref{eq-steeringmaps-cond1}) states that the assemblage is 
unsteerable, if one finds four positive semidefinite operators 
$\omega_{ij}$ with $i,j=\pm$ such that
$
\varrho_{+|1}=\omega_{++}+\omega_{+-}, 
$
$
\varrho_{+|2}=\omega_{++}+\omega_{-+}, 
$
$
\varrho_{-|1}=\omega_{-+}+\omega_{--}, 
$
and
$
\varrho_{-|2}=\omega_{+-}+\omega_{--}.  
$
These equations are not independent. If one takes $\omega_{++}$ as a free 
variable, one has the relations
$\omega_{+-}=\varrho_{+|1}-\omega_{++}$,
$\omega_{-+}=\varrho_{+|2}-\omega_{++},$ and 
$\omega_{--}=\varrho_B-\varrho_{+|1}-\varrho_{+|2}+\omega_{++}$.
Of course, this is only a valid solution if all $\omega_{ij}$ are 
positive semidefinite.

Then, one takes four positive definite operators $Z_{ij}$ with 
$i,j=\pm$, which obey the relation  $Z_{++}=Z_{+-}+Z_{-+}-Z_{--}$,
and considers the bipartite operator
\begin{equation}
\label{eq:SIGMA}
\Sigma_{AB} = \sum_{ij} Z_{ij} \otimes \omega_{ij}.
\end{equation}
This is, after appropriate normalization, a separable state as in 
Eq.~(\ref{eq-separabilitydefinition}). The point is that with the 
given relations on the $\omega_{ij}$ and $Z_{ij}$ this state can 
be written as
\begin{equation}
\label{eq:Z_sol}
\Sigma_{AB}=Z_{+-} \otimes \varrho_{+|1} + Z_{-+} \otimes \varrho_{+|2} 
+Z_{--} \otimes (\varrho_B-\varrho_{+|1}-\varrho_{+|2}) 
\end{equation}
as can be verified by direct inspection. Here, all the dependencies on 
the $\omega_{ij}$ dropped out, so $\Sigma_{AB}$ is uniquely determined 
by the assemblage and the $Z_{ij}$ only. Also the required normalization
follows directly from Eq.~(\ref{eq:Z_sol}).

From this the desired connection to the separability problem follows: 
Given an unsteerable assemblage and operators $Z_{ij}$ obeying the 
conditions from above, the state $\Sigma_{AB}$ in Eq.~(\ref{eq:Z_sol}) 
is separable. Moreover, one can show the opposite direction: If the 
assemblage is steerable, then there exist a set of operators  $Z_{ij}$
such that the state $\Sigma_{AB}$ 
is entangled. In this case, the entanglement of $\Sigma_{AB}$ can even 
be detected by a special entanglement witness, namely the flip operator.

The statement can be generalized to an arbitrary number of measurements
and outcomes \cite{moroder16}. In fact, it is related to the dual of 
the original SDP. 

This reformulation of the steerability problem can give insights in the 
detection of steering, if the measurements on Bob's side are not 
fully characterized, but only the dimension of the space where the 
measurements act on is known. The core idea is the following: For any 
bipartite state $\varrho_{AB}$ and sets of local orthonormal observables 
$G_k^A$ and $G_l^B$ [that is, $\tr(G_i^X G_j^X)=\delta_{ij}$ for $X\in\{A,B\}$] 
one can build the matrix $\Lambda_{kl}=\tr(\varrho_{AB} G_k^A\otimes G_l^B).$
Then, the computable cross-norm or realignment (CCNR) criterion
states that if $\varrho_{AB}$ is separable, then the trace norm is bounded 
by one, $\Vert \Lambda \Vert_1 \leq 1$ \cite{chen03, rudolph05}, see also
Section \ref{sec:full_information_inqualities}. 
This criterion has already been used to detect entanglement with uncharacterized 
devices, if the dimension is known: If Alice and Bob make uncharacterized 
measurements $A_k$ and $B_l$ they can build the expectation value matrix 
$\Delta_{kl}=\tr(\varrho_{AB} A_k\otimes B_l)$ and, using the dimension assumption, 
from that estimate the trace norm $\Vert \Lambda \Vert_1$ \cite{moroder12}. 

A similar approach can be used for steering \cite{moroder16}. For a choice of 
$Z_{ij}$ one considers the state $\Sigma_{AB}$. Then, on Alice's side one 
takes a set of local orthogonal
observables $G_k^A$ and for Bob's side uncharacterized measurements $B_l$ and builds 
an expectation value matrix, which can be used to estimate whether $\Sigma_{AB}$ violates
the CCNR criterion. If this is the case, then the original assemblage was steerable.
The resulting criteria are strong: For two-qubit Werner states 
$\varrho(p)= p \ketbra{\psi^-}{\psi^-} + (1-p)\openone/4$ and Pauli measurements 
$\sigma_x,
\sigma_y$ and $\sigma_z$ one can evaluate from the data the steering inequality in 
Eq.~(\ref{eq-steer-c2}). It detects steerability for $p > 1/\sqrt{3}$, which is the same
threshold as the steering inequality. So, for this case the approach 
allows to draw the same conclusion from 
the resulting data, but without assuming that the measurements were correct Pauli
measurements, the only assumption that is made is that Bob's space is a qubit.

%\subsubsection{Input text below Eq.~\ref{eq-sdp-ana}}

%A key observation for the practical implementation of the SDP is that
%the 

%As a possible extension of the steering protocol we discuss the scenario where Alice's 
%measurements are fixed and Bob has only a dimension bound on his system. Such a scenario
%is called dimension-bounded steering, introduced in~\cite{moroder16}. 
%In contrast to the above subsection, this subsection contributes to the other end of 
%the steering hierarchy. Namely, methods for detecting dimension-bounded steering are 
%heavily based on known entanglement detection methods.

%% file: Properties/Superactivation.tex
Let us come back to the formulation of quantum steering as a simulation task where Alice tries to convince Bob that she can steer his system from a distance as discussed in Section~\ref{sec-intro}. \new{Note that in this protocol, Alice has to prepare a large number of pairs of particles in a certain state. One of the particles in each pair is then sent to Bob. Note that it is crucial for Alice to prepare \emph{many copies} of the state, so that Bob later on can do tomography to verify the steered states on his side.} Alice then declares the set of measurements she can make, or equivalently the assemblage she can steer Bob's system to. To maximize her steering ability, Alice clearly should choose the largest set of measurements. Most often, Alice's measurements are assumed to be projective measurements (or POVMs) on separated particles on her side. This, however, is not yet the maximal set of measurements she can do. As Alice has prepared a large number of bipartite states, she can actually make collective measurements on several particles on her side. We will see that when such collective measurements are considered, the steerability of a state may change. More precisely, for an unsteerable (but entangled) state $\varrho_{AB}$, one asks whether there exists a finite number $n$ such that $\varrho_{AB}^{\otimes n}$ is steerable. In this case, we say that the quantum steerability of $\varrho_{AB}$ can be superactivated. 

For nonseparability, a similar question is answered trivially negative for any states, but for Bell nonlocality, it has been extensively investigated since~\cite{peres96a}. For Bell nonlocality, the confirmative answer was first obtained by~\citet{palazuelos12} and later refined by~\citet{cavalcanti13}. The authors showed that indeed for a certain state $\varrho_{AB}$ which admits an LHV model, for sufficiently large $n$, $\varrho_{AB}^{\otimes n}$ can violate some Bell inequality. Note that this is distinct from the notion of superactivation of bound entanglement~\cite{shor03}. While the superactivation of Bell nonlocality investigated by the mentioned authors also implies the ability to superactivate quantum steering, exact characterizations of quantum steering also significantly simplify the understanding of the \new{phenomenon~\cite{quintino16,hsieh16a}}. \new{In fact,~\citet{quintino16} and~\citet{hsieh16a}  extended the results of~\cite{cavalcanti13} to show that the steerability of all unsteerable states $\varrho_{AB}$ that satisfy the so-called reduction criterion for entanglement~\cite{horodecki99} can be superactivated. The reduction criterion states that if $\openone_A \otimes \varrho_B - \varrho_{AB}$ is not positive, then the state $\varrho_{AB}$ is nonseparable.} As satisfying the reduction criterion is a necessary and sufficient condition for a two-qubit or a qubit-qutrit state to be nonseparable~\cite{horodecki99}, the steerability of all entangled states of dimension $2\times 2$ or $2 \times 3$ can be superactivated. 

Their idea is based on the exact threshold for quantum steering of the isotropic state in Eq.~\eqref{eq:isotropic_state_unsteerable}. For convenience, here the isotropic state is reparametrized as
\begin{equation}
S^{f}_d= f \ketbra{\phi_+}{\phi_+} + (1-f) \frac{\openone - \ketbra{\phi_+}{\phi_+}}{d^2-1},
\label{eq:isotropic_state_2}
\end{equation}
with the same notation as defined in Eq.~\eqref{eq:isotropic_state_general} and $f=1-(1-1/d^2)(1-\eta)$. According to Eq.~\eqref{eq:isotropic_state_unsteerable}, the isotropic state in Eq.~\eqref{eq:isotropic_state_2} is unsteerable if and only if $f\le [(1+d)H_d - d]/d^2$. It is known~\cite{horodecki99} that a state that violates the reduction criterion can be brought into an entangled isotropic state with $f>1/d$ by local filtering on Bob's side and the so-called isotropic twirling operation. As we mentioned in Section~\ref{sec-filtering}, local filtering on Bob's side does not change the steerability of the state. The isotropic twirling operation consists of averaging the state under certain random local unitary transformations, thus also does not increase the steerability. So it is sufficient to show that the steerability of the isotropic state $S^f_d$ with $f>1/d$ can be superactivated. This again can be shown by observing that the isotropic twirling on $(S^{f}_d)^{\otimes n}$ yields the isotropic state $S^{f^n}_{d^n}$ of dimension $d^n \times d^n$. Thus $(S^f_d)^{\otimes n}$ is steerable if
\begin{equation}
f > \frac{\left[(1+d^n)H_{d^n}-d^n\right]^{1/n}}{d^2}.
\end{equation}
At large $n$, the right-hand side asymptotically approaches $1/d$. Therefore, whenever $f>1/d$, there exists $n$ such that the inequality is satisfied, or equivalently the steerability of $S^f_d$ can be superactivated. 

Beyond states that violate the reduction criterion, one may ask if quantum steerability, or more generally, Bell nonlocality can always be superactivated for arbitrary entangled states. This question \new{remains as} a challenge for future research. If this were the case, the hierarchy of quantum nonlocality would be unified into a single concept~\cite{cavalcanti13}.   

Besides the notion of superactivation of quantum nonlocality via collective measurements on multiple copies of the state as described above, there is also the notion of superactivation of quantum nonlocality via local filtering (on both sides). The phenomenon is dated back to~\cite{popescu95}, who showed that the Werner states in dimension $d \ge 5$ that admit LHS models for projective measurements can violate a Bell inequality after appropriate local filtering. Recently,~\citet{hirsch13} showed that there are states which admit an LHS model for POVMs but become Bell nonlocal after appropriate local filtering. However,~\citet{hirsch16a} later showed that there are also entangled states whose quantum nonlocality cannot be superactivated by local filtering.

%% file: Joint-measurability/Measurement-incompatibility.tex
Measurement incompatibility manifests itself in various operationally motivated forms in quantum theory. Maybe the best-known notion is that of non-commutativity. Here by non-commutativity we mean the mutual non-commutativity of the POVM elements of two POVMs, i.e. for POVMs $\{A_a\}_a$ and $\{B_b\}_b$ we ask whether $[A_a,B_b]=0$ for all $a,b$ or not. From text books on quantum mechanics we know that non-commutativity of observables places certain restrictions on the variances of the measured observables. Such restrictions do not, however, give any further operational insight into the involved measurements, they just follow from simple mathematics.

One possible operationally motivated extension of commutativity is that of joint measurability. Namely, one can ask whether two measurements can be performed simultaneously (or jointly), i.e. whether there exists a third measurement whose statistics can be classically processed to match those of the original pair. Further fine-tunings of measurement incompatibility have been presented in the literature, e.g. coexistence, broadcastability, and non-disturbance \cite{busch16,heinosaari10,heinosaari16a}, see also Section \ref{sec-Further}. Typically all the incompatibility related extensions of non-commutativity (on a single system) coincide with non-commutativity for projective measurements, but for the case of POVMs they form a strict hierarchy \cite{heinosaari10}. It is worth mentioning that in the process matrix formulation of POVMs, even commuting process POVMs can be incompatible \cite{sedlak16}.

For investigating steering from the measurement perspective, the notion of joint measurability appears most fitting. A set $\{A_{a|x}\}_{a,x}$ of POVMs (i.e. positive operators summing up to the identity for every $x$) is said to be jointly measurable if there exists a POVM $\{G_\lambda\}_\lambda$ together with classical post-processings $\{p(a|x,\lambda)\}_{a,x,\lambda}$ such that
\begin{align}
A_{a|x}=\sum_\lambda p(a|x,\lambda)G_\lambda.
\end{align}
The POVM $\{G_\lambda\}_\lambda$ is called a joint observable or a joint measurement of the set $\{A_{a|x}\}_{a,x}$.

To give an example of a set of jointly measurable POVMs, one could use a mutually commuting pair of POVMs in which case a joint measurement is given by a POVM whose elements are products of the original ones. For a more insightful example, we take a pair of noisy Pauli measurements defined as
\begin{align}
S^\mu_{\pm|x}&:=\frac{1}{2}(\openone\pm\mu\sigma_x)\\
S^\mu_{\pm|z}&:=\frac{1}{2}(\openone\pm\mu\sigma_z),
\end{align}
where $0<\mu\leq 1$. The question is now how to find candidates for a joint measurement. For the above pair, an educated guess [i.e. a candidate with similar symmetry as the pair $(S^\mu_{\pm|x}, S^\mu_{\pm|z})$] gives
\begin{align}
G^\mu_{i,j}:=\frac{1}{4}(\openone + i \mu\sigma_x + j \mu\sigma_z),
\end{align}
where $i,j\in\{-,+\}$. One notices straight away that
\begin{align}
S^\mu_{\pm|x}&=G^\mu_{\pm,+}+G^\mu_{\pm,-}\\
S^\mu_{\pm|z}&=G^\mu_{+,\pm}+G^\mu_{-,\pm},
\end{align}
i.e. there exist (deterministic) post-processings that give the original measurements. The last thing to check is that $\{G^\mu_{i,j}\}_{i,j}$ forms a POVM. As the normalisation follows from the definition, one is left with checking the positivity of the elements, which is equivalent to $\mu\leq 1/\sqrt 2$. It can be shown that this is indeed the optimal threshold for joint measurability in our example, i.e. beyond this threshold the POVMs $\{S^\mu_{\pm|x}\}$ and $\{S^\mu_{\pm|z}\}$ do not admit a joint measurement \cite{busch86}.

The above example shows that joint measurability is indeed a proper generalization of commutativity. In the literature, many such examples have been discussed in finite and continuous variable quantum systems \cite{busch16}. A typical question is as above: how much noise can be added until measurements become jointly measurable. For small numbers of measurements and outcomes, this can be efficiently checked with SDP \cite{wolf09,uola15}. For more complicated scenarios various optimal and semi-optimal analytical and numerical techniques have been developed \cite{kunjwal14,heinosaari16b,uola16,bavaresco17,designolle19}.

%% file: Joint-measurability/JM-Alice.tex
%In this subsection we will characterize the observables useful for steering as those which 
%don't allow a joint measurement~\cite{quintino14,uola14}. We will also show how steering 
%inequalities serve as a necessary condition for joint measurability~\cite{uola16}.

Comparing the definition of joint measurability with that of unsteerability, one recognizes similarities. Indeed, joint measurability is a question about the existence of suitable post-processings and a common POVM, whereas unsteerability asks the existence of suitable response functions and a common state ensemble. To make the connection exact, we recall the main result of \cite{quintino14,uola14}:

\textit{A set of measurements $\{A_{a|x}\}_{a,x}$ is not jointly measurable if and only if it can be used to demonstrate steering with some shared state}.

To be more precise, using a jointly measurable set of observables on Alice's side, i.e. $A_{a|x}=\sum_\lambda p(a|x,\lambda)G_\lambda$, and a shared state $\varrho_{AB}$ results in a state assemblage
\begin{align}
\varrho_{a|x}&=\sum_\lambda p(a|x,\lambda)\text{tr}_A[(G_\lambda\otimes\openone)\varrho_{AB}]\\
&=\sum_\lambda p(a|x,\lambda)\sigma_\lambda,
\end{align}
where we have written $\sigma_\lambda=\text{tr}_A[(G_\lambda\otimes\openone)\varrho_{AB}]$. Hence, the existence of a joint observable for Alice's measurements implies the existence of a local hidden state model. For the other direction, using a full (finite) Schmidt rank state $|\psi\rangle=\sum_i\lambda_i|ii\rangle$ one has
\begin{align}
\varrho_{a|x}:=\text{tr}_A[(A_{a|x}\otimes\openone)|\psi\rangle\langle\psi|]=CA_{a|x}^TC,
\end{align}
where $C=\sum_j\lambda_j|j\rangle\langle j|$ and $X^T$ is the transpose of the operator $X$ in the basis $\{|i\rangle\}_i$. Assuming that the assemblage $\{\varrho_{a|x}\}_{a,x}$ has a local hidden state model one gets
\begin{align}
A_{a|x}=\sum_\lambda p(a|x,\lambda)C^{-1}\sigma_\lambda C^{-1},
\end{align}
from which it is clear that $\{C^{-1}\sigma_\lambda C^{-1}\}_\lambda$ forms the desired joint measurement of $\{A_{a|x}\}_{a,x}$.

To demonstrate a possible use of this result, one can consider a steering scenario where Alice performs measurements on a noisy isotropic state. This noise in the state can be translated to Alice's measurements by writing
\begin{equation}\label{Noisechange}
\text{tr}[(A_{a|x}\otimes\openone)\varrho_{AB}^\mu]=\text{tr}[(A_{a|x}^\mu\otimes\openone)\varrho_{AB}],
\end{equation}
where $\varrho_{AB}^\mu=\mu|\psi^+\rangle\langle\psi^+|+\frac{(1-\mu)}{d^2}\openone$ and $A_{a|x}^\mu=\mu A_{a|x}+\frac{(1-\mu)}{d}\text{tr}[A_{a|x}]\openone$ with $\mu\in [0,1]$. For different sets of measurements on Alice's side one can either solve the steerability by using known incompatibility results or vice versa. To give an example, consider the known \cite{wiseman07} steerability threshold for the noisy isotropic state (with projective measurements) $\mu^*=(\sum_{n=1}^d\frac{1}{n}-1)/(d-1)$. Using Eq.~(\ref{Noisechange}) one sees that for any $\mu>\mu^*$ there exists a set of projective measurements that remains incompatible with the amount $\mu$ of white noise. On the contrary, any set of projective measurements with an amount $\mu\leq\mu^*$ of white noise results in an unsteerable state assemblage (with the isotropic state) and, hence, such a set is jointly measurable.

It is worth noting that the connection between incompatibility of Alice's measurements and steerability of the resulting assemblage is strongly motivated by a similar work on non-locality. In \cite{wolf09} incompatibility of Alice's measurements was proven to be equivalent to the ability of violating the CHSH inequality (when optimizing over Bob's measurements and the shared state). This connection, however, is known not to be true in general. In \cite{hirsch18,bene18} counterexamples for the non-locality connection in scenarios with more measurement settings are presented, i.e. there exist sets of measurements that are not jointly measurable but always lead to local correlations. {In contrast, joint measurability and steering can both be described in terms of operational contextuality \cite{tavakoli19}. The CHSH inequality is a criterion for this type of contextuality. It is an open question whether there are other contextuality inequalities that fully characterise incompatibility. In \cite{tavakoli19} 
numerical evidence is provided that a specific family of contextuality criteria generalising the CHSH inequality characterise the incompatibility of sets of binary qubit measurements. Such characterisation, among any other, is directly applicable to steerability of state assemblages by the use of the techniques presented in the following subsection.}

%Examples from incompatibility to steering include sets of MUBs [...], various sets of qubit measurements [...], and certain symmetric higher dimensional scenarios [...]. From steering to incompatibilty one can translate, for example, the well-known thresholds for steerability of an isotropic state with continuous symmetric sets of measurements (i.e. all PVMs or in qubit case the measurements on the equator of the Bloch sphere) [...].

%% file: Joint-measurability/JM-Bob.tex
%Here we will connect the steerability problem of state assemblages 
%one-to-one with the joint measurability problem of measurements~\cite{uola15}, thus highlighting a deep 
%connection between the rather new subfield of quantum steering and the much older 
%subfield of joint measurability. We will also show how joint measurement uncertainty 
%relations can be used as steering inequalities.

The connection between steering and joint measurements presented in the above section is based on the use of full Schmidt rank states (on finite-dimensional systems). To loosen the assumption on purity of the state, we recall the main result of \cite{uola15}:

\textit{The question of steerability (of a state assemblage) is a non-normalised version of the joint measurement problem}.

More precisely, by normalising a state assemblage $\{\varrho_{a|x}\}_{a,x}$ one gets abstract POVMs $\tilde B_{a|x}:=\varrho_B^{-1/2}\varrho_{a|x}\varrho_B^{-1/2}$, where $\varrho_B=\sum_a\varrho_{a|x}$ and a pseudo-inverse is used when necessary. Note that we use tilde to distinguish between Bob's actual measurements and the normalised state assemblage (which consists of abstract POVMs on a possibly smaller dimensional system than the one Bob's measurements act on).

It is straight-forward to show \cite{uola15} that the state assemblage $\{\varrho_{a|x}\}_{a,x}$ is steerable if and only if the abstract POVMs $\{\tilde B_{a|x}\}_{a,x}$ are not jointly measurable. Namely, as the normalisation keeps the post-processing functions fixed, the local hidden states map to joint measurements of the normalised assemblage and joint measurements map to local hidden states.

Such connection broadens the set of techniques that are translatable between the fields of joint measurability and steering. In general, joint measurability criteria map into steering criteria and vice versa. To give an example, we take a well-known joint measurability characterisation of two qubit POVMs \cite{busch86}. Namely, take two qubit POVMs of the form
\begin{equation}
A_{\pm|x}:=\frac{1}{2}(\openone\pm\vec a_x\cdot\vec\sigma),
\end{equation}
where $x=1,2$. This pair is jointly measurable if and only if
\begin{equation}\label{Paulunbias}
\|\vec a_1+\vec a_2\|+\|\vec a_1-\vec a_2\|\leq2.
\end{equation}
Note that this criterion is necessary for joint measurability in the more general case, i.e. for pairs of POVMs given as
\begin{equation}
A_{+|x}:=\frac{1}{2}((1+\alpha_x)\openone\pm\vec a_x\cdot\vec\sigma),\ A_{-|x}=\openone-A_{+|x},
\end{equation}
where $\alpha_x\in[-1,1]$ and $\|\vec a_x\|\leq 1+\alpha_x$. In \cite{yu10,0802.4167,0802.4248} a necessary and sufficient criterion for joint measurability of such pairs is given as
\begin{align}\label{Paulbias}
(1-F_1^2-F_2^2)\left(1-\frac{\alpha_1^2}{F_1^2}-\frac{\alpha_2^2}{F_2^2}\right)\leq(\vec a_1\cdot\vec a_2-\alpha_1\alpha_2)^2, 
\end{align}
with ${F_i=\frac{1}{2}(\sqrt{(1+\alpha_i)^2-\|\vec a_i\|^2}+\sqrt{(1-\alpha_i)^2-\|\vec a_i\|^2})}$, for $i=1,2$.

\begin{figure}
\includegraphics[width=.42\textwidth]{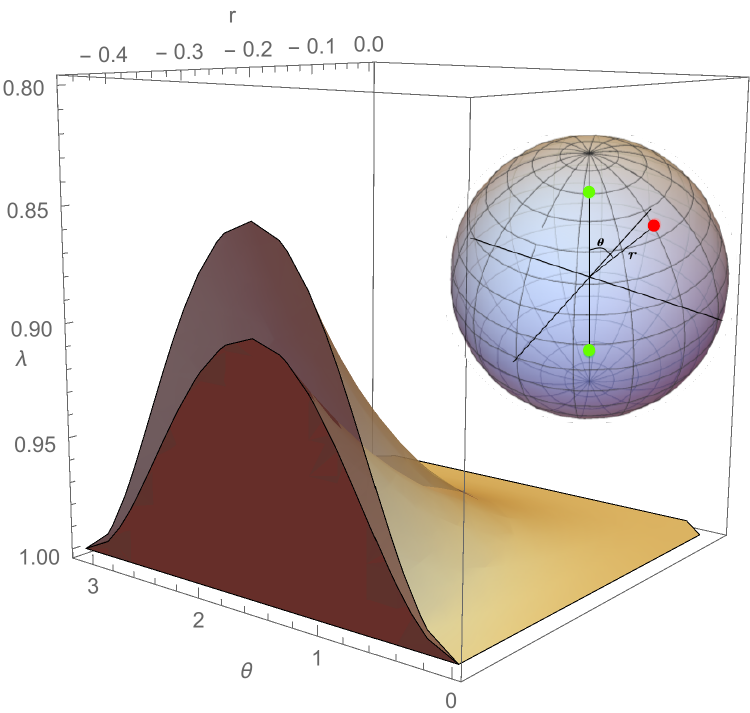}
\caption{\label{ineqplot}  Regions of the parameters $\lambda,r,\theta$ allowing for steering, detected by the inequality \eqref{Paulunbias} (inner region) and inequality \eqref{Paulbias} (outer region), with $r=\|\vec r^+\|$ and $\theta$ being the angle between $\vec r^+$ and the $z$ axis, and $\beta^+ = 0.45$ (fixed). { Inset}: representation in the Bloch sphere of the reduced states $\varrho_{\pm|1}$ (green points one upon the other) and $\varrho_{+|2}$ (red point on the right). The normalization factor $\beta^+=\tr[\varrho_{+|2}]$ is not represented. The figure is taken from \cite{uola15}.}
\end{figure}

The criteria in Eq.~(\ref{Paulunbias}) and Eq.~(\ref{Paulbias}) are both steering inequalities. The latter of them characterises all pairs of binary unsteerable assemblages in the qubit case. Labelling members of such assemblages by $\varrho_{\pm|1}=\frac{1}{4}(\openone\pm\lambda\sigma_z)$ and $\varrho_{\pm|2}=\beta^\pm\openone\pm r^\pm\sigma_z$ we present a comparison of these criteria in Fig.~\ref{ineqplot} \cite{uola15}.

As another example, we demonstrate how steering robustness defined in Eq.~(\ref{eq:steering_robustness}) translates to an incompatibility robustness \cite{uola15}. Recall that the steering robustness ${\rm SR}(\varrho_{a|x})$ of an assemblage $\{\varrho_{a|x}\}_{a,x}$ can be written as
\begin{align}
\min& \quad t\geq0 \nonumber \\
\textrm{s.t.}& \quad \frac{\varrho_{a|x} + t \gamma_{a|x}}{1+t}\text{ unsteerable} \quad \text{for all } a,x.
\end{align}
Here the optimization is over assemblages $\{\gamma_{a|x}\}_{a,x}$ and positive numbers $t$, see also Fig.~\ref{fig:sdp_quantification}.
Mapping a state assemblage into a set of POVMs, one can define an incompatibility robustness ${\rm IR}(\tilde B_{a|x})$ for a set $\{\tilde B_{a|x}\}_{a,x}$ as \cite{uola15}
\begin{align}
\min& \quad t\geq0 \nonumber \\
\textrm{s.t.}& \quad \frac{\tilde B_{a|x} + t T_{a|x}}{1+t}\text{ jointly measurable} \quad \text{for all } a,x.
\end{align}
Here the optimisation is performed over POVMs $\{T_{a|x}\}_{a,x}$ and positive numbers $t$. Note that this definition works also for a generic set of POVMs, i.e. the POVMs do not need to originate explicitly from a steering problem. To give the incompatibility robustness ${\rm IR}(A_{a|x})$ of a finite set of POVMs $\{A_{a|x}\}_{a,x}$ in the standard SDP form one writes \cite{uola15}
\begin{align}
\min& \quad \text{tr}[\frac{\sum_\lambda G_\lambda'}{d}] \nonumber \\
\textrm{s.t.}& \quad \sum_\lambda D(a|x,\lambda)G_\lambda'\geq A_{a|x}\quad \text{for all } a,x \nonumber \\
& \quad G_\lambda'\geq 0 \quad  \text{for all }\lambda \nonumber\\
& \quad \sum_\lambda G_\lambda'=\openone\ \text{tr}[\sum_\lambda G_\lambda']/d,
\end{align}
where $\{D(a|x,\lambda)\}_{a,x,\lambda}$ are deterministic post-processings.

To finish this section, we stress out that even though the normalised state assemblages appear as abstract POVMs that do not have a direct connection to the POVMs actually measured in the correlation experiment, the normalised state assemblages can be interpreted as the POVMs Alice should measure on the canonical purification of $\sum_a\varrho_{a|x}$ in order to prepare $\{\varrho_{a|x}\}_{a,x}$. Hence, we see that the results on steering and joint measurements stated in the Section \ref{sec-JMAlice} (i.e. non-jointly measurable POVMs allow steering with some shared state) are closely related to the connection presented here.

%% file: Joint-measurability/breaking-quantum-channels.tex
%In this subsection we will briefly comment on the connection of steering and joint measurements on continuous variable systems~\cite{heinosaari15,kiukas17}.

The extension of the connection between steering and joint measurements to infinite-dimensional systems and measurements with possibly continuous outcome sets has been done in \cite{kiukas17}. The point is to generalize the Choi-Jamio{\l}kowski isomorphism to arbitrary shared states (i.e. not only ones with maximally mixed marginals). The generalisation gives a one-to-one connection between states $\varrho$ with a fixed full-rank marginal $\varrho_B$ (on Bob's side) and channels $\Lambda$ from Bob to Alice through
\begin{align}
\varrho=(\Lambda\otimes\openone)(|\psi_{\varrho_B}\rangle\langle\psi_{\varrho_B}|),
\end{align}
where $|\psi_{\varrho_B}\rangle=\sum_{i=1}^d \sqrt{s_i}|ii\rangle$ is a purification of $\varrho_B=\sum_{i=1}^d s_i|i\rangle\langle i|$. To see a connection to incompatibility, one writes the channel corresponding to a given state (in the Heisenberg picture) as
\begin{align}\label{Steerincchan}
\varrho_B^{1/2}\Lambda^\dagger(A)\varrho_B^{1/2}=\text{tr}_A[(A\otimes\openone)\varrho]^T,
\end{align}
where $^T$ is a transpose in the eigenbasis of $\varrho_B$. Inputting sets of POVMs $\{A_{a|x}\}_{a,x}$ on the left-hand side of Eq.~(\ref{Steerincchan}) results in transposed state assemblages on the right-hand side. From this it is clear (at least in the finite-dimensional case in which $\varrho_B$ can be inverted) that the Heisenberg channel $\Lambda^\dagger$ sends Alice's POVMs to normalised state assemblages (i.e. POVMs) on Bob's side. Joint measurability of these POVMs is equivalent to the unsteerability of $\{\varrho_{a|x}\}_{a,x}$. It turns out that this correspondence can be extended to infinite-dimensional systems \cite{kiukas17} resulting in a fully general connection between steering and joint measurements:

\textit{A state assemblage $\{\varrho_{a|x}\}_{a,x}$ given by Alice's measurements $\{A_{a|x}\}_{a,x}$ and a state $\varrho$ is steerable if and only if the POVMs $\{\Lambda^\dagger (A_{a|x})\}_{a,x}$ are not jointly measurable.}

Note that although the notation here is adapted to the case of discrete POVMs (to avoid technicalities), the connection works also for POVMs with continuous outcome sets.

To demonstrate the power of the above result, we list some of its implications \cite{kiukas17}. First, the connection is quantitative in the sense that the incompatibility robustness of $\{\Lambda^\dagger (A_{a|x})\}_{a,x}$ coincides with the so-called consistent steering robustness (i.e. a special case of steering robustness: one allows mixing only with assemblages that have the same total state as the original assemblage) of $\{\varrho_{a|x}\}_{a,x}$. Second, for pure states the corresponding channel $\Lambda^\dagger$ is unitary, hence, extending the main results of \cite{quintino14,uola14} presented in Section \ref{sec-JMAlice} to the infinite-dimensional case. Third, the result characterises unsteerable states as those whose corresponding Choi-Jamio{\l}kowski channel is incompatibility breaking (i.e. outputs only jointly measurable observables). Finally, seemingly different steering problems (such as NOON states subjected to photon loss and systems with amplitude damping dynamics) can have the same channel $\Lambda$, hence, making it possible to solve many steering problems in one go.

%% file: Joint-measurability/Further-JM.tex
As mentioned in the beginning of this section, measurement incompatibility manifests itself in various ways in quantum theory. In this section we review briefly two well-known fine-tunings of commutativity and discuss their relation to quantum correlations.

First, in the case of joint measurability one asks for the existence of a common POVM and a set of post-processings. One can relax this concept by dropping the assumption on post-processings. Namely, we say that a set of POVMs $\{A_{a|x}\}_{a,x}$ is \textit{coexistent} if there exists a POVM $\{C_\lambda\}_\lambda$ such that
\begin{align}
A_{X|x}=\sum_{\lambda\in\tau_{X|x}}C_\lambda,
\end{align}
where $\tau_{X|x}$ is a subset of outcomes of $\{C_\lambda\}_\lambda$ for every pair $(X,x)$. Here we have used the notation $X$ to emphasize that the definition is not only required to hold for all POVM elements of $\{A_{a|x}\}_{a,x}$, but also for sums of outcomes, e.g. for $X=\{a_1,a_2\}$ one has $A_{X|x}=A_{a_1|x}+A_{a_2|x}$. To give the concept a physical interpretation, in \cite{heinosaari16b} the authors noted that the definition is equivalent to the joint measurability of the set of all binarizations (i.e. coarse-grainings to two-valued ones) of the involved measurements. Whereas it is clear that joint measurability implies coexistence (by the use of deterministic post-processings), the other direction does not hold in general \cite{reeb13,pellonpaa14}.

As coexistence is closely related to joint measurability, one can ask if anything can be learned from using this concept in the realm of steering. As pointed out in \cite{uola14} one can reach steering with coexistent measurements on the uncharacterised side (provided that the measurements are not jointly measurable). What has not appeared in the literature so far, but we wish to point out here, is that when using this concept on the characterized side, one can find examples of steerable assemblages that nevertheless form one ensemble. Consider the example of coexistent but not jointly measurable POVMs given in \cite{reeb13} by defining a vector $|\varphi\rangle=\frac{1}{\sqrt3}(|1\rangle+|2\rangle+|3\rangle)$ and the POVMs
\begin{align}
A_{i|1}&:=\frac{1}{2}(\openone -|i\rangle\langle i|),\ i=1,2,3\\
A_{+|2}&:=\frac{1}{2}|\varphi\rangle\langle\varphi|,\ A_{-|2}:=\openone - A_{+|2}.
\end{align}
To see that these POVMs are coexistent, one can define a POVM $\{C_\lambda\}_\lambda$ through the elements
\begin{align}
\Bigg(\frac{1}{2}|1\rangle\langle1|,\frac{1}{2}|2\rangle\langle2|,\frac{1}{2}|3\rangle\langle3|,\frac{1}{2}|\varphi\rangle\langle\varphi|,\frac{1}{2}(\openone -|\varphi\rangle\langle\varphi|)\Bigg).
\end{align}
For the proof that these POVMs are not jointly measurable we refer to \cite{reeb13}. Applying the mapping between measurement assemblages and state assemblages (with a full rank state $\varrho_B$, see also Section \ref{sec-JMBob}) to the above (or any similar) example, one ends up with a steerable state assemblage that nevertheless fits into a single ensemble.

As another example, we consider the concept of measurement disturbance. A POVM $\{A_a\}_a$ is called \textit{non-disturbing} with respect to a POVM $\{B_b\}_b$ if there exists an instrument $\{\mathcal I_a\}_a$ implementing $\{A_a\}_a$ (i.e. $\text{tr}[\mathcal I_a(\varrho)]=\text{tr}[A_a\varrho]$ for all states $\varrho$) such that
\begin{align}
\sum_a\text{tr}[\mathcal I_a(\varrho)B_b]=\text{tr}[\varrho B_b]
\end{align}
holds for all states $\varrho$ and all outcomes $b$.

Non-disturbance is located in between commutativity and joint measurability. Clearly commutativity implies non-disturbance by the use of the L\"uders rule and non-disturbance implies joint measurability by defining for a non-disturbing scenario $G_{a,b}=\mathcal I_a^\dagger(B_b)$, where the dagger refers to the Heisenberg picture. For a proof that the implications can not be reversed in general, and for more detailed analysis on when the implications are reversible, we refer to \cite{heinosaari10}.

As some disturbing measurements can be jointly measurable, measurement disturbance is only necessary but not sufficient for steering. One could, however, ask if there exist other types of quantum correlations or tasks for which disturbance is necessary and sufficient. It turns out that the question can be answered in positive and one answer is given by violations of typical (i.e. choose between measuring or not measuring) models of macrorealism \cite{uola19a}. More precisely, the authors of \cite{uola19a} have shown that when all classical disturbance (i.e. clumsy measurement implementation) is isolated from a quantum system, the system can violate macrorealism with some initial state if and only if the involved measurements do not fulfil the definition of non-disturbance.

Motivated by the strong connections between quantum measurement theory and quantum correlations presented in this section (see also Section \ref{sec-temporal-steering} and Section \ref{sec-selftesting}), it will be an interesting question for future research to isolate the measurement resources behind other quantum tasks. Conversely, it will be of interest to see if other concepts of incompatibility such as broadcastability \cite{heinosaari16a}, incompatibility on many copies \cite{carmeli16}, and measurement simulability \cite{oszmaniec17} will find counterparts in the realm of quantum correlations. To conclude, we note that whereas further connections between measurement theory and correlations remain unknown, jointly measurable sets \cite{skrzypczyk19,carmeli19}, or more generally all convex subsets of measurements \cite{uola19b}, can be characterized through state discrimination tasks.

%% file: Applications/Multipartite-steering.tex
The extension of steering to multipartite systems is an emerging field of 
research and different approaches for defining multipartite steering exist. 
Before explaining them, we would like to point out some peculiarities of the multipartite 
scenario, if one considers steering across a bipartition.

%%%%%%%%%%%%%%%%%%%%%%%%%%%%%%%%%%%%%%%%%%%%%%%%%%%%%%%%%%%%%%%%%%%%
\subsubsection{Steering across a bipartition}
%%%%%%%%%%%%%%%%%%%%%%%%%%%%%%%%%%%%%%%%%%%%%%%%%%%%%%%%%%%%%%%%%%%%

In order to discuss the different effects that play a role for 
steering in the multipartite scenario, consider a tripartite state 
$\varrho_{ABC}$ and investigate steering  across a given bipartition, 
say $AB|C$ for definiteness. Then, there are different scenarios that 
have to be distinguished, where in all of them Alice and Bob want to 
steer Charlie.

\begin{itemize}

\item Global steering: In the simplest case, Alice and Bob make global 
measurements on their two particles and steer Charlie. This reduces to 
a bipartite steering problem for $\varrho_{AB|C}$ and all the usual 
methods can be applied.

\item Reduced steering: Another simple case arises, if Alice (or Bob) 
try to steer Charlie, without needing the help of the other. If the 
action of one of them is not required, this reduces to the bipartite 
steerability of the reduced state $\varrho_{A|C}$ or $\varrho_{B|C}$. 
Again, all the bipartite steering theory can be applied.

\item Local steering: The interesting case arises, if Alice and Bob try 
to steer Charlie by local measurements on their respective parties. In 
this case, one has to consider the assemblage $\varrho_{ab|xy}$ and ask 
whether its elements can be written as
\begin{equation}
\varrho_{ab|xy} = \int d\lambda p(\lambda) p(a,b|x,y,\lambda) \sigma_\lambda^C. 
\label{eq-sec-multi-1}
\end{equation}
Here, we can distinguish among several cases, depending on the properties of 
$p(a,b|A,B,\lambda)$. It may be a general probability distribution, it may 
obey the non-signaling constraint, or it may factorize,
\begin{equation}
p(a,b|x,y,\lambda) = p(a|x,\lambda)p(b|y,\lambda).
\end{equation}
As $p(a,b|x,y,\lambda)$ has the interpretation of a simulation strategy 
[see Eq.~(\ref{eq-lhsmodel})] the latter means that Alice and Bob play 
an independent strategy. 
\end{itemize}

A simple example of the difference between global and local
steering can be constructed from the phenomenon of superactivation 
of steering~\cite{quintino16} (see also Section~\ref{sec-superactivation}). 
For certain states, one copy of the 
state is unsteerable, but many copies of the same state may become steerable. 
So one can consider a state $\varrho_{ABCC'} = \varrho_{AC}\otimes \varrho_{BC'}$, 
where $\varrho_{BC'}$ is a copy of $\varrho_{AC}$, and $\varrho_{AC}$ is unsteerable, 
but its steerability can be superactivated [where only two copies are already 
enough~\cite{quintino16}]. For this state, local measurements give an unsteerable 
state assemblage as
\begin{eqnarray}
\varrho_{ab|xy}^{CC'} &=& 
\textrm{Tr}_{AB}[(A_{a|x}\otimes B_{b|y}\otimes \mathbbm{1}_{CC'})\varrho_{ABCC'}] 
\nonumber \\
&=& \textrm{Tr}_{AB} [(A_{a|x}\otimes \mathbbm{1}_C)\varrho_{AC}\otimes (B_{b|y}\otimes \mathbbm{1}_{C'})\varrho_{BC'}] 
\nonumber \\
&=& \int d\lambda d\mu p(\lambda) p(\mu) p(a|x,\lambda) p(b|y,\mu)\sigma_\lambda^C\otimes \tilde \sigma_\mu^{C'} 
\nonumber \\
&=& \int d\nu p(\nu)p(a,b|x,y,\nu)\hat\sigma_{\nu}^{CC'},
\end{eqnarray}
and the state is locally unsteerable even with the restriction to a factorizing 
$p(a,b|x,y,\lambda)$. However, because of the superactivation phenomenon,
this state is steerable with global measurements.

Another possibility is to consider the bipartition $A|BC$, where Alice wants to 
steer Bob and Charlie. Here, one has to consider the ensemble $\varrho_{a|x}^{BC}$ 
and ask whether it can be written as
$
\varrho_{a|x} =\int d\lambda p(\lambda)p(a|x,\lambda)\sigma_\lambda^{BC}.
$
In this case, two possible scenarios emerge, one where the state of Bob and 
Charlie is a single system, reducing to a bipartite steering problem, and 
another one where the local hidden state of Bob and Charlie factorizes, 
i.e. $\sigma_\lambda^{BC} = \tilde\sigma_\lambda^{B}\otimes \hat\sigma_\lambda^{C}$. 

From the above picture, one can see that the extension of steering to multipartite 
systems leads to different scenarios, making its characterization even more difficult 
than the ones for entanglement and nonlocality. 

%%%%%%%%%%%%%%%%%%%%%%%%%%%%%%%%%%%%%%%%%%%%%%%%%%%%%%%%%%%%%%%%%5
\subsubsection{Different approaches towards multipartite steering}
%%%%%%%%%%%%%%%%%%%%%%%%%%%%%%%%%%%%%%%%%%%%%%%%%%%%%%%%%%%%%%%%%%%
The existing works on multipartite steering can be divided into two
different approaches. The first approach sees steering as an \new{one-sided device-independent} entanglement verification and translates this to the multipartite
scenario. The second approach asks for a multipartite system whether steering 
is possible for a given bipartition.

To discuss the first approach, we need to recall the basic definitions of the 
different entanglement classes for multipartite systems \cite{guehne09}. For a 
three-partite system $\varrho_{ABC}$ one calls the state fully separable, if it 
can be written as 
\begin{equation}
\varrho_{ABC}^{\rm fs} = \sum_k p_k \varrho_k^A \otimes \varrho_k^B \otimes \varrho_k^C,
\end{equation}
where the $p_k$ form a probability distribution. If a state is not of this form, 
it is entangled, but not all particles are necessarily entangled. For instance, 
a state of the form
$
\varrho_{A|BC}^{\rm bs} = \sum_k p_k \varrho_k^A \otimes \varrho_k^{BC}
$
may contain entanglement between $B$ and $C$, but it is separable for the 
bipartition $A|BC$ and therefore called biseparable. More generally, mixtures
of biseparable states for the different partitions are also biseparable
\begin{equation}
\varrho_{ABC}^{\rm bs} = 
p_1 \varrho_{A|BC}^{\rm bs} +
p_2 \varrho_{B|AC}^{\rm bs} +
p_3 \varrho_{C|AB}^{\rm bs},
\label{eq-gme}
\end{equation}
and states which are not biseparable are genuine multipartite entangled. These
definitions can straightforwardly be extended to more than three particles. 

\new{One-sided device-independent} entanglement detection in the multipartite scenario 
has first been discussed by \cite{cavalcanti11}. There criteria for full 
separability in the form of Mermin-type inequalities have been given, which
hold for $k$ trusted sites and $N-k$ untrusted sites. These criteria can be 
violated in quantum mechanics, and the possible violation increases exponentially 
with the number of parties. Inequalities for higher-dimensional systems have 
been derived in~\cite{he11}.

In general, one if has a quantum network of $N$ parties where some of the parties 
perform untrusted measurements, the parties which trust their measurement apparatus 
can perform quantum state tomography, and reconstruct the conditional state after 
the untrusted parties announce their measurement choices and outcomes. For three 
parties there are two \new{one-sided device-independent} scenarios: when only one party's 
device is untrusted, with state assemblage
\begin{equation}
\label{sec-mp-assem-1}
\varrho_{a|x}^{BC} = \text{Tr}_A(A_{a|x}\otimes \mathbbm{1}_B\otimes\mathbbm{1}_C \varrho_{ABC}),
\end{equation}
and when two of them are untrusted
\begin{equation}
\label{sec-mp-assem-2}
\varrho_{ab|xy}^{C} = \text{Tr}_{AB}(A_{a|x}\otimes B_{b|y}\otimes\mathbbm{1}_C \varrho_{ABC}).
\end{equation}
If $\varrho_{ABC}$ is biseparable, this condition imposes constraints on the assemblages. Then, 
for a given state, to test whether the assemblages of the form~\eqref{sec-mp-assem-1} 
or~\eqref{sec-mp-assem-2} obey the conditions, one can use SDPs \cite{cavalcanti15a}. 
Also entropic conditions for this scenario have been studied \cite{riccardi18, costa18}.

The second approach uses steering between the bipartitions to define genuine multipartite steering \cite{he13}. First, for two parties one can say that
they share steering if the first one can steer the other or vice versa.
Then, for three parties one can define genuine multipartite steerability 
as the impossibility of describing a state with a model where steering is shared
between two parties only. This means that the state cannot be  described
by mixtures of bipartitions as in Eq.~(\ref{eq-gme}), where for each partition
(e.g., $A|BC$) the two-party state (e.g., $BC$) is allowed to be steerable.

One can then directly see that for checking this criterion, it is sufficient 
to consider the bipartitions $AB|C$, $AC|B$, and $BC|A$, where the two-party 
sites are uncharacterized and the single-party site obeys quantum mechanics. 
In addition, on the two-party site only local measurements are allowed, but
the results only have to obey the non-signaling condition. For proving 
genuine multipartite steering in this sense, several methods are possible. If
the state is pure, it suffices to check the steerability for the mentioned
bipartitions, as a pure state convex combinations into different bipartitions
are not possible \cite{he13}. Otherwise, one may derive a linear (or convex) 
inequality that holds for unsteerable states of all relevant bipartitions. Due to linearity, it also holds for convex combinations and violation rules out the model
mentioned above. This approach has been experimentally used in 
\cite{armstrong15,li15}.

%% file: Applications/Steering-ellipsoid.tex
\new{
Note that the definition of quantum steering Eq.~\eqref{eq-lhsmodel} requires considering the ensembles of unnormalized conditional states at Bob's side. However, one can expect that important insights can be gained by simply studying the normalized version of these conditional states. Note that in doing so, two things are lost: the steering ensemble to which a conditional state belongs, and the probability with which the conditional state is steered to.} 

\new{For two-qubit states, the normalized conditional states Alice can steer Bob's system to form an ellipsoid inside Bob's Bloch sphere, referred to as the steering ellipsoid~\cite{verstraete02a,shi11a,shi12a,jevtic14}. Detailed analysis of their geometry has led to the proposal to use them as a tool to represent two-qubit quantum states, in a way similar to the Bloch representation of states of a single qubit. In particular, given the reduced states of both parties, a steering ellipsoid on one side allows recovering of the density operator upto a certain local unitary or anti-unitary operation on the other side~\cite{jevtic14}. Special attention later on was attracted to the volumes of the steering ellipsoids~\cite{jevtic14,milne14a,cheng16a,paternostro17,zhang19a}. In particular,~\citet{milne14a,milne15a} showed that the volumes of the steering ellipsoids give upper bounds for the entanglement of the state in terms of its concurrence. Even more interestingly, it is shown that the volumes of the steering ellipsoids obey certain monogamy relations~\cite{milne14a,milne15a,cheng16a}, which will be discussed in the following.

Consider a system of three qubits $ABC$. Denote the volume of the steering ellipsoids for steering from $A$ to $B$ and $A$ to $C$ by $V_{B|A}$ and $V_{C|A}$, respectively.~\citet{milne14a,milne15a} showed that for all pure states of the system of three qubits $ABC$, one has
\begin{equation}
\sqrt{V_{B|A}} +\sqrt{V_{C|A}} \le \sqrt{4\pi/3}. 
\label{eq:milne_monogamy}
\end{equation}
The authors also showed that the famous Coffman-Kundu-Wootters monogamous inequality for entanglement~\cite{Coffman00a} can be derived from this inequality.
However,~\citet{cheng16a} showed that the monogamy relation Eq.~\eqref{eq:milne_monogamy} is violated when the three qubits are in certain mixed states. Instead, the authors showed that a weaker monogamy relation can be derived for all possible states over the three qubits,
\begin{equation}
{V_{B|A}}^{2/3} + {V_{C|A}}^{2/3} \le (4\pi/3)^{2/3}. 
\label{eq:cheng_monogamy}
\end{equation}
Recently, both the monogamy relation Eq.~\eqref{eq:cheng_monogamy} and the violation of Eq.~\eqref{eq:milne_monogamy} were illustrated experimentally~\cite{zhang19a}.

It is in fact the analysis of the geometry of the steering ellipsoids for Bell-diagonal states that leads to the exact characterization of quantum steering for this family of states~\cite{jevtic15,chau16b}; see also Section~\ref{sec-full-information}. Beyond the Bell-diagonal states, little is known about the extent to which the steering ellipsoids, or their volumes, can characterize the quantum steerability of the state. In particular, an interesting question for future research might be whether the monogamy relations Eq.~\eqref{eq:milne_monogamy} and Eq.~\eqref{eq:cheng_monogamy} can induce certain monogamy relations between some measures of steering such as the critical radii as defined in Section \ref{sec-full-information}.
}

%% file: Applications/Gaussian-steering.tex
%In this part we review the main results on steering of Gaussian states with 
%Gaussian observables~\cite{wiseman07,jones07,kogias15a}. Moreover, we summarize the results on monogamy of Gaussian steering. Namely, we show that in the Gaussian regime steering is monogamous~\cite{reid13,xiang17} and that outside this regime the monogamy breaks~\cite{ji16}.

Before discussing Gaussian steering some preliminary notions are needed. First, Gaussian systems refer to a special class of continuous variable scenarios. Hence, one deals with infinite-dimensional Hilbert spaces $\otimes_{j=1}^N\mathcal L^2(\mathbb R)$, where the index $N$ refers to the number of modes. Every Gaussian state is described by a real symmetric matrix, the so-called covariance matrix $V$ satisfying
\begin{equation}\label{GaussCovState}
V + i\Omega\geq 0,
\end{equation}
where $\Omega = \oplus_{j=1}^N \left( \begin{smallmatrix} 0 & 1 \\ -1 & 0\end{smallmatrix}\right)$. More precisely, the covariance matrix of a quantum state $\varrho$ is given as $(V)_{ij}=\text{Tr}[\varrho\{R_i-r_i,R_j-r_j\}]$, where $R=(Q_1,P_1,\dots,Q_n,P_n)^T$ with quadrature operators $Q_i$ and $P_j$ satisfying $[Q_i,P_j]=i\delta_{ij}\openone$ and $[Q_i,Q_j]=[P_i,P_j]=0$, and $r_j=\text{tr}[\varrho R_j]$. Moreover, every real symmetric matrix satisfying Eq.~(\ref{GaussCovState}) defines a Gaussian state. The use of the word Gaussian in this context originates from the fact that the above described states correspond to the ones whose characteristic function $\hat\varrho(x):=\text{tr}[W(x)\varrho]$ is Gaussian. Here $W(x)=e^{-ix^T R}$ with $x=(q_1,p_1,...,q_n,p_n)^T$ and
\begin{align}
\hat\varrho(x)=e^{-\frac{1}{4}x^TVx-ir^Tx}.
\end{align}

Second, a Gaussian measurement is a POVM $M_a$ (with values in $a\in\mathbb R^d$) whose outcome distribution for any Gaussian state is Gaussian. Such POVMs correspond to triples $(K,L,m)$ satisfying
\begin{equation}
L - iK^T\Omega K\geq 0,
\end{equation} 
where $K$ is an $N\times d$ matrix, $L$ is an $d\times d$ matrix, and $m$ is a displacement vector. The correspondence between the POVM $M_a$ and the triple $(K,L,m)$ is given through the operator-valued characteristic function as
\begin{align}
\hat M(p):=&\int da e^{ip^T a}M_a \nonumber \\
=&W(Kp)e^{-\frac{1}{4}p^TLp-im^Tp}.
\end{align}

With these definitions we are ready to state the characterisation of steerable states in Gaussian systems originally given in \cite{wiseman07}.

\textit{A bipartite Gaussian state with covariance matrix $V_{AB}$ and displacement $r_{AB}$ is unsteerable with Gaussian measurements if and only if}
\begin{equation}
V_{AB}+i(0_A \oplus \Omega_B)\geq 0.
\end{equation}
Here $0_A$ is a zero matrix on Alice's side and $\Omega_B$ is the matrix $\oplus_{j=1}^N \left( \begin{smallmatrix} 0 & 1 \\ -1 & 0\end{smallmatrix}\right)$ on Bob's side.

In contrast to other steering scenarios, the Gaussian case appears special in that the steerability of a state can be characterized through an easy to evaluate inequality. This is, however, not the only special feature for Gaussian steering. Namely, within the Gaussian regime one can also prove monogamy relations for steering with more than two parties \cite{reid13,ji15b,adesso16,lami16}. One should note that the monogamy can break when one is allowed to perform non-Gaussian measurements \cite{ji16}.

%% file: Applications/EPR-revisited.tex
%As a special case of interest we will bring up the setup of the original EPR argument, 
%i.e. steerability with position and momentum observables. We focus on Gaussian as well as on 
%specific non-Gaussian (i.e. lossy singlet) states~\cite{wiseman07,kiukas17,kogias15b}. Whereas the Gaussian case can be completely resolved, the non-Gaussian case of photon loss 
%will be briefly reviewed through the joint measurability connection~\cite{kiukas17} and 
%the moment matrix approach~\cite{kogias15b}.

As a special case of interest in the Gaussian regime we discuss steering with canonical quadratures. It was shown in \cite{kiukas17} that steerability of a given state in the Gaussian scenario can be readily detected by a pair of quadrature observables.

To be more concrete, we sketch the construction of the quadratures from \cite{kiukas17}. First, a channel is called Gaussian if it maps Gaussian states to Gaussian states. Gaussian channels between systems of $n$ and $m$ degrees of freedom correspond to triples $(M,N,c)$ with $M$ being a real $2n\times 2m$ matrix, $N$ being a real $2m\times 2m$ matrix, and $c$ being the displacement, that satisfy
\begin{equation}\label{GaussChannelCov}
N - iM^T\Omega M+i\Omega\geq 0.
\end{equation}
The transformation of Gaussian states on the level of covariance matrices is given as
\begin{align}
V\mapsto M^T VM + N,\quad r\mapsto M^Tr+c.
\end{align}
Given that a bipartite Gaussian state has a corresponding Choi-Jamio{\l}kowski channel with parameters $(M,N,c)$, one first notes that the state is unsteerable with Gaussian measurements if and only if the channel parameters define also a Gaussian measurement \cite{kiukas17}. Hence, for a steerable state there exists two vectors $x$ and $y$ such that $(y^T-ix^T)(N - iM^T\Omega M)(y+ix)<0$. As the triple $(M,N,c)$ also fulfils Eq.~(\ref{GaussChannelCov}), we have $r:=x^T\Omega y>0$ and
\begin{equation}
(M\tilde x)^T\Omega M\tilde y>\frac{1}{2}(\tilde x^T N\tilde x + \tilde y^T N\tilde y),
\end{equation}
where $\tilde x=r^{-1/2} x$ and $\tilde y=r^{-1/2}y$. From here one can construct two canonical quadratures as $Q_{\tilde x}=\tilde x^T R$ and $P_{\tilde y}=\tilde y^T R$. These are canonical as by definition $\tilde x^T \Omega\tilde y=1$. To see that the state is indeed steerable with these measurements we refer to \cite{kiukas17}. To summarise, we state the following refined characterisation of Gaussian steering \cite{kiukas17}:

\textit{For a bipartite Gaussian state $\varrho_{AB}$ with a covariance matrix $V_{AB}$ and displacement $r_{AB}$ the following are equivalent:}\\
(i) $\varrho_{AB}$ is steerable with Gaussian measurements\\
(ii) $\varrho_{AB}$ is steerable with some pair of canonical quadratures\\
(iii) $V_{AB}+i(0_A \oplus \Omega_B)$ is not positive semi-definite.

%To enter the more specific question about steerability with position and momentum observables, one could use the above list of techniques to build a wide range of necessary criteria. We wish, however, to provide a specific (non-Gaussian) example in which the techniques provide a necessary and sufficient criterion for steerability. The example is a noisy NOON state given as
%\begin{align}
%\rho_\eta=\eta|N00N\rangle\langle N00N|+(1-\eta)|00\rangle\langle 00|,
%\end{align}
%where $0\leq\eta\leq 1$, which we wish to steer using position and momentum observables. In [...] it was shown using the moment matrix approach that the state is steerable for $\eta\geq 2/3$. Using incompatibility related techniques, the authors of [...] showed how techniques for building joint observables can be translated to a local hidden state model for the example in hand for $\eta\leq 2/3$. As the connection between joint measurability and steering is one-to-mane, one should notice that despite the specific nature of the example, the solution has also been used in probing non-Markovianity through steering [...].

%To start with, let us list the typical methods for steering detection in continuous variable systems: Gaussian covariance matrix criterion [...], correlation matrix method [...], binning the outcomes [...] and incompatibility related techniques [...]. 

%% file: Applications/Temporal-channel-steering.tex
%Here we will discuss a temporal analogue of steering \cite{chen14} together with its applications~\cite{chen16a,chen17}. Also, we will explain the concept of steerability of a channel, first introduced in \cite{piani15a}. We discuss how the channel framework can be used to combine all three steering scenarios and map them one-to-one to joint measurability~\cite{uola17}. Moreover, this approach gives a simple way to prove a hierarchy of temporal correlations similar to that introduced in Section~\ref{sec-hierarchy} ~\cite{uola17}.

So far we have concentrated on steering in spatial scenarios, i.e. scenarios where Alice and Bob are space-like separated. Some efforts for defining similar concepts in temporal scenarios, i.e. scenarios where Alice and Bob form a prepare-and-measure type scenario, and on the level of quantum channels have also been pursued in the literature \cite{piani15a,chen14,chen16a,chen17}. In a temporal scenario (consisting of two measurement times), one can ask whether a state assemblage resulting from measurements at the first time step allows a local hidden state model on the second time step. Of course, in temporal scenarios signalling is possible and, hence, such models are sometimes trivially violated. Despite signalling, temporal steering has found applications in non-Markovianity \cite{chen16a} and in QKD \cite{bartkiewicz16}, and some criteria \cite{chen14} and quantifiers \cite{bartkiewicz16} have been developed. As the criteria and quantifiers resemble strongly those of spatial steering presented in Sections \ref{sec-correlations} and \ref{sec-assemblages}, we don't wish to go through them in detail.

In channel steering \cite{piani15a} one is interested in instrument assemblages instead of state assemblages. Namely, given a quantum channel $\Lambda^{C\rightarrow B}$ from Charlie to Bob and its extension $\Lambda^{C\rightarrow A\otimes B}$, one asks if an assemblage defined as
\begin{align}
\mathcal I_{a|x}(\cdot):=\text{tr}_A[(A_{a|x}\otimes\openone)\Lambda^{C\rightarrow A\otimes B}(\cdot)],
\end{align}
where $\{A_{a|x}\}_{a,x}$ is a set of POVMs, can be written as
\begin{align}
\mathcal I_{a|x}(\cdot)=\sum_\lambda p(a|x,\lambda)\mathcal I_\lambda(\cdot)
\end{align}
for some instrument $\{\mathcal I_\lambda\}_\lambda$ (i.e. a collection of CP maps summing up to a quantum channel) and classical post-processings $\{p(a|x,\lambda)\}_{a,x,\lambda}$. Whenever this is the case, the instrument assemblage $\{\mathcal I_{a|x}\}_{a,x}$ is called unsteerable.

The concept of channel steering relates to the coherence of the channel extension. Namely, a channel extension $\Lambda^{C\rightarrow A\otimes B}$ is coherent if it can not be written as
\begin{align}
\Lambda^{C\rightarrow A\otimes B}(\cdot)=\sum_\lambda\mathcal I_\lambda(\cdot)\otimes\sigma_\lambda
\end{align}
for some instrument $\{\mathcal I_\lambda\}_\lambda$ and states $\{\sigma_\lambda\}_\lambda$. Any extension that is of this form is called incoherent. One can show that incoherent extensions always lead to unsteerable instrument assemblages and any unsteerable instrument assemblage can be prepared through some incoherent extension \cite{piani15a}. Note that in spatial steering any separable state leads to an unsteerable assemblage and any unsteerable assemblage can be prepared with a separable state \cite{moroder16}.

In the original paper defining channel steering \cite{piani15a}, the concept is mainly probed through channel extensions as above. This leads to some connections with state-based correlations. For example, an extension can lead to a steerable instrument assemblage if and only if its Choi state allows Alice to steer Bob, and an extension is incoherent if and only if the Choi state is separable in the cut $A|BC'$, where $C'$ is the extra input system from the isomorphism.

One can investigate the channel protocol by replacing the extension with a (minimal) dilation. For completeness, we note that a minimal dilation of a channel $\Lambda:\mathcal L(\mathcal H)\mapsto\mathcal L(\mathcal H)$ can be written as $\Lambda(\cdot)=\text{tr}_A[V(\cdot)V^\dagger]$, where $V|\psi\rangle=\sum_{k=1}^n|\varphi_k\rangle\otimes(K_k|\psi\rangle)$ for all $|\psi\rangle\in\mathcal H$, $\{K_k\}_{k=1}^n$ form a linearly independent Kraus decomposition of $\Lambda$, and $\{|\varphi_k\rangle\}_k$ is an orthonormal basis of the ancillary system. In this case, the correspondence between instruments and POVMs on the dilation is one-to-one and is given through
\begin{align}
\mathcal I_{a|x}(\cdot)=\text{tr}_A[(A_{a|x}\otimes\openone)V(\cdot)V^\dagger].
\end{align}
This generalises directly the connection between joint measurements and spatial steering to the level of channel steering \cite{uola18}. Namely, a measurement assemblage $\{A_{a|x}\}_{a,x}$ on the minimal dilation is jointly measurable if and only if the corresponding instrument assemblage is unsteerable. By noticing, furthermore, that channel steering with trivial inputs (i.e. one-dimensional input system) corresponds to spatial steering, and that in this case a dilation corresponds to a purification of the total state of the assemblage, one recovers the connection between joint measurements and spatial steering.

The dilation technique can also be used to prove that any non-signalling state assemblage originates from a set of non-signalling instruments \cite{uola18}, hence, showing that channel steering captures non-trivial (i.e. non-signalling) instances of temporal steering (with two time steps), and that in this case a connection between temporal steering and joint measurements follows directly from the one between channel steering and incompatibility. Moreover, using the channel framework one can translate concepts from spatial to temporal scenarios. One example of this is given in \cite{uola18} showing that temporal steering and violations of macrorealism respect a similar strict hierarchy as spatial steering and non-locality. Note that in \cite{ku18a} the hierarchy was proven independently.

%% file: Applications/Quantum-key-distribution.tex
In quantum key distribution (QKD) two main types of protocols can be distinguished 
\cite{scarani09}. In prepare \& measure (PM) schemes, such as the BB84 protocol, Alice 
prepares some quantum states and sends them to Bob, who performs measurements on 
them. Using classical communication, Alice and Bob can then try to generate a 
secret key from the measurement data. In entanglement-based (EB) schemes, such 
as the E91 protocol, an entangled quantum state is distributed to Alice and Bob, 
and both make measurements on their part of the state. The source of the state might
be under control of an eavesdropper Eve. Again, the measurement data
are then used to generate a secret key. 

A central result concerns the role of entanglement for security. In \cite{curty04}
it was proved that entanglement is a necessary precondition for security. For
EB schemes, this means that if the measurement data can be explained by a separable
state, then no secret key can be distilled. For PM schemes, one can consider an 
equivalent EB scheme, then the same statement holds. It should be noted, however, 
that the provable presence of entanglement was not shown to be sufficient for 
secret key generation. The question of whether entanglement can be verified depends on
the measurement data taken and the assumptions made on the measurements. In a 
device-independent scheme, where no assumptions about the measurements are made, 
only Bell inequalities can be used to test the presence of entanglement. Still,
device-independent QKD can be proved to be secure against certain
attacks \cite{acin07}.

One can also consider an asymmetric situation, where one party trusts its devices 
and the other one does not. This can be realistic, e.g. if Alice corresponds to a 
client of a bank having only a cheap device, while Bob represents the bank itself.
Clearly, in such a situation QKD can only work if the underlying state is steerable.
In \cite{branciard12} this problem has been considered for the BBM92 protocol 
\cite{bennett92}. In this protocol, Alice and Bob share a two-qubit Bell state
and measure either $A_1= B_1 =\sigma_z$ or $A_2 =B_2=\sigma_x.$ The correlations for
the measurement $A_1 \otimes B_1$ are used for the key generation, while the 
correlations in the $A_2 \otimes B_2$ measurement are used to estimate Eve's 
information.

\begin{figure}[t!]
\includegraphics[width=0.9\columnwidth]{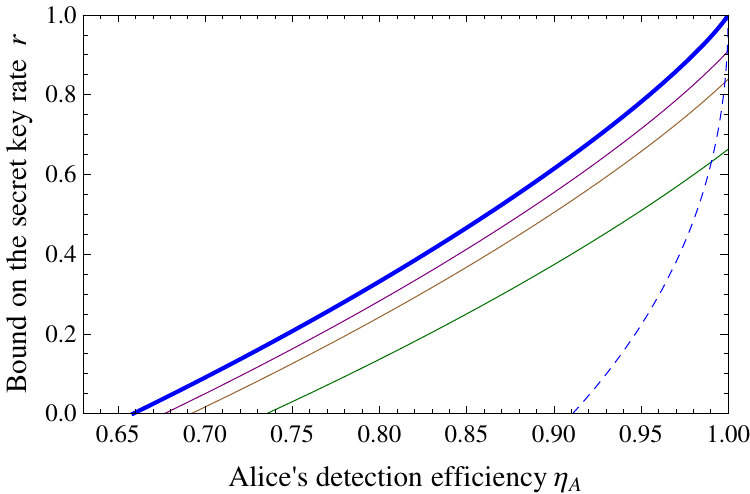}
\caption{Key rates for the QKD based on steering. For different visibilities
of the initial state ($V \in \{1, 0.99,0.98,0.95\})$ lower bounds on the key 
rate are shown. For perfect visibility (solid blue line) a key can be 
extracted for detector efficiencies of $\eta \geq 0.659$ for Alice. The 
dashed line shows a bound (obtained with the same methods) for the fully 
device-independent scenario for the case of perfect visibility. The figure 
is taken from \cite{branciard12}.}
\label{fig:steeringqkd}
\end{figure}

In \cite{branciard12} the security of one-sided device-independent QKD using
this protocol against attacks where Eve has no quantum memory has been studied
(see also Fig.~\ref{fig:steeringqkd}). It has been shown that only the detector
efficiency of the untrusted party matters and that already for detector 
efficiencies $\eta \geq 0.659$ a secret key can be distilled, and a non-zero key 
rate proves that the underlying states are steerable. The obtainable key rates are 
higher and the required detector efficiencies are lower than in the fully 
device-independent case. 

In addition, results on steering and PM schemes for QKD have been obtained 
in \cite{branciard12, ma12}. An analysis for finite key length can be found
in \cite{wang13, zhou17}, and upper bounds on the key rate in one-sided device-independent 
QKD have been obtained in \cite{kaur18}. Finally, it should be noted that 
similar ideas have also been studied and implemented for QKD with continuous 
variables \cite{gehring15, walk16}.

%% file: Applications/Randomness-verification.tex
The task of randomness certification can be defined as follows \cite{acin12, law14}. 
On a quantum system $\varrho$ a measurement labeled by $z$ is made and the result $c$ 
is obtained. Depending on the situation, the measurement may be a joint measurement 
on two parties of an entangled state; then the labels for the measurement and result 
can be written as $z=(x,y)$ and $c=(a,b)$, as in the Bell scenario in Eq.~(\ref{eq-lhvmodel}). 
The task is to quantify the extent to which an external adversary Eve can predict the outcome 
$c$ of the probability distribution $p(c|z)$. Clearly, this depends on the assumptions made
about Eve: For instance, one can distinguish the case where the state $\varrho$ is fixed and 
only known to Eve from the case where Eve indeed provides the state. In the former case, one can 
furthermore distinguish the knowledge Eve has. She may have only classical information 
about the state or she may hold a purification of it, see \cite{law14} for a detailed 
discussion.

In the simplest case, the state $\varrho=\ketbra{\psi}{\psi}$ is pure and the measurement
$z$ is characterized. Then, the best strategy for Eve is to guess the $c$ with the
maximal probability, and the probability of guessing correctly is given by
\begin{equation}
G(z, {\psi}) = \max_{c} p(c|z,{\psi}).
\end{equation}
If the state $\varrho$ is mixed then Eve may hold a purification of it and she may 
know the exact decomposition $\varrho= \sum_k p_k \ketbra{\phi_k}{\phi_k}$ into pure states.
Consequently, the maximal guessing probability is
\begin{equation}
G(z, \varrho) = \max_{p_k, \phi_k} \sum_k p_k G(z, \phi_k), 
\end{equation}
where the maximization runs over all decompositions  of $\varrho$. If the measurements 
$z$ are not characterized, one has to optimize over all possible quantum realizations
of the classical probability distribution $p(c|z)$. So, the maximal guessing probability 
is
\begin{equation}
G(z, p(c|z)) = \max_{\varrho, M_{c|z}} G(z, \varrho),
\end{equation}
where the maximization runs over all quantum realizations, described by a state $\varrho$ 
and measurement operators $M_{c|z}$ with $p(c|z)=\tr(M_{c|z}\varrho).$ Optimizations over
this set can be carried out by hierarchies of SDPs \cite{navascues07, navascues08}. In all 
cases, the number of random bits that one can extract from $p(c|z)$ is given by the 
min-entropy, $H_{\rm min}(G)=-\log_2(G).$

Initially, the task of randomness certification was mainly studied in the Bell scenario, 
where $z=(x,y)$ and $c=(a,b)$ describe measurements on an entangled state \cite{acin12}. 
Here, the devices are not characterized and as soon as a Bell inequality is violated,  
one can prove that the results of a fixed setting cannot be predicted, so the randomness 
is certified. In \cite{law14} randomness certification has been studied for the steering 
scenario: Again, one makes local measurements on an entangled state, but 
this time the devices on Bob's side are characterized. This leads to additional constraints
in the SDP hierarchy \cite{navascues07} and consequently more randomness can be extracted.
Interestingly, also for states that are not steerable the randomness can be certified; so
the
violation of a steering inequality is not necessary for randomness certification in the 
one-sided device-independent scenario.

Also in \cite{passaro15} the task of randomness certification in the distributed
one-sided device-independent scenario between two parties was studied. But here mainly 
the randomness for a single measurement setting of Alice was considered. 
It has been shown that this can directly be computed with an SDP, without the need of
a convergent hierarchy of SDPs. This is then shown to hold also for the scenario 
considered in \cite{law14}. In \cite{skrzypczyk18} the problem has been considered for
two $d$-dimensional systems. A steering inequality has been derived, such that the 
maximal violation guarantees $\log(d)$ random bits for Alice's outcomes in the 
one-sided device-independent scenario. Furthermore, any pure entangled state with 
full Schmidt rank can be used to generate this amount of randomness.

Finally \cite{curchod17} showed that if one considers the Bell scenario, then 
sequential measurements on one party can lead to an unbounded generation of 
randomness. The extension of this to the steering scenario is discussed in 
\cite{coyle18}.

%As another semi-device independent application of steering we will mention randomness verification \cite{law14}.

%% file: Applications/Subchannel-discrimination.tex
%As a direct application of the steering protocol we will review subchannel discrimination, i.e. the discrimination of branches in a quantum evolution. Steerable states can be shown to provide a quantum advantage in a subchannel discrimination task \cite{piani15}. This provides naturally an implicit characterisation of steerable states.

In \cite{piani15b} an operational characterisation of steerable quantum states is provided. The idea is similar to the main result of \cite{piani09} stating that every entangled state provides an advantage over separable ones in some channel discrimination task to the realm of steering. In the case of steering, the related task turns out to be that of subchannel discrimination, i.e. discriminating different branches of a quantum evolution. Namely, take an instrument $\mathcal I=\{\mathcal I_a\}_{a}$ (i.e. a collection of completely positive maps summing up to a quantum channel), a POVM $B=\{B_b\}_b$, an input state $\varrho$ and define the probability of correctly identifying the subchannel (i.e. instrument element) as
\begin{align}
p_{\text{cor}}(\mathcal I,B,\varrho):=\sum_a\text{tr}[\mathcal I_a(\varrho)B_a].
\end{align}
To find the best strategy for the task, one maximises over input states and POVMs on the output.

As mentioned above, entanglement provides an advantage in channel discrimination tasks, i.e. tasks of discriminating between subchannels of the form $\mathcal I_a=p(a)\Lambda_a$, where $\{\Lambda_a\}_a$ are quantum channels. To prove a similar result for general subchannels, the authors of \cite{piani15b} limit the set of allowed measurements between the system (i.e. outputs of the instruments) and the ancilla to local measurements supported by forward communication from output to ancilla (i.e. one-way LOCC measurements). Such measurements have POVM elements of the form $C_a^{\text{out}\rightarrow\text{anc}}=\sum_x A_{a|x}\otimes B_x$, where $\{B_x\}_x$ is a POVM on the output system and $\{A_{a|x}\}_a$ is a POVM on the ancilla for every $x$. The probability of correctly identifying the branch of the evolution with such measurements and a shared state $\varrho_{AB}$ is given as
$p_{\text{cor}}(\mathcal I,1\text{-LOCC},\varrho_{AB})=\sum_{a,x}\text{tr}_{\text{out}}[\mathcal I_a^\dagger(B_x)\varrho_{a|x}]$.
Note that any unsteerable state can perform at most as good as some single system state. One sees this by using an LHS model for the assemblage in the above equation and by choosing the best performing hidden state as the single system state.

To prove the main result of the paper, the authors define a quantity called steering robustness of a bipartite state $\varrho_{\text{AB}}$ by maximising the steering robustness of all possible assemblages the state allows. More formally
\begin{align}
R_{\text{steer}}^{A\rightarrow B}(\varrho_{AB})=\text{sup}\{R(A)|\{A_{a|x}\}_{a,x}\},
\end{align}
where $R(A)$ is the steering robustness of the assemblage $\varrho_{a|x}=\text{tr}_A[(A_{a|x}\otimes\openone)\varrho_{AB}]$. Clearly the quantity $R_{\text{steer}}^{A\rightarrow B}(\varrho_{AB})$ is zero if and only if the state $\varrho_{AB}$ is unsteerable. The main result now reads:

\textit{For any steerable state there exists a subchannel discrimination task (with forward communication from the output to the ancilla) in which the state performs better than any unsteerable one, i.e.}
\begin{align}
\text{sup}\frac{p_{\text{cor}}(\mathcal I,1\text{-LOCC},\varrho_{AB})}{p_{\text{cor}}^{NE}(\mathcal I)}=1+R_{\text{steer}}^{A\rightarrow B}(\varrho_{AB}).
\end{align}
Here the supremum is taken over all instruments and one-way LOCC measurements from the output to the ancilla. Note that this result gives the set of steerable states an operational characterisation. Experimental demonstration of this result has been presented in \cite{sun18}.

We note that the result on steering robustness can be generalised. One can define a robustness measure for any convex and closed subset of assemblages, and reach a similar conclusion as above using conic programming \cite{uola19b}. Moreover, one can show that a related measure called convex weight or free fraction has a similar interpretation: whereas robustness measures discrimination power, the free fraction is a measure of exclusivity \cite{uola19c}.

%% file: Applications/Self-testing.tex
As steering is closely related to joint measurability and non-locality, some works \cite{cavalcanti16,chen16b} have pursued ways of deriving one-sided device-independent and device-independent bounds on measurement incompatibility. The idea is to show that quantifiers of incompatibility (e.g. incompatibility robustness) are lower bounded by quantifiers of steering (e.g. steering robustness), which in turn are lower bounded by non-locality quantifiers (e.g. non-locality robustness). Hence, on top of the one-sided device-independent and fully device-independent lower bounds on incompatibility, one gets also a device-independent lower bound on a quantifier of steering.

We follow \cite{cavalcanti16} to make the aforementioned hierarchy more concrete. It is worth mentioning that the hierarchy presented here corresponds to one choice of quantifiers. Analogous results are possible for various fine-tuned quantifiers.

To write down the result, recall the definitions of incompatibility, steering and non-locality robustness. Steering robustness is defined in Eq.~(\ref{eq:steering_robustness}) and analogously to that one defines incompatibility robustness ${\rm IR}(A_{a|x}) $ of a set $\{A_{a|x}\}_{a,x}$ of POVMs as
\begin{align}
\min& \quad t \nonumber \\
\textrm{s.t.}& \quad \frac{A_{a|x} + t N_{a|x}}{1+t} = \sum_\lambda D (a|x,\lambda)G_\lambda \quad\text{for all } a,x \nonumber \\
& \quad t\geq 0 \nonumber \\
& \quad N_{a|x}\geq 0 \quad \text{for all } a,x, \quad \sum_a N_{a|x} = \openone \quad \text{for all } x \nonumber \\
& \quad G_\lambda\geq 0 \quad \text{for all } \lambda, \quad \sum_\lambda G_\lambda = \openone.
\end{align}
Note that here $D(\cdot|x,\lambda)\in\{0,1\}$ is a deterministic assignment for every $x$ and $\lambda$. The interpretation of this robustness is that one mixes the POVMs $\{M_{a|x}\}_{a,x}$ with $\{N_{a|x}\}_{a,x}$ until they become jointly measurable.

Now, if Alice's measurements $\{A_{a|x}\}_{a,x}$ in a steering scenario with a state $\varrho_{AB}$ have incompatibility robustness $t$, then replacing Alice's measurements with the jointly measurable POVMs $\frac{A_{a|x} + t N_{a|x}}{1+t}$ shows that $t$ is an upper bound for the steering robustness of $\sigma_{a|x}:=\text{tr}_A[(A_{a|x}\otimes\openone)\varrho_{AB}]$. In other words, steering robustness of a given assemblage lower bounds the incompatibility robustness of the measurements on the steering party. This bound, moreover, is one-sided device-independent. Notice that with a fine-tuned steering quantifier called consistent steering robustness the aforementioned inequality is tight for full Schmidt rank states \cite{cavalcanti16}, see also \cite{kiukas17}.

For the device-independent quantification of steering and incompatibility one can use the non-locality robustness ${\rm NR}[p(a,b|x,y)]$ of a probability table $\{p(a,b|x,y)\}_{a,b,x,y}$ given as \cite{cavalcanti16}
\begin{align}
\min& \quad r \nonumber \\
\textrm{s.t.}& \quad \frac{p(a,b|x,y) + r q(a,b|x,y)}{1+r} \nonumber \\
=& \quad \sum_{\lambda,\mu} p(\lambda,\mu) D (a|x,\lambda) D(b|y,\mu) \ \text{for all } a,b,x,y \nonumber \\
& \quad r\geq 0, \quad p(\lambda,\mu)\geq 0 \nonumber \\
& \quad q(a,b|x,y)\in Q,
\end{align}
where $Q$ is the set of all possible quantum correlations defined as
\begin{align}
Q=\{&\text{tr} [(A_{a|x} \otimes B_{b|y})\varrho_{AB}]| \nonumber\\
&\{A_{a|x}\}_{a,x},\ \{B_{b|y}\}_{b,y} \text{ POVMs }, \varrho_{AB} \text{ a state} \}.
\end{align}
Similarly to the one-sided device-independent quantification of incompatibility above, one sees that for a given state assemblage $\{\sigma_{a|x}\}_{a,x}$ the non-locality robustness of any probability table originating from this assemblage, i.e. $p(a,b|x,y)=\text{tr}[\sigma_{a|x}B_{b|y}]$ with $\{B_{b|y}\}_{b,y}$ being POVMs on Bob's side, gives a lower bound for the steering robustness of the assemblage. In other words
\begin{align}
{\rm IR}(A_{a|x})\geq {\rm SR}(\sigma_{a|x})\geq {\rm NR}[p(a,b|x,y)].
\end{align}
Hence, using the machinery of \cite{cavalcanti16,chen16b} one finds one-sided device-independent and fully device-independent lower bounds for quantifiers of measurement incompatibility and device-independent lower bounds for quantifiers of steering.

%As a further application of steering we discuss self-testing of quantum devices and their properties \cite{supic16,gheorghiu17,chen16b,cavalcanti16}.

%% file: Applications/Secret-sharing.tex
\label{sec:secret_sharing}

Secret sharing is a cryptography protocol that allows a dealer (Alice) to send a message to players (Bob and Charlie) in a way such that the message can only be decoded if when the players work together---neither of them can decode it by himself. If Alice shared a secret key (see Section~\ref{sec-qkd}) with Bob and another with Charlie, she can simply encode the message twice with the two keys to ensure that only Bob and Charlie together can decode the message. Thus,  normal quantum key distribution protocols already provide one with protocols for quantum secret sharing. But one can do it more straightforwardly with multipartite entanglement~\cite{hillery99}; see also~\cite{Zukowski1997a} for a related protocol. Take the case where Alice prepares a large number of the Greenberger-Horne-Zeilinger (GHZ) states~\cite{hillery99},
\begin{equation}
\ket{\mathrm{GHZ}}=\frac{1}{\sqrt{2}}(\ket{000}+\ket{111}).
\end{equation}
Alice keeps one particle, and sends the other two to Bob and Charlie. Each then measures their particles in random directions, $x$ or $y$. After communicating via a classical public channel, they can identify the triplets where they have measured in directions $xxx$, $xyy$, $yxy$, $yyx$; the other triplets are discarded. As one can check, the GHZ state is an eigenstate of the retained measurement operators. In these triplets, Bob and Charlie can use their outcomes to predict Alice's outcomes.  However, the outcomes at Bob's or Charlie's sides separately are not enough to infer her outcomes. Thus, Alice can use the series of outcomes of her measurements to encode the message in her secret sharing protocol.

The fact that Bob and Charlie have to collaborate to infer the measurement outcomes at Alice's side resembles the distinction between the concepts of local and global steering in the multipartite steering scenarios~\cite{he13,xiang17}; see also Section~\ref{sec-multipartite-steering}.  
This similarity has been made precise in analyzing the security of secret sharing~\cite{he13,kogias17,xiang17}. Specifically,~\citet{xiang17} computed the secrete key-rate bound which guarantees unconditional security of the protocol against eavesdroppers and dishonest players~\cite{kogias17} for three-mode Gaussian states and found it to be essentially the quantification of the difference between collective steering and local steering from Bob and Charlie to Alice~\cite{kogias15a}. To our knowledge, whether this quantitative relation between quantum steering and quantum secret sharing extends beyond Gaussian states is at the moment unknown.

%% file: Applications/Quantum-teleportation.tex
\label{sec:quantum_teleprotation}

Steering shares a close conceptual similarity with state teleportation~\cite{bennett93}. We follow~\cite{cavalcanti17a} and consider the following abstract teleportation protocol. Alice and Bob share a bipartite quantum state $\varrho^{AB}$.  Charlie, perceived as the verifier, draws a pure state $\omega_x$ from a certain set of $\abs{x}$ states indexed by $x$, gives it to Alice and asks her to teleport it to Bob. Without knowing the state $\omega_x$, Alice makes a measurement with POVM elements $\{E_a^{CA}\}_a$ jointly on the received state and her particle that is entangled with Bob's system. Depending on Alice's outcome $a$, Bob's system is then `steered' to a conditional state,  
\begin{equation}
\varrho_{a}^B (\omega_x)= \frac{\tr_{CA} \left[(E^{CA}_a \otimes \openone^B)(\omega_x \otimes \varrho^{AB}) \right]}{p(a \vert \omega_x)},
\end{equation}
where the normalization $p(a\vert \omega_x)= \tr \left[(E^{CA}_a \otimes \openone^B)(\omega_x \otimes \varrho^{AB}) \right]$ is the probability for Alice to get outcome $a$ in her protocol given she received state $\omega_x$ from Charlie. Alice communicates her measurement outcome $a$ to Bob, who then makes some appropriate local unitary operation; by the end of the procedure, the state of his system is $U_a \varrho_{a \vert \omega_x}^B U^\dagger_a$. The design of Alice's measurement and Bob's local unitary operations is such that the final state at Bob's side resembles Charlie's original state $\omega_x$ as much as possible. The quality of the teleportation protocol can be assessed by the so-called average fidelity,
\begin{equation}
\bar{F}_{\mathrm{tel}} = \frac{1}{\abs{x}} \sum_{a,x} p(a \vert \omega_{x}) \tr [\omega_{x} U_a \varrho^B_{a} (\omega_{x}) U_a^\dagger].
\label{eq:teleportation_fidelity}
\end{equation}
If the teleportation is perfect, the average fidelity is $1$ for pure states $\omega_{x}$.

\begin{figure}
\includegraphics[width=0.3\textwidth]{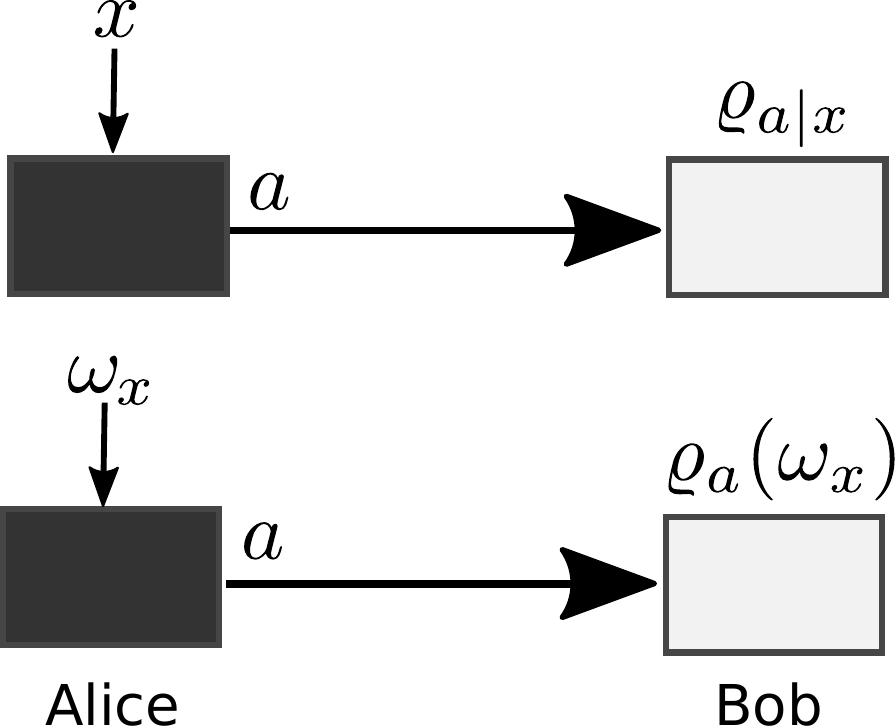}
\caption{Quantum steering and quantum teleportation. In quantum steering (top) Alice receives a classical input $x$, performs a measurement and communicates the output $a$ to Bob; Bob's system is steered to $\varrho_{a|x}$. In teleportation (bottom) Alice receives a quantum state $\omega_x$ as input, performs a measurement and communicates the output to Bob; Bob's system is steered to $\varrho_{a|x}$. Bob further makes a local unitary evolution on his system depending on the outcome he received to obtain the final state. }
\label{fig:teleportation}
\end{figure}

One can easily observe the similarity of the teleportation protocol with that of quantum steering: instead of receiving a classical input $x$, Alice received a quantum state from Charlie $\omega_x$ as an input; see Fig.~\ref{fig:teleportation}. \new{The idea of receiving quantum inputs from a verifier (Charlie) instead of a classical input has been previously considered for entanglement~\cite{buscemi12,branciard13}, and later for quantum steering both theoretically and experimentally~\cite{cavalcanti13a,kocsis15a}. The benefit of allowing for the quantum inputs is that the verifier can now verify that Alice and Bob share a quantum correlation (entanglement, quantum steering) without trusting their measurement devices or their actual measurements~\cite{hall16}; see also Section~\ref{sec-intro} and further discussion in~\cite{hall19a}.} Utilizing the similarity,~\citet{cavalcanti17a} showed that all entangled states can demonstrate nonclassical teleportation in certain sense. One should, however, note that the introduced notion of nonclassical teleportation does not imply high teleportation average fidelity, which has been a standard figure of merit.       

There is another line of works which attempt to relate quantum steering with the security of quantum teleportation. In teleporting a state to Bob, Alice does not want an eavesdropper to also obtain some version of the state. Certain security is guaranteed when the average fidelity of teleportation in Eq.~\eqref{eq:teleportation_fidelity} is high enough~\cite{pirandola15}. It is then shown that for certain family of bipartite states, to obtain the required fidelity, the state is not only entangled but necessarily two-way steerable~\cite{he15}. 

Another way to investigate the security of teleportation is to study its sister protocol known as entanglement swapping. In this protocol, the state given to Alice by Charlie is priorly entangled with another particle which Charlie keeps. By the end of the teleportation protocol performed by Alice and Bob, the entanglement between Charlie and Alice is transferred to that between Charlie and Bob. In this case, the teleportation can be secured by the monogamy of entanglement: if Charlie is sufficiently entangled with Bob, an eavesdropper cannot be entangled with Charlie. Instead of monogamy of entanglement,~\citet{reid13} then used the monogamy of a certain steering inequality to demonstrate the security of quantum teleportation.

%% file: Applications/Resource-theory.tex
%Here we will briefly comment the resource theory of steering~\cite{gallego15}.

A resource theory is typically seen as consisting of two basic components: free states and free operations. Free states constitute a set which remains unchanged under the actions of free operations. In this sense, one could define a resource theory merely from the free operations. Consequently, any state that is not free has some resource in it, as it can not be created from the set of free states with free operations. As an example, in the case of entanglement, free states are given by the set of separable states and free operations are local operations assisted by classical communication (LOCC). Another important aspect of a resource theory are resource measures or monotones. A proper measure should not increase under free operations, i.e. free operations can not create the resource, should be faithful, i.e. equal to zero only for free states, and should be convex, i.e. randomisation should not create resource either.

In the case of steering a resource theory has been proposed \cite{gallego15}. The free states in this theory are the unsteerable assemblages and free operations can be any operations on the assemblages that do not map unsteerable assemblages into steerable ones. In \cite{gallego15} one-way stochastic LOCC operations are shown to be free operations. In order to introduce one-way (stochastic) LOCC operations, we adapt the notation of \cite{gallego15} for state assemblages. Namely, any state ensemble $\{\varrho_a\}_a$ can be embedded into larger space via the correspondence $\{\hat\varrho_{a}\}_a:=\sum_a |a\rangle\langle a|\otimes\varrho_a$. To make a similar correspondence for state assemblages, one can define a map $\hat\varrho_{A|X}(x):=\sum_a |a\rangle\langle a|\otimes\varrho_{a|x}$, where $X$ and $A$ label the sets of Alice's inputs and outputs respectively. Now, a one-way (stochastic) LOCC operation $\mathcal M$ is defined on the assemblage $\hat\varrho_{A|X}$ as
\begin{align}
\mathcal M(\hat\varrho_{A|X}):=\sum_\omega (\openone \otimes K_\omega)\mathcal W_\omega(\hat\varrho_{A|X}) (\openone \otimes K^\dagger_\omega),
\end{align}
where $\{K_\omega\}_\omega$ are Kraus operators and $\{\mathcal W_\omega\}_\omega$ are wiring maps defined pointwise as
\begin{align}
\mathcal W_\omega(\hat\varrho_{A|X})(x_f):=&\sum_x p(x|x_f,\omega) \sum_{a_f,a} p(a_f|a,x,x_f,\omega)\nonumber\\
&(|a_f\rangle\langle a|\otimes\openone)\hat\varrho_{A|X}(x)|a\rangle\langle a_f|\otimes\openone.
\end{align}
Here $a_f$ and $x_f$ refer to the inputs and outputs of the final assemblage. Physically such transformations correspond to performing an operation on the characterized party, communicating the information about which operation (i.e. $\omega$) was performed, and the uncharacterized party applying the corresponding classical pre- and post-processings [i.e. $p(\cdot|x_f,\omega)$ and $p(\cdot|a,x,x_f,\omega)$] on their side. Note that the use of one-way (stochastic) LOCC operations as free operations has also an interesting physical motivation: they can be seen as safe operations in one-sided device-independent quantum key distribution \cite{gallego15}.

A typical resource theory aims at quantifying the resource in hand. For this purpose, one wishes to find a mapping (or monotone) $f$ from the set of states of the resource theory to the set of non-negative real numbers that fulfils certain requirements. In the case of steering the following requirements are considered \cite{gallego15}
\begin{itemize}
\item $f(\hat\varrho_{A|X})=0$ if and only if $\hat\varrho_{A|X}$ is unsteerable
\item $f$ is non-increasing on average under deterministic one-way LOCC
\end{itemize}
If in addition the mapping $f$ is convex, it is called a convex steering monotone. In \cite{gallego15} the typical steering quantifiers, i.e. steerable weight and robustness of steering (see Section \ref{sec-assemblages}), are shown to be convex steering monotones. Furthermore, the authors introduce a novel convex steering monotone called relative entropy of steering, see also \cite{kaur17a,kaur17b} for further monotones and alternative definitions of relative entropy of steering.

%% file: Applications/Post-quantum-steering.tex
Post-quantum steering is the phenomenon that certain assemblages $\{\varrho_{a|x}\}$
may not be realizable by quantum mechanics, although no signaling between the parties 
is possible. For the case of Bell inequalities, it is known that there are probability 
distributions which are non-signaling, but cannot come from a quantum state. The most 
prominent example is the Popescu-Rohrlich (PR) box \cite{popescu94}, which is a non-signaling 
distribution for two parties with two measurements having two outcomes, which leads 
to a violation of the CHSH inequality with a value 
$\langle \mathcal S_{\rm CHSH} \rangle=4$ while in quantum mechanics only values 
$\langle \mathcal S_{\rm CHSH} \rangle \leq 2 \sqrt{2}$ can occur. The analogous
question for steering highlights the difference between steering in the bipartite 
and the multipartite case.

For the bipartite case, one may consider an assemblage $\{\varrho_{a|x}\}$, 
obeying the no-signaling constraint $\sum_a \varrho_{a|x} = \sum_a \varrho_{a|x'} = \varrho_B$ 
for all $x,x'$. As already mentioned in Sections \ref{sec-assemblages} and \ref{sec-history}, any such 
assemblage can be 
realized by quantum mechanics. This means that there is a state $\varrho_{AB}$ 
and measurements $E_{a|x}$ such that $\varrho_{a|x} = \tr_A(E_{a|x} \varrho_{AB}).$

This is not the case for the tripartite scenario \cite{sainz15}. Here, one considers 
the scenario where Alice and Bob make local measurements in order to
steer Charlie's state. So, Charlie has an assemblage $\{\varrho_{ab|xy}\}$, where
$x$ and $a$ ($y$ and $b$) denote the measurement setting and outcome of Alice (Bob).
Besides being positive and the normalization constraint 
$\tr(\sum_{ab}\varrho_{ab|xy})=\tr(\varrho_C)=1$ this assemblage should fulfill 
that neither Alice nor Bob can signal to the other parties, that is
\begin{align}
\sum_a \varrho_{ab|xy} &= \sum_a \varrho_{ab|x'y} \mbox{ for all } x,x',
\nonumber
\\
\sum_b \varrho_{ab|xy} &= \sum_b \varrho_{ab|xy'} \mbox{ for all } y,y'.
\end{align}
One can directly check that these conditions imply that also Alice and Bob jointly
cannot signal to Charlie by the choice of their measurements.

Contrary to the bipartite case, an assemblage obeying these constraints does 
not need to have a quantum realization. A simple counterexample can be derived
from the PR box mentioned above: If the conditional states are of
the form $\varrho_{ab|xy} = p(ab|xy)\ketbra{0}{0}_C$, where $p(ab|xy)$ is 
the probability table of the PR box, the assemblage is clearly non-signaling, 
but cannot be realized within quantum mechanics. In \cite{sainz15} more 
interesting examples of this behavior were provided. Using iterations of SDPs
the authors found an example of a qutrit assemblage $\varrho_{ab|xy}$ with
the properties that for any possible measurement $E_{c|z}$ of Charlie, the resulting
probability distribution can be explained by a fully local hidden variable model, 
which is even a stronger requirement than being non-signaling. Still, the 
assemblage has no quantum realization, so there is no state $\varrho_{ABC}$ such 
that $\varrho_{ab|xy} = \tr_{AB}(E_{a|x} \otimes E_{b|y} \varrho_{ABC}).$

In further works the theory of post-quantum steering has been extended. 
\cite{sainz18} provided general methods to construct examples of post-quantum
steering and  defined a quantifier of this phenomenon, \cite{hoban18}
established a connection to the theory of quantum channels, and \cite{chenbudroniliang18}
used moment matrices to characterize the phenomenon.

%% file: Applications/History-steering.tex
%%%%%%%%%%%%%%%%%%%%%%%%%%%%%%%%%%%%%%%%%%%%%%%%%%%%%
\subsubsection{Discussions between Schrödinger and Einstein}
%%%%%%%%%%%%%%%%%%%%%%%%%%%%%%%%%%%%%%%%%%%%%%%%%%%%%%%%%%%%%

As mentioned already in the introduction, the first observation of 
the steering phenomenon dates back to Schrödinger's discussions 
of the EPR argument with Einstein. Schrödinger corresponded with 
several physicists on this problem, the letters have been edited
by \cite{meyennbook}.

To understand the origin of Schrödinger's idea, it is important
to note that Einstein did not like the way the EPR paper was written
and how the argument was formulated, for detailed discussions see 
\cite{kieferbook}. Instead, Einstein preferred a somehow simpler 
version. He published this version much later \cite{einstein48}, 
but he also explained the basic idea in a letter to Schrödinger 
on June 19th, 1935.

The argument (in the formulation of \cite{einstein48}) goes as follows. 
First, one considers position $X$ and momentum $P$ as non-commuting 
observables. Then there are, according to Einstein, two possibilities:

(i) One can assume that position and momentum have definite values 
before a measurement of them is carried out. Then one has to admit
that the wave function $\ket{\psi}$ is not a complete description.

(ii) One can assume that the values of position or momentum are 
created during a measurement. This is compatible with the assumption 
that the wave function $\ket{\psi}$ is a complete description. If 
$\ket{\psi}$ is a complete description, it follows, according to 
Einstein, that two different wave functions describe two different 
physical situations. 

These two ways of thinking cannot be distinguished without additional 
assumptions. Here, Einstein introduces a locality principle, stating 
that if one considers a bipartite system, the real physical situation 
at one side is independent of what happens on the other side.

In order to conclude the incompleteness of the quantum mechanical
description, one can consider a pure entangled wave function, as
in the usual EPR argument. Then, the conditional wave function 
$\ket{\phi}_B$ on Bob's side depends on the choice of the measurement 
on Alice's side. According to the locality principle, however, the 
physical reality on Bob's side cannot change. So, one arrives at a 
contradiction to (ii) and the incompleteness follows. It is interesting 
to note that for this argument the (perfect) correlations between 
measurements on both sides are not relevant. As Einstein formulated 
it: ``I couldn't care less whether or not $\ket{\phi}_B$ and 
$\ket{\bar \phi}_B$ are eigenstates of some observables.''

In the direct reply to this letter (on July 13th, 1935) 
Schrödinger spelled out that the dependence of the conditional 
state $\ket{\phi}_B$ includes some ``steering'' from a distance. 
Although this phenomenon does not allow signaling between the 
parties, he considers it to be magic. Recalling discussions with 
Einstein and colleagues in Berlin during the 1920s, he writes:

{``All the others told me that there is no incredible magic 
in the sense that the system in America gives $X=6$ if I perform 
in the European system {\it nothing} or a {\it certain} action 
(you see, we put emphasis on spatial separation), while it gives 
$X = 5$ if I perform {\it another} action; but I only repeated 
myself: It does not have to be {\it so} bad in order to be silly. 
I can, by maltreating the European system, steer the American system 
{\it deliberately} into a state where either $X$ is sharp, or into a 
state which is certainly {\it not} of this class, for example where 
$P$ is sharp. This is {\it also} magic!''}

It must be added, of course, that the view of the steering phenomenenon
as ``nonlocal'' is based on a certain interpretation of the wave function, 
which is not shared by everyone, see \cite{griffiths19} for a discussion.
In any case, the question remains for which states this phenomenon can be 
observed and which states on Bob's system can be reached by 
performing measurements on Alice's side. This was also part of the 
discussion between Schrödinger and other physicists (such as 
von Laue) and Schrödinger presented his results in two 
subsequent papers.

%%%%%%%%%%%%%%%%%%%%%%%%%%%%%%%%%%%%%%%%%%%%%%%%%%%%%%%%%5
\subsubsection{The two papers by Schrödinger}
%%%%%%%%%%%%%%%%%%%%%%%%%%%%%%%%%%%%%%%%%%%%%%%%%%%%%%%%%%

The first paper entitled ``Discussion of probability relations 
between separated systems'' \cite{schroedinger35discussion} was submitted 
in August 1935. Schrödinger states that he finds it ``rather 
discomforting'' that quantum mechanics allows a system to be 
steered by performing measurements in a different location. He
then presents several results on this phenomenon.

First, he shows that every bipartite pure state can be written
as
\be
\ket{\psi} = \sum_k s_k \ket{a_k}\ket{b_k}
\label{eq-erwinfirst}
\ee
where the vectors $\ket{a_k}$ and $\ket{b_k}$ form orthogonal
sets. This is nowadays called the Schmidt decomposition.
He proves that this is unique if the coefficients $s_k$ are 
different. He also recognizes that this implies that for 
generic states there is one measurement for Alice (defined by 
the eigenvectors $\ket{a_k}$) which is perfectly correlated 
with a measurement on Bob's side (defined by the  $\ket{b_k}$). 
 
Then, he discusses in some more detail the EPR state from the 
1935 argument \cite{epr}, which is not a generic state, 
as all the Schmidt coefficients coincide. He proves that for any 
observable $F(X_2, P_2)$ on Bob's side the value can be predicted 
by making a suitable measurement $\hat{F}(X_1, P_1)$ on Alice's 
side. This fact appeared already in a letter from Schrödinger to 
von Laue, and it demonstrates the surprising effect that one 
system seems to know the answers to all possible questions
on the other system.

The second paper entitled ``Probability relations between separated 
systems'' was submitted in April 1936 \cite{schrodinger36}. Schrödinger first states that
the essence of the previous work was the observation that in quantum 
mechanics one can not only determine the wave function at one party
by making measurement on the other, but one can also control the state
at one side by choosing the measurements on the other side. So the 
question arises, to which extent the wave function can be controlled. 

In order to answer this, he first proves a statement on density matrices.
A given density matrix may have different decompositions into pure states
\be
\varrho = \sum_k p_k \ketbra{\psi_k}{\psi_k} =  \sum_i q_i \ketbra{\phi_i}{\phi_i},
\label{eq-rhodecompositions}
\ee
and the question arises, which conditions the $\ket{\phi_i}$
and $\ket{\psi_k}$ have to fulfill. Schrödinger proves, that
two ensembles give the same density matrix, if and only if 
there is a unitary matrix $U$ such that
\be
\sqrt{p_k} \ket{\psi_k} = \sum_{i} U_{ki} \sqrt{q_i} \ket{\phi_i}.
\label{eq-sresult1}
\ee
This implies that any $\ket{\phi_i}$ in the range of the space 
spanned by the $\ket{\psi_k}$ can be an element of a suitable 
ensemble.

Then, Schrödinger applies this to the bipartite state in 
Eq.~(\ref{eq-erwinfirst}). Here, the reduced state 
\be
\varrho_B = \sum_k s_k^2 \ketbra{b_k}{b_k} {=}  \sum_i q_i \ketbra{\beta_i}{\beta_i}
\label{eq-sresult2}
\ee
has different decompositions. As already mentioned, the ensemble 
$\{s_k^2, \ket{b_k}\}$ can be reached by making the measurement 
defined by the orthogonal states $\ket{a_k}$ on Alice's side, and the question arises, 
whether any other ensemble $\{q_i, \ket{\beta_i}\}$ can be reached. 
Schrödinger proves that this is the case. Especially, if the state has 
full Schmidt rank and $\ket{b_k}$ span the whole space, {\it any} state 
$\ket{\beta_i}$ on Bob's side can be prepared by making a suitable measurement on 
Alice's side.

Finally, Schrödinger stresses again that he finds the phenomenon of 
controlling a distant state repugnant and suggests that quantum
mechanics may be modified to avoid it. As a potential modification, 
he suggests that for a state as in Eq.~(\ref{eq-erwinfirst}) the
phase relations between the $s_k$ may be lost. This means that 
instead of taking the pure state $\varrho=\ketbra{\psi}{\psi}$
the two-particle system should be described by a diagonal mixed 
state 
\be
\varrho = \sum_k s_k^2 \ketbra{a_k b_k}{a_k b_k},
\ee
which he considers to be a possible modification, not contradicting
the experimental evidence at that time.

%%%%%%%%%%%%%%%%%%%%%%%%%%%%%%%%%%%%%%%%%%%%%%%%%%%%%%%%%%%%%%%%%%%%%%%%%5
\subsubsection{Impact of these papers}
%%%%%%%%%%%%%%%%%%%%%%%%%%%%%%%%%%%%%%%%%%%%%%%%%%%%%%%%%%%%%%%%%%%%%%%%%%%
In the following years, Schrödinger's ideas on steering were not 
further considered in the literature. His mathematical results 
from the second paper, however, were several times rederived 
without any reference to him. In the following, we give a 
short overview, a detailed discussion can be found in 
\cite{kirkpatrick06}.

A first rediscovery was presented by Jaynes \cite{jaynes57}. He 
derived the first statement in Eq.~(\ref{eq-sresult1}) while 
studying general properties of density matrices. Based on 
Jaynes' paper, Hadjisavvas presented later a simplified proof 
and an extension to infinite-dimensional systems \cite{hadjisavvas81}. 

The second mathematical statement (below Eq.~(\ref{eq-sresult2}))
was derived by Gisin in the context of modifications of the 
Schrödinger dynamics \cite{gisin89}. Here, the question
arises whether the modified dynamics for pure states extends 
uniquely to mixed states. If it were different for two 
decompositions such as the ones in Eq.~(\ref{eq-rhodecompositions}), 
then ensembles will become distinguishable at some point. Given
the fact that both ensembles can be prepared by measurements on
a distant system, this would lead to a violation of the non-signaling
condition, enforced by special relativity. Finally, both mathematical
statements from Schrödinger have also been rederived independently by
\cite{hughston93}.

Besides these mathematical results, the notion of steering as a kind
of quantum correlation was not discussed for a long time. The situation
changed in the eighties of the last century. Then, Bell inequalities 
started to attract more attention \cite{clauser78} and the mathematical notions of 
entangled and separable states were studied \cite{primas83, werner84, 
werner89}. 

The paper \cite{vujicic88} was the first to give a clear
summary of Schrödinger's ideas and an extension of his results 
for continuous variable systems. Also, Vuji{\v{c}}ić and Herbut 
argue that steering is different from Bell nonlocality, as it is 
based on the formalism of quantum mechanics. Independently, 
\citet{reid89} presented quantitative conditions for continuous 
variable systems to lead to an EPR-type argument. Verstraete 
noted in his dissertation \cite{verstraete02a} the connection 
of Schrödinger's ideas to quantum teleportation and entanglement 
transformations, as in both cases one aims at preparing a quantum 
state on one side by making measurements on the other. Also, the 
notion of the steering ellipsoid was introduced there. Shortly 
thereafter, steering was recognized to be relevant for foundational
questions of quantum mechanics \cite{spekkens07, clifton03}.
Finally, Wiseman
and coworkers introduced the notion of local hidden state models 
\cite{wiseman07}, laying the foundation for the modern notion of 
quantum steering.

%% file: Conclusion/Conclusion.tex
The notion of quantum steering is motivated by the Einstein-Podolsky-Rosen 
argument and it took seven decades until a precise formulation was given. Since 
then, quantum steering has initiated a new surge of results in quantum 
information and the foundations of quantum mechanics: Old concepts were 
put into the new light; long-standing problems gained progress and some 
were resolved; connections between areas were established, and novel 
problems were formulated. In this review, we have sketched the dynamic 
development of the field over the last ten years. Yet, future research 
is facing many challenges. So, to close the review, we summarize
some of the open problems:

\begin{itemize}

\item 
As a complete characterization of quantum steerability has been obtained for 
two-qubit systems and projective measurements, it is desirable to extend such 
a characterization to higher-dimensional systems. Although there is an indication 
that such an extension is possible, much remains to be worked out in details.

\item
The question, whether there are states that are unsteerable with projective measurements, but are steerable 
with POVMs is also relevant. The analogous question in the context of Bell nonlocality 
has been a long-standing problem without any evidence whether such a state exists. With 
quantum steering, one now has evidence indicating that there might be no such state 
for a two-qubit system. Yet, to date there is no rigorous proof of their non-existence. 
In particular, one might still expect that such a state exists in high dimensions.   

\item 
We have discussed in Section~\ref{sec-superactivation} that the operational definition 
of steerability requires multiple copies of the considered state, which implies the 
possibility of making collective measurements on Alice's side. Thus, apart from being 
fundamental, the question whether all entangled states become steerable upon making 
collective measurements on Alice's side is also important to the intrinsic 
consistency of the concept.

\item
The connection between quantum steering and incompatibility has initiated 
further questions: are there other connections between the different notions 
of incompatibility and different forms of quantum correlations?

\item
A further open question asks whether there are other physically motivated properties 
of state assemblages than that of having a local hidden state model. For instance, 
one may want to deduce something more than entanglement and incompatibility from 
such properties. Examples are the preparability from states with a given Schmidt number 
or properties motivated by incompatibility, such as compatibility on many copies, 
coexistence or simulability.

\item
The study of multipartite steering is in its infancy. A systematic investigation and 
comparison between different definitions is necessary in the future.  

\item
A closely related research direction is to study steering of 
parties who are connected in networks. For a given directed 
network one may ask whether there is a quantum state allowing
steering along the directed edges.

\item
\new{For applications, it would be interesting to identify tasks 
in quantum information processing, where the assumptions that can
be made are highly asymmetric. Then, the methods developed in
steering theory may be useful to study the role of correlations therein.
}

\item
In experiments, quantum steering is verified by a finite number of measurement 
settings. There are only few works on optimizing the measurement settings (when 
having a fixed number of inputs/outputs) and clearly more research is needed 
to serve as inputs for experiments.

\end{itemize}

This is only a small list of problems, and clearly further interesting challenges 
remain. Given the current interest in the steering phenomenon, we expect that the
old observations from Schr\"odinger can still influence current and future 
discussions on the foundations of quantum mechanics. 

We thank 
R.~M.~Angelo,
N.~Brunner,
C.~Budroni, 
S.-L. Chen,
\new{B.~Coyle,}
S.~Designolle, 
\new{U.~R.~Fischer,}
O.~Gittsovich, 
\new{R.~B.~Griffiths,}
E.~Haapasalo,
M.~J.~W.~Hall,
T.~Heinosaari,
F.~Hirsch, 
\new{Y.~Huang,}
M.~Huber, 
S.~Jevtic, 
J.~Kiukas,
T.~Kraft,
\new{L.~Lami,}
F.~Lever,
\new{Y.-C.~Liang,}
K.~Luoma, 
A.~Milne, 
T.~Moroder, 
H.-V.~Nguyen, 
J.-P.~Pellonp\"a\"a, 
M.~Piani, 
\new{T.~Pramanik,}
\new{Z.~Qin,}
M.~T.~Quintino, 
J.~Shang,
\new{G.~T\'oth,}
\new{F.~Verstraete,}
T.~Vu,  
\new{M.~M.~Wilde,}
H.~M.~Wiseman,
\new{J.-S.~Xu,}
and
X.-D.~Yu
for discussions and collaborations on the topic of steering 
or {comments on the manuscript. In addition, we thank the 
anonymous referees for their help.}

This work has been supported by the DFG, the ERC (Consolidator 
Grant 683107/TempoQ), the Finnish Cultural Foundation, the CNPq-Brazil (process number: 153436/2018-2), and the CAPES-Brazil.